\documentclass{article}

\usepackage[english]{babel}

\usepackage[letterpaper,top=2cm,bottom=2cm,left=3cm,right=3cm,marginparwidth=1.75cm]{geometry}

\usepackage{amsmath}
\usepackage[symbol]{footmisc}
\usepackage{amsfonts} 
\usepackage{graphicx}
\usepackage{subcaption}
\usepackage{multirow}
\usepackage{hhline}
\usepackage{bm}
\usepackage{tikz}
\usepackage{makecell}
\usetikzlibrary{tikzmark, shapes.arrows}
\usepackage[colorlinks=true, allcolors=blue]{hyperref}
\usepackage{caption} 
\newcolumntype{M}{>{\linespread{1}\selectfont\centering}m{2cm}}
\newcolumntype{Q}{>{\linespread{1}\selectfont\centering}m{2.5cm}}
\newcolumntype{P}[1]{>{\centering\arraybackslash}p{#1}}
\newcolumntype{C}[1]{>{\centering\arraybackslash}p{#1}}
\usepackage{natbib}
\usepackage{titlesec}
\usepackage{paralist}
\usepackage{multicol}
\usepackage{dirtytalk}
\usepackage{enumitem}
\usepackage{booktabs} 

\usepackage[colorinlistoftodos,textwidth=1.8cm]{todonotes}

\usepackage{soul}

\titleformat{\paragraph}
{\normalfont\normalsize\bfseries}{\theparagraph}{1em}{}
\titlespacing*{\paragraph}
{0pt}{3.25ex plus 1ex minus .2ex}{1.5ex plus .2ex}

\title{Bayesian Handwriting Evidence Evaluation using MANOVA via Fourier-Based Extracted Features}
\date{} % clear date

\author{\textbf {Lampis Tzai}\\
School of Criminal Justice,
    University of Lausanne, Switzerland and \\
    Department of Statistics Athens University of Economics and Business, Greece \\
    Email: lampis.tzai@unil.ch
    \hspace{2mm}\\
    \hspace{2mm}\\
    \and \textbf {Ioannis Ntzoufras}\\
    Department of Statistics Athens University of Economics and Business, Greece\\
    E-mail: ntzoufras@aueb.gr
    \hspace{2mm}\\
    \hspace{2mm}\\
    \and \textbf {Silvia Bozza} \\
    Department of Economics, Ca' Foscari University of Venice, Italy and\\
    School of Criminal Justice, University of Lausanne, Switzerland\\
        Email:  silvia.bozza@unive.it}

\begin{document}
\maketitle

\begin{abstract}

This paper proposes a novel statistical approach that aims at the identification of valid and useful patterns in handwriting examination via Bayesian modeling. Starting from a sample of characters selected among 13 French native writers, an accurate loop reconstruction can be achieved through Fourier analysis. 
The contour shape of handwritten characters can be described by the first four pairs of Fourier coefficients and by the surface size. Six Bayesian models are considered for such handwritten features. These models arise from two likelihood structures: (a) a multivariate Normal model, and (b) a MANOVA model that accounts for character-level variability. For each likelihood, three different prior formulations are examined, resulting in distinct Bayesian models: (i) a conjugate Normal-Inverse-Wishart prior, (ii) a hierarchical Normal-Inverse-Wishart prior as proposed by \citet{bozza2008probabilistic}, and (iii) a Normal-LogNormal-LKJ prior specification. 
The hierarchical prior formulations are of primary interest because they can incorporate the between-writers variability, a distinguishing element that sets writers apart. 
These approaches do not allow calculation of the marginal likelihood in a closed-form expression. Therefore, bridge sampling is used to estimate it.
 The Bayes factor is estimated to compare the performance of the proposed models and to evaluate their efficiency for discriminating purposes. Bayesian MANOVA with Normal-LogNormal-LKJ prior showed an overall better performance, in terms of discriminatory capacity and model fitting. Finally, a sensitivity analysis for the elicitation of the prior distribution parameters is performed.

\end{abstract}
\textbf{Keywords} :
\begin{inparaenum}
\item Handwriting Evidence
\item Forensic Science
\item Fourier Analysis
%\item Normal - Inverse - Wishart
\item Bayesian Hierarchical Modeling
\item Conjugate Analysis
\item Bridge sampling
\item Bayes Factor
\item Sensitivity
\end{inparaenum}

\newpage
\section{Introduction}

Handwriting examination is a domain of forensic science in which document examiners are often asked to inform the actors of a legal process who are confronted with handwritten documents whose origin is contested or unknown. The evaluation of handwriting evidence is still an open problem in forensic science. Although numerous studies have been conducted to support the individualization process, it still depends to a large extent on experts assessing handwriting characteristics qualitatively and subjectively. This is so even though much thoughtful judicial literature has pointed out that current categorical conclusions (statements of certainty) based exclusively and without reference to contextual information on so-called evidential material, cannot be justified on logical grounds \citep{Aitkenetal2021,izenman2020comparing}.

%\subsection{State of the art}
%\label{literature_review}

A number of recent studies have explored novel methods for inspecting handwritten documents and assessing the evidence on writership problems. \cite{johnson2022handwriting} developed a method that quantifies the similarity between handwritten documents using machine learning and statistical techniques. They applied score-based likelihood ratios (SLRs) to assess the value of the evidence in two scenarios: common source and specific source. \cite{wydra2022likelihood} proposed a method that evaluates handwriting evidence using the likelihood ratio approach. They measured the similarity between handwriting samples by the Jaccard index, which is a statistic that captures the overlap between two sets of features. \cite{crawford2023rotation}  proposed a statistical model to compute the probability of authorship when the writer of a questioned document is part of a closed set of writers. They utilized a rotation-based feature to extract measurements from handwritten documents, and a Bayesian hierarchical model to estimate the posterior predictive probability of authorship.

This paper investigates and extends the Bayesian probabilistic approach proposed by \cite{bozza2008probabilistic}. In that paper, specific handwritten features originating from an ad-hoc image analysis procedure proposed by \cite{marquis2005quantification} were used. This technique is based on Fourier analysis and enables a precise reconstruction of the contour shape of characters' loops. Following this methodology, each character loop can be described by means of  Fourier coefficients, which can be used to characterize shape complexity and other geometric attributes. Preliminary studies by \citet{marquis2006quantitative} have shown that this feature characterization has good discriminating power. The value of the evidence is subsequently assessed by means of the Bayes factor (BF for short), which can be interpreted as a measure of the strength of support provided by the evidence for the competing hypotheses put forward by opposing parties in legal proceedings. The use of the Bayes factor as a metric to assess the probative value of forensic findings is largely supported by operational standards and recommendations in different forensic disciplines (see, for example, the guidelines of the European Network of Forensic Science Institutes \citep{Willisetal2015}). A general review of the use of the Bayes factor for forensic decision analysis from an operational perspective is given by \citet{Bozzaetal2022}.

In particular, \cite{bozza2008probabilistic} proposed a hierarchical two-level random effects model, which has the advantage of modeling both within-writer and between-writers variability. 
%This model was used to deal with different investigative purposes (see, e.g. \cite{taroni2012use, taroni2014bayes}). 
However, as the authors themselves pointed out, the main limitation of this proposal is that it does not model the variability that characterizes each type of character. 
Consequently, the model can be applied either without distinguishing between different character types, thereby losing an important source of information, or separately for each character type.
This approach can be problematic, particularly when attempting to combine conflicting evidence (i.e., evidence supporting different hypotheses; for example, one character may support \( H_1 \), while another supports \( H_2 \)). To illustrate, measurements associated with one character (e.g., {\em a}) might support the hypothesis that the compared materials originate from the same writer, whereas measurements from a different character (e.g., {\em d}) might support the opposing hypothesis that the materials come from different writers.

To reduce the impact of this model constraint, a Bayesian Multivariate Analysis of Variance (MANOVA)  model is proposed in this work, implemented using the loop characters as predictors. The indicator of the loop character is transformed into a dummy variable (corner-point representation, see Appendix \ref{dummy_variables}), so that it is possible to model variables describing the handwriting characters of different types jointly, taking into account the variability between characters, the variability between-writers for every character, and within-writer variability. 

Another open issue highlighted by the authors concerns the sensitivity of the Bayes factor to the elicitation of the prior distribution, as well as to the specification of the prior distribution modeling within-writer handwriting variability.

The paper is structured as follows. The available data, and in particular the Fourier-based image analysis procedure, are described in Section~\ref{data}. The scenario of interest, as well as the notion of evaluation of forensic findings, the definition of BF, and, in particular, its use in forensic science, are introduced in Section~\ref{sn_methodologies}. The compared probabilistic models are illustrated in Sections~\ref{normal-inverse-wishart} and~\ref{bayesian_manova}, respectively. Three prior specification setups are investigated, starting from a conjugate approach (see Sections~\ref{normal-inverse-wishart-conjugate} and~\ref{bayesian_manova_conjugate}). 
 The main limitation of this approach is that it does not allow for the separate modeling of within-writer and between-writers variability. Its key advantage, however, is that the marginal distributions required for Bayes factor computation can be obtained analytically.
Two additional hierarchical approaches are proposed. These approaches enable separate modeling of both sources of variability; see Sections~\ref{normal-inverse-wishart-2level}, ~\ref{normal-LogNormal-LKJ} and~\ref{bayesian_manova_2level},~\ref{MANOVA-normal-LogNormal-LKJ}. 
Since, in contrast to the conjugate approach, the marginal likelihoods are not analytically available, bridge sampling is used to estimate the required marginal likelihoods. 
Experimental results are presented in Section~\ref{experimental_results}, with emphasis on model comparison (Section~\ref{comparison_of_models}) and on the evaluation of each model's efficiency in supporting the correct hypothesis; see Section~\ref{models_efficiency}. 
Finally, in Section~\ref{sensitivity_analysis_prior}, a sensitivity analysis is conducted with respect to the elicitation of the prior distribution parameters. The analysis also examines the sensitivity in the specification of the prior parameters for the Inverse-Wishart degrees of freedom and the $\eta$ parameter of the LKJ distribution, both of which model within-writer handwriting variability. 
Section~\ref{conclusions} concludes the paper by presenting the main conclusions, discussing the main findings of the paper, and highlighting directions for future work.

\section{Engineering and Structure of Handwriting Data}
\label{data}

The data used for this study refer to a sample of 13 writers collected for a previous study \citep {marquis2006quantitative}, who were selected from a population of native French writers from the School of Criminal Justice of the University of Lausanne (Switzerland) because of their habit of closing loops. The contour shape of characters {\em a}, {\em d}, {\em o} and {\em q} was processed according to the image analysis procedure proposed by \citet{schmittbuhl1998shape} for the polymorphism analysis of the piriform aperture of given skeleton bones of some primates (i.e., the mandible) and adapted by \cite{marquis2005quantification} for handwriting examination purposes. 

This approach has a practical limitation, as it cannot be directly applied to writers who do not complete or fully form their loop characters. 
Nevertheless, techniques such as interpolation and filtering can facilitate the reconstruction of open or incomplete loop characters \citep{thiel2011elasticurves}. However, further research is required to establish their efficacy and to optimize the application of these techniques. Another issue is that the proposed method is language-dependent, since looped characters differ substantially across languages—for instance, between Roman, Chinese, or Slavic scripts. 
%For language-independent handwriting features, see \cite{miller2017set}. % we do not agree. 
Finally, the identification and labeling of looped characters in this dataset were performed manually, based on the handwriting expert. 
As a result, some evaluator-induced errors may be present; however, these are assumed to be negligible given the evaluator’s expertise; see \cite{marquis2006quantitative} for details.

\subsection{Fourier Coefficients as Loop Character Features}
\label{fourie_harmonics}
\begin{figure}[!ht]
    \centering
    \includegraphics[width=\textwidth, height = 4cm]{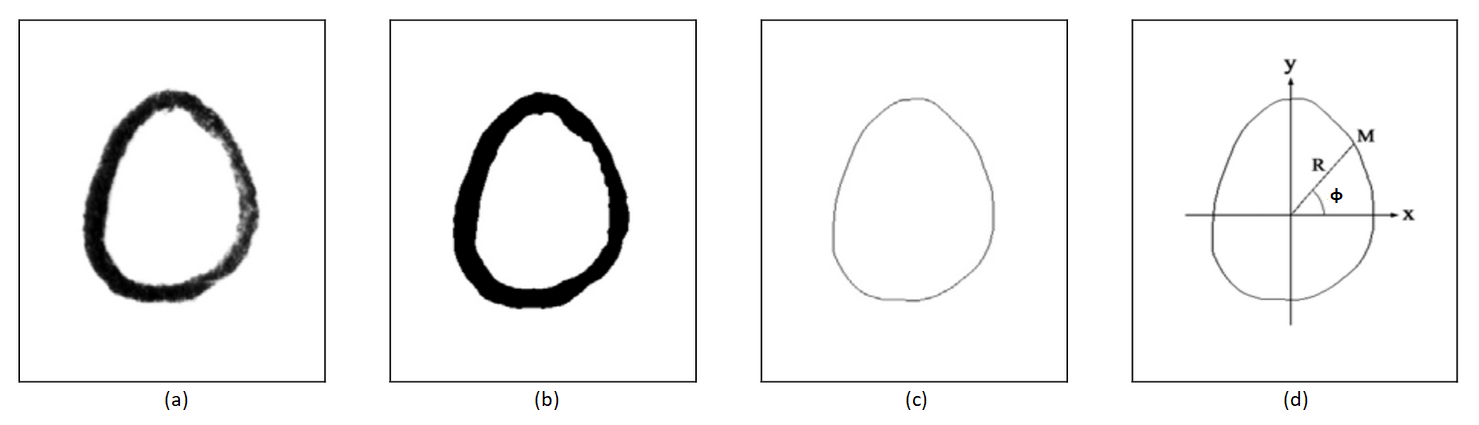}
    \caption{Image analysis procedure: from the original loop to polar coordinates. Adapted from \citet{marquis2005quantification}. }
    \label{loop_to_polar}
\end{figure}

The image analysis procedure, which is sketched in Figure~\ref{loop_to_polar}, can be summarized as follows:

\begin{enumerate}[label={(\alph*)}]
    \item A character is digitized into an image; 
    \item The image is binarized;
    \item The skeleton of the image is obtained to extract the contour;
    \item The contours are expressed in polar coordinates.
\end{enumerate}

%\textcolor{red}{In this way}, each character \textcolor{red}{loop} can be described by a discrete function R($\theta$) representing the length of a line joining a point of the contour to the centroid, where $\theta$ is the angle made by this line with the horizontal axis\textcolor{red}{, as it will be detailed in Section~\ref{fourie_harmonics}.}
%,  as represented in Figure \ref{loop_to_polar}. 
Following this procedure, the contour shape of each handwritten character was reconstructed by means of 128 pairs of polar coordinates. Note that, to eliminate the influence of size on shape analysis, the contour was normalized. All contour coordinates were adjusted so that the areas enclosed by them were equal to $1\,cm^2$. Consequently, the surface size measurement (S) was extracted prior to normalization as an additional general feature of interest, measuring the approximated area  (using, e.g., the trapezoidal rule or Simpon's rule) of each loop character.

%Finally, the size of the surface (S) of each character loop was measured as an additional feature. 

%\subsection{Fourier Coefficients as Loop Character Features}
%\label{fourie_harmonics}

%fourier analysis
Starting from the polar coordinates, each loop character can be reconstructed by means of a Fourier series:

\begin{equation}
    R(\phi) = a_0 + \sum_{h=1}^H[a_h cos(h\phi) + b_h sin(h\phi)], \qquad \qquad\ \phi \in [0,2\pi).
    \label{sine_cosine_form}
\end{equation}
%In this formula, 
In particular, each character loop can be described by a discrete function R($\phi$) representing the length of a line joining a point of the contour to the centroid, where $\phi$ is the angle made by this line with the horizontal axis.
%,  as represented in Figure \ref{loop_to_polar}. 
The contour shape can therefore be described by a series of harmonics ($h=1, \dots, H$), each characterizing a specific contribution to the shape. Each harmonic is constructed by a pair of Fourier coefficients, i.e., $a_h$ and $b_h$ ($\in R$). 
%However, the first analyses of this approach implemented
Note that the original proposal by \cite{marquis2005quantification} envisaged the amplitude-phase form of the Fourier series:
\begin{equation}
    R(\phi) = A_0 + \sum_{h=1}^H[A_hcos(h\phi - \phi_h)], \qquad \qquad\ \phi \in [0,2\pi)
    \label{amplitude_phase_form}
\end{equation}
%
%which can be shown that 
 where the amplitude $A_h$ ($\in R^+$) and the phase $\phi_h$ (degrees or radians) are just the polar coordinates of the coefficients $a_h$ and $b_h$, and 
 %the computed quantities are 
 $ a_h = A_h cos(\phi_h)$, $b_h = A_h sin(\phi_h)$ and  $A_0 = a_0$. 
 %This variant was necessary for this study because we apply probabilistic modeling to Fourier harmonics. However, such 
 Fourier descriptors ($A_h$ and $\phi_h$), however, take non-negative values and cannot be modeled directly with a multivariate Gaussian distribution, which is chosen as a probabilistic model in this work. Furthermore, the Fourier coefficients ($a_h$ and $b_h$) are not only defined over the whole set of the real numbers, but also have better Mardia's skewness and kurtosis values per writer than the Fourier descriptors \citep{mardia1970measures}. This is true even if logarithmic or square root transformations of the Fourier descriptors are implemented. We therefore proceed with our analysis using Fourier coefficients ($a_h$ and $b_h$).

\begin{figure}[!ht]
    \centering
\includegraphics[width=8cm, height=5cm]{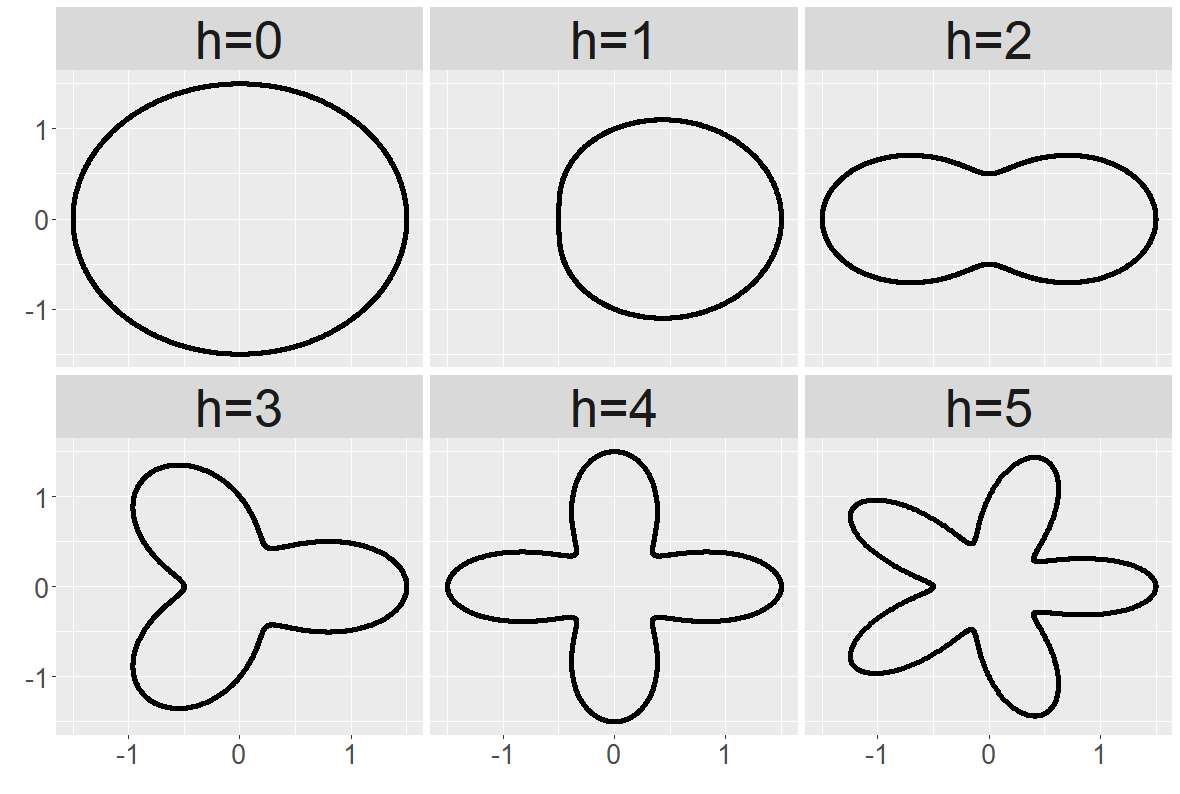}

\caption{Harmonics contribution obtained by the sum of the unit circle (h=0) and the specific harmonics of interest, $\alpha_h = 0.5$ and  $b_h = 0$}. 
\label{harmonic_contribution}
\end{figure}

The harmonic contribution to the contour shape of characters is represented in Figure \ref{harmonic_contribution}: 
%for  harmonics contribution presentation, 
the first harmonic ($h=1$) informs about the ovate contribution to the shape, the second ($h=2$) about the ellipticity of the shape, the third ($h=3$) about the triangularity, the fourth ($h=4$) about the quadrangularity, and the fifth ($h=5$) about the pentagonality.
A more detailed illustration of this image analysis procedure can be found in \cite{schmittbuhl1998shape} and \cite{marquis2005quantification}.

\begin{figure}[!ht]
\centering
\includegraphics[width=0.96\textwidth ]{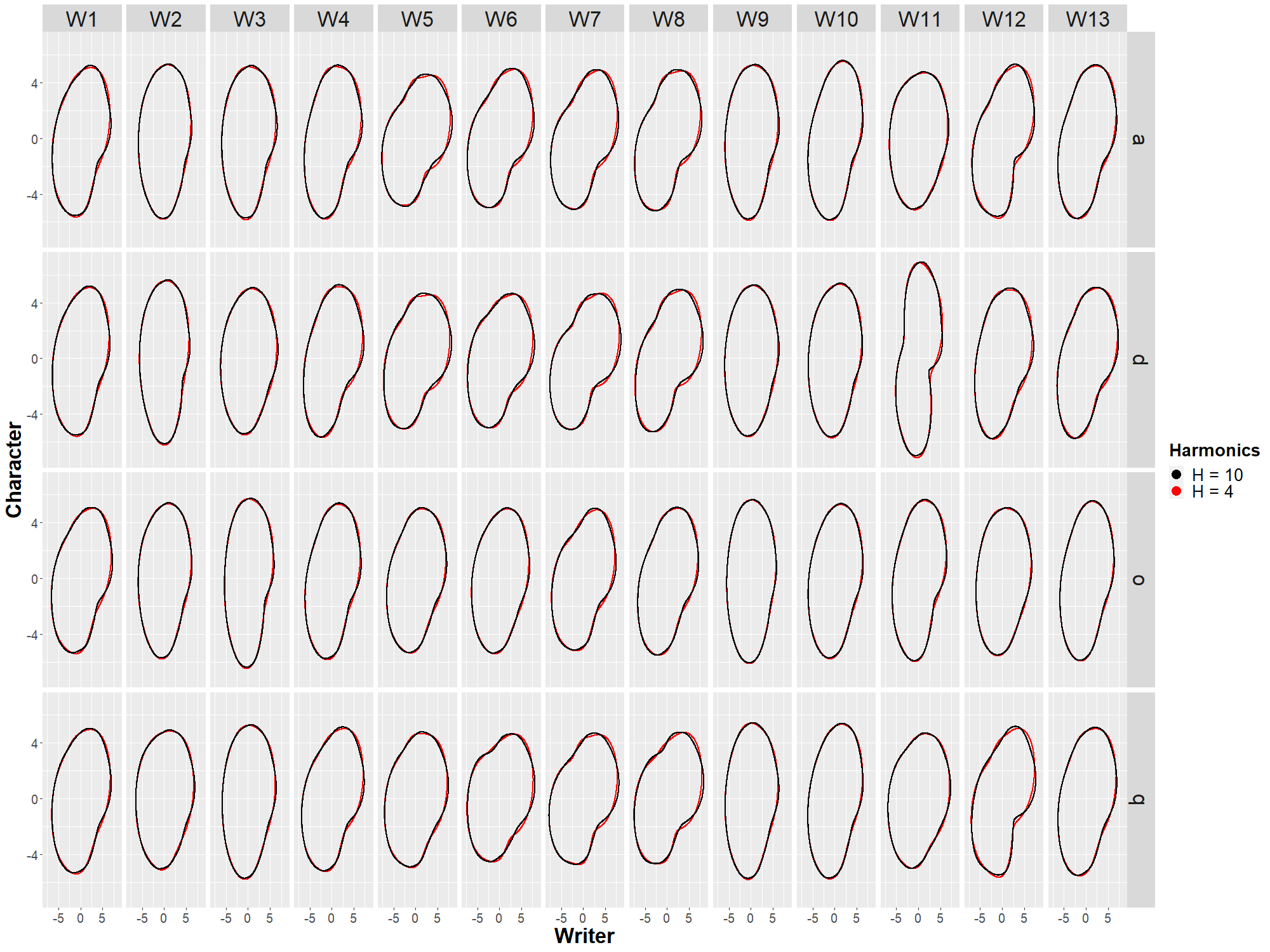}\\
\begin{minipage}{0.9\textwidth}
{\scriptsize \it All characters were normalized to 1 $cm^2$ of surface size. 
%The original surface size (S) is not depicted. 
H refers to the number of harmonics.} 
\end{minipage} \\ 
\caption{Reconstructed loop characters shown for the average Fourier coefficients per writer and character using the first four pairs of Fourier coefficients (H = 4; red curve) and the first ten pairs (H = 10; black curve).}
\label{fourier_reconstruction}
\end{figure}

%images per character
%By reconstructing the Fourier series, the real contour of the loop character can be simulated. In 
The reconstruction of the original contours obtained by means of such a Fourier-based procedure is represented in
Figure \ref{fourier_reconstruction}, for each writer and each character. 
For each analysed loop, the average of the Fourier parameters is considered, with a total number of $H=4$ and $H=10$ harmonics. %Fourier descriptors. 
The final image represents the average contour of loop characters. As can be observed, while some writers present loop structures characterized by marked peculiarities, others are more similar. This illustration does not account for within-writer variability, which will be addressed in the data modeling section (see Section \ref{modeling}).
The size is not illustrated because the original image had been normalized.

%important fourier descriptors
In this paper, only the first four harmonics %($h = 1,..,4$) 
were 
%considered 
retained, as suggested by \cite{marquis2006quantitative}.
%, \cite{bozza2008probabilistic}
The global shape of each character contour can, in fact, be reconstructed without the contribution of further harmonics, as can be observed in Figure \ref{fourier_reconstruction}. Furthermore, it must be underlined that the constant term $\alpha_0$ (or $A_0$ in the amplitude-phase form) 
%is not influential for 
does not have an impact on the morphological characteristics of the shape of the characters under study, as discussed by \cite{marquis2006quantitative}. Therefore, this term is considered a nuisance parameter and is not further considered in the subsequent analyses. Thus, each character loop can be described by means of $p = 9$  variables representing the surface size (S) of the character, the $\alpha_h$ and the $b_h$ of the first four harmonics, $h=1,\dots,4$ ($H = 4$). Hence, the available  data contain the measurements of $L=4$ %different types of character loops 
characters (i.e., {\em a}, {\em d}, {\em o} and {\em q}) collected in correspondence of $m=13$ writers, and 
%$m = 13$ writers, $L = 4$ characters, $n_l$
%\footnote{The number of each character that is available for each writer is not constant.} 
 %\textcolor{red}{each character, $l=1,\dots,L$}, and $p = 9$ variables and are 
 can be denoted by a four-dimensional array 
 ${\bm{\mathcal{D}}}$ with elements 
 ${\cal D}_{i\ell j \kappa}$, where $i = 1,\dots,m$ (writers), $\ell = 1,\dots,L$ (characters), $j = 1,\dots,n_{i\ell}$ (repetition), 
 and $\kappa=1, \dots ,p$ (Fourier coefficients and the surface size). 
 %Finally, the values of the coefficients are very small; they 
 Note that Fourier coefficients and the surface size have been standardized by dividing each value by the overall standard deviation of each Fourier coefficient.  For a detailed, comprehensive analysis of the features of the available characters, please refer to Appendix \ref{descriptive_analysis}.

%A closer examination of the Fourier coefficients reveals that the second harmonic, which quantifies ellipticity, is the most informative measurement for discriminative purposes. Furthermore, while some writers exhibit notable similarities, reflected in smaller distances (e.g., writers 6, 7, and 8), others display more distinct characteristics, resulting in larger distances (e.g., writers 10 and 13). These pronounced peculiarities are expected to facilitate discrimination. 

%--------------------------------------------------------------------

%\section{Methodology}\label{sn_methodologies}
%\section{Bayesian testing for handwriting authentication}
\section{Bayesian Probabilistic Approach for Handwriting Evidence Evaluation}
\label{sn_methodologies}
\label{handwriting_authentication}

Forensic document examiners frequently need to cope with situations involving handwritten documents whose writership is questioned. Consider the following scenario involving a handwritten document whose origin is contested. Written material originating from an individual who is suspected to be the actual source of the disputed item is collected and examined for comparative purposes. The available evidence should be evaluated under a set of propositions within a given framework of information put forward by opposing parties (i.e., the prosecutor and the defense attorneys, respectively) at trial. The propositions of interest can therefore be formulated as follows:
%Let us now consider the following propositions:
%that are of prime interest for examining documents:
 \begin{itemize}
\item[$H_1$:] the person of interest (PoI) is the writer of the questioned document;
\item[$H_2$:] the person of interest (PoI) is not the writer of the questioned document.
\end{itemize}
%For comparison reasons, written material from the individual who is considered to be the actual source of the disputed item is chosen.
%Thus, the key issue 
%in this case is whether the suspect wrote the handwritten evidence. 
%\textcolor{red}{is to inform about the value of the evidence in support of the competing propositions.}
%Thus, it is crucial to provide information about the value of the evidence \textcolor{red}{(i.e., questioned and control material)} in support of the \textcolor{red}{competing} propositions. 

First of all, $n_1>0$ characters are selected from the anonymous manuscript document; these are referred to as questioned material. Then, $n_2>0$ characters are selected from the handwritten material originating from the person of interest; these are referred to as control material or reference material. Measurements (i.e., Fourier coefficients and the surface size) made on questioned and control material are denoted by: $\bm{y}_{w\bullet\bullet\bullet} = (\bm{y}_{w\ell j\bullet},w=1,2,\;\ell=1,\dots, L,\;j=1,\dots,n_{w\ell})$, with $\bm{y}_1$ denoting the measurements made on the questioned material and $\bm{y}_2$ the measurements made on the control material. Thus, it is crucial to provide information about the value of the evidence $(\bm{y}_1,\bm{y}_2)$ in support of the competing propositions. This is given by the ratio of the marginal likelihoods under the competing propositions:

\begin{equation}\label{eq_BF}
     BF=\frac{m(\bm{y}_1,\bm{y}_2|H_1)}{m(\bm{y}_1,\bm{y}_2|H_2)} 
    = \frac{m(\bm{y}_1,\bm{y}_2|H_1)}{m(\bm{y}_1|H_2)m(\bm{y}_2|H_2)}
\end{equation}

The marginal likelihoods in \eqref{eq_BF} can be obtained using the same probabilistic model that will be adopted to describe the handwriting data (and that will be illustrated in Section~\ref{modeling}). Specifically, the marginal likelihood in the numerator, $m(\bm{y}_1,\bm{y}_2|H_1)$, is calculated assuming that all observations of the combined dataset $\{\bm{y}_1,\bm{y}_2\}$ share a common parameter vector, since it is assumed that $H_1$ holds and the data originate from the same writer. On the other hand, if $H_2$ holds and the competing materials originate from different writers, the measurements $\bm{y}_1$ and $\bm{y}_2$ can be considered independent. Consequently, the marginal likelihood $m(\bm{y}_1,\bm{y}_2|H_2)$ can be obtained as the product of two independent marginal likelihoods, $m(\bm{y}_1|H_2)$ and $m(\bm{y}_2|H_2)$. The latter marginal likelihoods are calculated using the same probabilistic model, but by fitting this model separately for each dataset $\bm{y}_1$ and $\bm{y}_2$ and by taking different parameter vectors. Note that the implemented Bayes Factor assumes independence between the questioned and control material under $H_2$. This assumption implies that possible disguised behaviour is not considered (see Appendix \ref{assumption_BF} for the considered BF assumptions). For further definition in identification problems in forensic science, see \cite{ommen2017characterization}.

In this paper, six different Bayesian modeling approaches are analysed and compared. The first set of models assumes that the available data follow a multivariate Normal distribution (Section~\ref{normal-inverse-wishart}) with three different prior specifications: (a) the standard conjugate prior (Section~\ref{normal-inverse-wishart-conjugate}), (b) a hierarchical prior structure in which the variance-covariance matrix is modeled independently of the prior distribution of the mean vector (Section~\ref{normal-inverse-wishart-2level}), and (c) a covariance decomposition approach, where the within-writer variability a priori follows a Log-Normal-LKJ distribution (Section~\ref{normal-LogNormal-LKJ}). 
Under this model, the character-level variability is not modeled, since one common variance-covariance matrix is considered for all observed characters. 
Therefore, all measurements are evaluated jointly without distinction by character type. Hence, to take into consideration the impact of character type, the models will be fitted for each character separately.

To address this limitation, a Bayesian MANOVA model was proposed in Section~\ref{bayesian_manova}. Under this approach, the character type is recorded by a set of dummy (indicator) variables (see Appendix~\ref{dummy_variables} for details), allowing the model to account for between-character variability.
Similarly to the first modeling approach, three prior specification setups are considered: a conjugate prior (Section~\ref{bayesian_manova_conjugate}), a hierarchical Normal-Inverse-Wishart prior (Section~\ref{bayesian_manova_2level}), and a Normal-LogNormal-LKJ prior (Section~\ref{MANOVA-normal-LogNormal-LKJ}). 
Bridge sampling is used to estimate the marginal likelihoods required for the  Bayes factor (\ref{eq_BF}) for the hierarchical models; see Section~\ref{marginal_likelihood_estimators_summary} for details.

%--------------------------------------------------------------------
\section{Modeling Fourier Coefficients and Surface size}
\label{modeling}
 
Let us consider the available measurements on questioned and control material denoted as $\bm{y}_{1}$ and $\bm{y}_{2}$, respectively, described in Section \ref{sn_methodologies}, $\bm{y}_{w \bullet \bullet \bullet} = (\bm{y}_{w\ell j \bullet},w=1,2,\;\ell=1,\dots, L,\;j=1,...,n_{w\ell})$. Following the Fourier-based features described in Section \ref{data}, these data form a vector of length $p$, corresponding to the surface size, along with four pairs of Fourier coefficients, and are denoted as ${\bf y}_{w\ell j {\bullet}}$. 
By $\boldsymbol{\theta}_w \in \mathbb{R}^p$ we denote the mean vector within %the $w$-th 
material $w$, and by $\bm{W}_w \in \mathbb{R}^{p \times p}$ the 
variance-covariance matrix within material $w$ with elements 
$W_{w\:\!\kappa_1\:\!\kappa_2}$ for $(\kappa_1,\kappa_2) \in \{1,2, \dots, p\}^2$. Then, given $\boldsymbol{\theta}_w$ and $\bm{W}_w$, the distribution of ${\bf y}_{w\ell j \bullet}$ is taken to be $p$-variate Normal $N_p$, with 
\begin{equation}
    {\bm y}_{w\ell j \bullet}%\mid \boldsymbol{\theta}_w,\bm{W}_w
    \sim N_p\left({\boldsymbol\theta}_w,\bm{W}_w\right).
    \label{multiv_normal_model}
\end{equation}

Let us further assume that we have a dataset (or database) of manuscripts from different writers, unrelated to the case, denoted by $\bm{X}$. We refer to this as background data, which will later be used to specify model parameters (i.e., to elicit the prior distributions) where required.
We further assume that the background data follow the same feature engineering process as $\bm{y}_{1}$ and $\bm{y}_{2}$. Accordingly, our modeling framework involves three distinct datasets:
(i) the questioned data, $\bm{y}_{1}$,
(ii) the control data, $\bm{y}_{2}$, and
(iii) the background data, $\bm{X}$.
The questioned and control data, $\bm{y}_{1}$ and $\bm{y}_{2}$, are directly used to evaluate hypotheses $H_1$ and $H_2$ via the Bayes factor \eqref{eq_BF}, whereas the background data $\bm{X}$ are used indirectly (via the prior) to facilitate parameter estimation in each model.

\subsection{Bayesian Normal Models}
\label{normal-inverse-wishart}

In this first modeling approach, a Normal-Inverse-Wishart (NIW) prior distribution is chosen for the parameters $(\bm{\theta}_w,\bm{W}_w)$. Under this general prior set-up, and considering the sampling model distribution \eqref{multiv_normal_model} for $ {\bm y}_{w\ell j \bullet}$, the following Bayesian model is specified:
\begin{eqnarray}
    \bm{y}_{w\ell j \bullet}&\sim& N_p( \bm{\theta}_w,\bm{W}_w) \nonumber\\
    \bm{\theta}_w|\bm{X} &\sim& N_p( \bm{\mu},\bm{G}) \label{general_model1}\\
    \bm{W}_w|\bm{X} &\sim& IW(\bm{U},\nu). \nonumber
\end{eqnarray}
In this model formulation, IW denotes the Inverse-Wishart, $\boldsymbol\mu$ is the prior mean vector of $ \bm{\theta}_w$, 
$\bm{W}_{w}$ is the within-writer covariance matrix, 
$\bm{G}$ is the covariance matrix of $\bm{\theta}_w$, 
while $\bm{U}$ and $\nu$ represent the scale matrix and the degrees of freedom of the Inverse-Wishart distribution that models the within-writer variability.

Depending on the specification of the prior variance-covariance matrix $\bm{G}$, two distinct variants of the Bayesian model in~(\ref{general_model1}) are obtained. First, by setting $\bm{G}=\bm{W}_w k_0^{-1}$, a conjugate prior is obtained (Section \ref{normal-inverse-wishart-conjugate}), and the prior mean vector is denoted as $\bm{\theta}_w|\bm{X}, \bm{W}_w$. Conversely, whenever the parameter $\bm{G}$ is set as the between-writers covariance matrix $\bm{B}$, a model with hierarchical prior setup is obtained (Section \ref{normal-inverse-wishart-2level}). To upgrade the latter prior approach, we perform a covariance decomposition of the within-writer covariance matrix $\bm{W}_{w}$, employing the LogNormal-LKJ prior approach as specified in Section \ref{normal-LogNormal-LKJ}.

\subsubsection{Conjugate Prior: Normal-Inverse-Wishart}
\label{normal-inverse-wishart-conjugate}

By assuming data follow a multivariate Normal distribution, $N_p( \bm{\theta}_w,\bm{W}_w)$, the implied conjugate prior distribution for $(\bm{\theta}_w,\bm{W}_w)$ is the Normal-Inverse-Wishart (NIW) distribution, 
$$
(\boldsymbol{\theta}_w,W_w)|\bm{X}\sim NIW(\boldsymbol{\mu},k_0,\bm{U},\nu),
$$
where the prior parameter $k_0$ is a parameter that controls the volume of the prior precision and the variance of  $\bm{\theta}_w$. The higher $k_0$, the more informative the prior distribution for $\bm{\theta}_w$. Moreover, $\bm{U}$ and $\nu$ represent the scale matrix and the degrees of freedom of the Inverse-Wishart distribution, respectively; see  \citet[Chap. 3.6]{gelman2013bayesian} for more details. Hence, by combining the sampling model distribution \eqref{multiv_normal_model} for $ \bm{y}_{w\ell j}$ with the NIW prior distribution, the Bayesian model  \eqref{general_model1} is obtained with $\bm{G}=\bm{W}_w k_0^{-1}$. 
%\begin{eqnarray*}
%\bm{X}_{i\ell j}&\sim& N_p(\boldsymbol\theta_i,\bm{W}_i)\\
%\bm{\theta}_i|\bm{W}_i &\sim& N_p\left(\bm{\mu},\bm{W}_i k_0^{-1}\right)\\
%    \bm{W}_i &\sim& IW(\bm{U},\nu)
%    %\end{aligned}
%\end{eqnarray*}

Since the NIW prior is conjugate, the marginal likelihood of the available measurements $\bm{y}$ is readily available in closed form and is given by:

\begin{equation*}
m(\bm{y}) = \frac{1}{\pi^{np/2}}\frac{\Gamma_p(\nu_n/2)}{\Gamma_p(\nu/2)}\frac{|\bm{U}|^{\nu/2}}{|\bm{U}_n|^{\nu_n/2}}\left(\frac{k_0}{k_n}\right)^{p/2},
\end{equation*}
where $n$ is the sample size, $p$ is the number of variables, $\Gamma_p$ is the multivariate gamma function, and
$$
k_n = k_0 + n, ~\nu_n = \nu + n, ~\bm{U}_n = \bm{U} + \bm{S} + \frac{k_0n}{k_0+n} (\bar{\bm{y}}-\bm{\mu} )(\bar{\bm{y}}-\bm{\mu} )^T,~ \bm{S} = \sum_i (\bm{y}_i - \bar{\bm{y}})(\bm{y}_i - \bar{\bm{y}})^T ~; 
$$ 
where $\bar{\bm{y}}$ the average measurements of questioned and control materials, see \cite{murphy2007conjugate} for more details. 
In this work, the prior parameters $\boldsymbol\mu$ and $\bm{U}$ are 
%specified 
elicited by using information from readily available background data $\bm{X}$; see Appendix~\ref{prior_estimation_normal_iw}.  

The degrees of freedom $\nu$ are set to be equal to $p + 2$, which is the smallest value for which the 
mean of the Inverse-Wishart distribution is available  \citep{press2005applied}. Finally, the prior parameter $k_0$  is selected based on a grid search in the (0,1) interval for all writers available in the background data. Specifically, the choice falls on the value of $k_0$ that maximizes the marginal likelihood in the background data.

Although the conjugate prior is the natural choice because it is computationally convenient, it has the disadvantage of not modeling the variability within-writer and between-writers separately. This can be a major limitation in this specific context, since the fundamental laws of handwriting state that (1) no one writes the same word exactly the same way twice (i.e., variability within writers), and (2) no two people write exactly the same way (i.e., variability between writers). For this reason, a second model with an independent specification of the prior is considered.
 
%\subsubsection{Model considering between-writers variability}
\subsubsection{Hierarchical Extension: Normal-Inverse-Wishart}
% \subsubsection{Independent prior setup}
\label{normal-inverse-wishart-2level}

\citet{bozza2008probabilistic}  chose an hierarchical prior set up for the model parameters $(\bm{\theta},\bm{W})|\bm{X}$ of the following form 
\begin{equation}
\bm{\theta}_w|\bm{X} \sim  N_p( \bm{\mu},\bm{B}) \mbox{~and~} \bm{W}_w|\bm{X} \sim IW(\bm{U},\nu).
\label{niw}   
\end{equation}
The proposed model is then given by \eqref{general_model1} with $\bm{B}=\bm{G}$ to be fixed and independent of $W_w$. 

In particular, the prior parameter $\bm B$ represents the variance-covariance matrix between writers. The prior parameters $\boldsymbol\mu$, $\bm{B}$, and $\bm{U}$  are estimated in a relatively straightforward manner from the information of the background writers (see Appendix \ref{prior_estimation_normal_iw}). %Moreover, 
Empirical evidence has shown that models incorporating the between-writers variability are performing better than models that do not have this flexibility \citep{bozza2008probabilistic}.  It must be said that without the background information, the ability to effectively discriminate between writers is greatly reduced.

Unfortunately, the marginal likelihood for this model cannot be obtained analytically.
Hence, bridge sampling is applied for estimation based on the output of an MCMC algorithm as it is presented and discussed in Section~\ref{marginal_likelihood_estimators_summary}.
In particular, MCMC algorithm No-U-Turn Sampler (NUTS) is implemented in {\it Stan} \citep[see][]{carpenter2017stan} version 2.26.1, via the {\it RStan} interface and {\it R} version 4.2.1.

\subsubsection{Covariance Decomposition: Normal-LogNormal-LKJ}
\label{normal-LogNormal-LKJ}

A widely used and flexible approach for modeling covariance matrices in Bayesian hierarchical models is to decompose the covariance matrix into standard deviations and a correlation matrix. Specifically, the covariance matrix $\bm{W}_w$ can be expressed as:
\begin{equation}
\bm{W}_w = \bm{D}_w\bm{R}_w \bm{D}_w  \mbox{~and~} \bm{D}_w = Diag\left( W_{w11}^{1/2}, W_{w22}^{1/2}, \dots,  W_{wpp}^{1/2} \right);
\end{equation}
where $\bm{D}_w$  is a diagonal matrix containing the standard deviations of the covariates of material $w$ (for $w=\{1,2\}$), while 
 $\bm{R}_w$ is the corresponding correlation matrix. 
 The diagonal elements of  $\bm{W}_w$ specify the corresponding elements of $\bm{D}_w$ while $\bm{R}_w$ is 
 obtained by standardizing $\bm{W}_w$, that is, by simply  setting $\bm{R}_w=\bm{D}_w^{-1}\bm{W}_w\bm{D}_w^{-1}$.

In this decomposition, the standard deviations $\bm{D}_w$ are typically assigned independent prior, such as LogNormal distribution, or Half-Cauchy distribution for less informative prior specifications. Therefore, this alternative parametrization allows for a more flexible modeling of marginal variances. The correlation matrix $\bm{R}_w$ is modelled separately using the Lewandowski-Kurowicka-Joe (LKJ) distribution \citep{lewandowski2009generating}, which provides a non-conjugate prior specifically designed for correlation matrices. The LKJ distribution has a shape parameter $\eta>0$ controlling the strength of the prior towards the identity matrix (independence) and probability function given by:
\begin{eqnarray*}
f( \bm{R} ; \eta ) &=& \mathcal{C} \times \left| \bm{R} \right|^{\eta-1} \mbox{~with} \\ 
\log(\mathcal{C}) &=& \sum_{\kappa=1}^p (2\eta-2+p-\kappa) \log 2 + \sum_{\kappa=1}^{p-1} (p-\kappa) \log \mathcal{B}( \eta + \tfrac{p-\kappa-1}{2},~\eta+ \tfrac{p-\kappa-1}{2} ); 
\end{eqnarray*}
where $\bm{R}$ is a $p \times p$ positive-definite matrix with unit diagonal, $|\bm{R}|$ denotes the determinant of the matrix $\bm{R}$, the normalizing constant $\mathcal{C}$ ensures that the $f( \bm{R} ; \eta )$ integrates to one over the space of $p \times p$ correlation matrices,  $p$ is the dimension of the correlation matrix $\bm{R}$, $\kappa$ is an index running from $1$ to $p$ (or $p-1$ in the second sum) and $\mathcal{B}()$ is the Beta function.

Hence, the Bayesian model is specified:

\begin{eqnarray}
    \bm{y}_{w\ell j \bullet}&\sim& N_p( \bm{\theta}_w,\bm{W}_w),  \quad  \bm{W}_w =\bm{D}_w\bm{R}_w \bm{D}_w\nonumber\\
    \bm{\theta}_w|\bm{X} &\sim& N_p( \bm{\mu},\bm{B}) \label{normal-ln-lkj_eq}\\
    D_{w\:\! \kappa\:\! \kappa}|\bm{X} &\sim& \operatorname{LogNormal}(\upsilon, \sigma) 
    \mbox{~for~} \kappa \in \{ 1, 2, \dots, p\}, \nonumber\\
    %\bm{D}_{w\kappa\kappa}|\bm{X} &\sim& \textit{LogNormal}(\mu, \sigma) \mbox{~for~} \kappa=1,2,\dots, p\nonumber\\
    \bm{R}_w &\sim& LKJ(\eta)\nonumber
\end{eqnarray}
 The $\operatorname{LogNormal}(\upsilon, \sigma)$ distribution is parameterized by the location parameter $\upsilon$ and the scale parameter $\sigma$, which are estimated based on prior knowledge about standard deviations derived from the background data enabling the prior elicitation (see Appendix~\ref{prior_estimation_normal_iw}) and the LKJ parameter $\eta$ is set equal to~1 for non-informative prior for the correlation matrix. The $\operatorname{LogNormal}$ prior was selected based on prior knowledge derived from background data, enabling the elicitation of prior parameters $\upsilon$ and $\sigma$. The same approach for prior specification is applied to the other model parameters, as described in Section~\ref{normal-inverse-wishart-2level}. The marginal likelihood for the Normal-LogNormal-LKJ prior setup is not analytically available in closed form. Hence, bridge sampling is implemented for the estimation of the marginal likelihood, as described in Section \ref{marginal_likelihood_estimators_summary}. For the extraction of posterior samples of the parameters, the MCMC algorithm No-U-Turn Sampler (NUTS) is implemented in {\it Stan} \citep[see][]{carpenter2017stan} version 2.26.1, via the {\it RStan} interface and {\it R} version 4.2.1. 

According to \cite{huang2013simple}, the LKJ prior approach can offer better performance over the classical Inverse-Wishart prior approach, particularly in scenarios where a more flexible prior for the correlation matrix is required or when limited prior information about the covariance structure is available. 
Additionally, the LKJ prior is the most popular approach in the context of sparse covariance matrix estimation.

%It should be emphasized that the conjugate approach (Section~\ref{normal-inverse-wishart-conjugate}) and the hierarchical approaches (Sections ~\ref{normal-inverse-wishart-2level} and ~\ref{normal-LogNormal-LKJ})   do not model the variability at character level. 
All the modeling approaches considered in Section \ref{normal-inverse-wishart} do not account for character variability since characters of different types must be either treated without any distinction, or analysed separately. Both of the above approaches, analyzing all data but treating all characters
without any distinction or analyzing each character separately, has major disadvantages. 
Specifically, in the first approach, treating all characters without distinction ignores not only the variability of the handwriting of characters within each writer but also the differences of characters across different writers.
Alternatively, analyzing each character separately raises the challenge of aggregating individual results into a single measure of evidence for $H_1$ or $H_2$. Character-specific Bayes factors cannot be directly combined, and the situation becomes more complicated when conflicting evidence arises, that is, when different character types support different hypotheses.
Hence, in Section \ref{bayesian_manova} we proceed in proposing models which take into account the within characters' variability.

\subsection{Bayesian MANOVA}
\label{bayesian_manova}

%$\theta \uptheta \vartheta \Theta \theta{\Theta}^* \theta{\Theta}^\star \theta^{\bullet}$
 
%With Bayesian MANOVA, the information of all characters can be contained in the model. In this manner, we can model the variability between characters. 
In this section, a second modeling approach is provided in an attempt to overcome the critical issues mentioned above. The Bayesian MANOVA model that will now be described has the great advantage of allowing the character-level variability to be modeled as well. Specifically, the indicators of the characters are transformed into dummy variables by selecting the character {\em a} as a reference group (see Appendix \ref{dummy_variables}). Hence, a regression model is built with dummy variables of characters as predictors. This model, in its general form, can be written as 
\begin{eqnarray}
    \bm{y}_{w \ell j \bullet} &\sim& N_p(\bm{\Theta}_w^T \bm{C}_{w \bullet j},\bm{W}_w)\nonumber \\
    \bm{\Theta}_w|\bm{W}_w,\bm{X} &\sim& MN_{L,p}( \bm{\mathcal{M}},\bm{K}_0^{-1},\bm{W}_w)
    \label{conjugate_bayesian_regression_notation}\\
    \bm{W}_w|\bm{X} &\sim& IW(\bm{U},\nu), \nonumber
\end{eqnarray}

where $MN_{L,p}$ is denoted the Matrix Normal with $L$=4 characters and $p$ = 9, the 4 pairs of Fourier coefficients, and the surface size,  $\bm C$ is a three-dimensional $m \times L \times n_L$ array of the explanatory dummy variables for ${\bm y}$. 
The sub-vector ${\bm C}_{w \bullet j}$ has length $L$ and contains the dummy variables for each writer and for each repetition of the retained $L$ characters. Each element $C_{w \ell j}$ of $\bm C$ is defined as $C_{w l j}=1$, for $l=1$ and for all $w,j$, while for $l=2,\dots,L$ it is defined as
$$
\begin{array}{cc}
C_{w \ell j}=1, & \text{if}\; \ell= \ell_{wj}^{obs}; \\
C_{w \ell j}=0, & \text{otherwise},
\end{array}
$$
where $\ell_{wj}^{obs}$ is the observed character for questioned or control material $w$ under repetition $j$ (i.e. (1,0,0,0) for \emph{a}, (1,1,0,0) for
\emph{d}, (1,0,1,0) for
\emph{o} etc.); see at Appendix \ref{dummy_variables} for a more detailed presentation. Moreover, $\bm{\Theta}_w$ is the $L \times p$ coefficient matrix and can be re-written as follows
\begin{eqnarray}\label{eq_vartheta}
{\bm \Theta}_w = \left( 
\begin{array}{c}
     {\bm \vartheta}_{w1}^T \\
     {\bm \vartheta}_{w2}^T \\ 
     \vdots\\
     {\bm \vartheta}_{wL}^T 
\end{array}
\right),
\end{eqnarray}
where each ${\bm \vartheta}_{wl}$, $l=1,\dots,L$, is a vector of length $p$. 
${\bm W}_{w}$ is the within-writer covariance matrix, 
$\nu$ denotes the degrees of freedom, 
$\bm{U}$ is the scale matrix of the Inverse-Wishart distribution,  
$\bm{\mathcal{M}}$ is the prior mean matrix (of dimension $L\times p$ prior mean matrix), and 
$\bm{K}_0$ is a $L \times L$ scale matrix tuning the variance of $\bm{\Theta}_{w}$. 
The matrix Normal distribution % of
in \eqref{conjugate_bayesian_regression_notation} can be re-written as 
$$
{\rm vec}(\bm{\Theta}_w)|\bm{X} \sim N_{L\times p}({\rm vec}(\bm{\mathcal{M}}),\bm{W}_w \otimes \bm{K}_0^{-1})
$$
where ${\rm vec()}$ vectorize of a matrix and $\otimes$ Kronecker product. Therefore, the overall variance of ${\rm vec}(\bm{\Theta}_{w})$ is simply given by ${\bm W}_w \otimes {\bm K}_0^{-1} $.
However, since the focus here is pointed toward the distribution of each different type of character (i.e., each row of ${\bm \Theta}_w$),  ${\bm W}_w$ will be taken as the main variance component, while ${\bm K}_0^{-1}$ will act as a variance multiplicator which will be elicited using background data.

%\subsubsection{Conjugate Analysis}
\subsubsection{Conjugate Prior}
\label{bayesian_manova_conjugate}

As for the conjugate Normal-Inverse-Wishart prior setup in Section \ref{normal-inverse-wishart-conjugate}, a conjugate approach is initially considered for the MANOVA model in \eqref{conjugate_bayesian_regression_notation} for computational convenience. Under this approach, the natural conjugate prior using the vectorized parameter $\bm{\Theta}_w$ is of the form:
\begin{eqnarray*}
    {\rm vec}(\bm{\Theta}_w)|\bm{W}_w,\bm{X}, {\bm K}_0 &\sim& N_{L\times p}({\rm vec}(\bm{\mathcal{M}}),\bm{W}_w \otimes \bm{K}_0^{-1}))\\
    \bm{W}_w |\bm{X}  &\sim& IW(\bm{U},\nu),
\end{eqnarray*}
where ${\bm K}_0$ is considered here as a fixed parameter elicited from the background data. Under the above Bayesian formulation, the posterior distribution can be expressed as a result of the same family. Hence, the marginal likelihood can be expressed in closed form and is given by:

\begin{equation*}
m(\bm{y}) = \frac{1}{2\pi^{np/2}}\frac{\Gamma_p(\nu_n/2)}{\Gamma_p(\nu/2)}\frac{|\bm{U}/2|^{\nu/2}}{|\bm{U}_n/2|^{\nu_n/2}}\left(\frac{|K_0|}{|K_n|}\right)^{p/2},
\end{equation*}

where $n$ is the sample size, $p$ is the number of variables (i.e., Fourier coefficients and the surface size),  %$\Gamma_p$ is the multivariate gamma function, 
\begin{eqnarray*}
\nu_n &=& \nu + n \\ 
 K_n  &=& C^TC+K_0 \\ 
 \bm{U}_n &=& \bm{U} + \bm{y}^T\bm{y} + \bm{\mathcal{M}}^TK_0\bm{\mathcal{M}} - \bm{\mathcal{M}}_n^TK_n\bm{\mathcal{M}}_n, \mbox{~and~} \\
 \bm{\mathcal{M}}_n &=& K_n^{-1}\left( C^T\bm{y} + K_0\bm{\mathcal{M}}\right); 
\end{eqnarray*}
see \citet[][Chapter 8.4]{rowe2002multivariate} and \citet[][Chapter 2.4]{JoramSoch}. A detailed illustration of prior parameter elicitation can be found in Appendix \ref{prior_estimation_bayes_regression}.

%While the conjugate approach for the MANOVA model has computational benefits, it ignores the heterogeneity among writers. Therefore, in the following section, we present a non-conjugate version of this model, which resolves this problem. 
However, as pointed out previously, although the conjugate approach may be attractive for computational convenience, it has the important limitation of not modeling heterogeneity between writers (i.e, between-writers variability). Therefore, to overcome this problem, a non-conjugate version of the MANOVA model is introduced in the following section.

%-------------------------------------------------------------------------
%\subsubsection{Model considering between-writers variability per character}
\subsubsection{Hierarchical Extension: MANOVA Normal-Inverse-Wishart}
\label{bayesian_manova_2level}

Let us consider a simplified version of Model \eqref{conjugate_bayesian_regression_notation}, where $\bm{K}_0^{-1}$ is set equal to the identity matrix and ${\bm W}_w$ is replaced by the between-writers covariance matrix ${\bm B}$ in the sampling distribution of ${\bm \Theta}_w$. This allows us to write
$$
{\bm \Theta}_w | {\bm W}_w,\bm{X} \sim MN_{L,p} ( \bm {\mathcal{M}}, {\bm I}_{L}, {\bm B}),
$$
which simplifies to

\begin{eqnarray}
%\bm{X}_{i \ell j \bullet} &\sim& N_p(\bm{\Theta}_i^T \bm{C}_{i \bullet j},\bm{W}_i) \nonumber \\
        \bm{\theta}_{w\ell}|\bm{X} &\sim& N_p( \bm{\mu}_\ell,\bm{B}), \qquad \text{for}~~ \ell \in \{1,\dots,L\}
        \nonumber 
 %   \bm{W}_i &\sim& IW(\bm{U},\nu)\nonumber,
\label{bayesian_regression_heter_notation}
\end{eqnarray}

where $\bm{\mu}_\ell$ is the  prior mean vector of $ \bm{\theta}_{w\ell}$. Furthermore, the common variance-covariance matrix ${\bm B}$ is replaced by ${\bm B}_\ell$, $l=1,\dots, L$, to consider a heteroscedastic version of the above model where, per the character level $l$, the between-writer variability differs. Under these setups, the final model is given by
\begin{eqnarray}
\bm{y}_{w \ell j \bullet} &\sim& N_p(\bm{\Theta}_w^T \bm{C}_{w \bullet j},\bm{W}_w) \nonumber \\
\bm{\theta}_{w\ell}|\bm{X} &\sim& N_p( \bm{\mu}_\ell, \bm {B}_\ell ) ~~ \text{for}~~ \ell \in \{1,\dots,L\} \label{bayesian_regression_notation}\\
 \bm{W}_w|\bm{X} &\sim& IW(\bm{U},\nu)\nonumber.
\end{eqnarray}

The elicitation procedure described in Section \ref{normal-inverse-wishart} has been slightly adapted to specify the prior distribution parameters for the mean vector $\bm{\theta}_{w\ell}$ and the covariance matrix $\bm{W}_{w}$, based on the available background data. In particular, the values of the hyperparameters characterizing the prior distribution of $\bm{\theta}_\ell$ for $\ell=2,\dots, L$ were elicited by taking the differences between the estimated $\widehat{\bm{\mu}}_a$ which is the sample mean of the background data of character \emph{a} and $\widehat{\bm{\mu}}_l$, for $l=2,\dots, L$ is the sample mean difference of character $l$ from \emph{a}. This approach effectively captures the mean differences between each character $l$ and the reference character \emph{a}. Further details can be found in Appendix~\ref{prior_estimation_bayes_regression}. As far as the scale matrix $\bm{U}$ characterizing the prior distribution of $\bm{W}_{w}$, the average of the variance-covariance matrix of every writer in the background data was considered. Hence, this is an initial a priori estimate of the within-writer variability which is a posteriori estimated via  $\bm{W}_{w}$ using the handwriting data of material $w$. In the same way, $\bm{B}_\ell$ is the variance-covariance matrix of character $\ell$, and $\bm{B}_\ell$ is elicited from the sample variance-covariance matrices of the measurements of character $\ell$ for all writers in the background data. Similarly, as noted in Section \ref{normal-inverse-wishart-2level} (Model independent prior setup of NIW), if the background information is not incorporated in the prior, then the ability to effectively discriminate between writers is significantly reduced.

The marginal likelihood in this case is not analytically available in closed form expression as in the conjugate approach of Section \ref{bayesian_manova_conjugate}. Hence, MCMC samples are used for the bridge sampling estimation of the marginal likelihood, as it will be described in  Section~\ref{marginal_likelihood_estimators_summary}. In particular,  the No-U-Turn Sampler (NUTS) is implemented in {\it Stan} \citep[see][]{carpenter2017stan} version 2.26.1, via the {\it RStan} interface and {\it R} version 4.2.1.

\subsubsection{Covariance Decomposition: MANOVA Normal-LogNormal-LKJ}
\label{MANOVA-normal-LogNormal-LKJ}

Similar to Section~\ref{normal-LogNormal-LKJ}, we decompose the within-writer variability in model~\eqref{bayesian_regression_notation} by representing the covariance structure as a function of standard deviations and the correlation matrix. Specifically, we assign an independent LogNormal prior to the standard deviations and employ the LKJ distribution as a prior for the correlation matrix. The resulting model can be expressed as follows:

\begin{eqnarray}
\bm{y}_{w \ell j \bullet} &\sim& N_p(\bm{\Theta}_w^T \bm{C}_{w \bullet j},\bm{W}_w),  \quad  \bm{W}_w =\bm{D}_w\bm{R}_w \bm{D}_w\nonumber\\
\bm{\theta}_{w\ell}|\bm{X} &\sim& N_p( \bm{\mu}_\ell, \bm {B}_\ell ) ~~ \text{for}~~ \ell \in \{1,\dots,L\} \label{bayesian_regression_notation_LKJ}\\
 D_{w\:\! \kappa\:\! \kappa}|\bm{X} &\sim& \operatorname{LogNormal}(\upsilon, \sigma) 
    \mbox{~for~} \kappa \in \{ 1, 2, \dots, p\}, \nonumber\\
    %\bm{D}_{w\kappa\kappa}|\bm{X} &\sim& \textit{LogNormal}(\mu, \sigma) \mbox{~for~} \kappa=1,2,\dots, p\nonumber\\
    \bm{R}_w &\sim& LKJ(\eta)\nonumber
\end{eqnarray}
%The $\operatorname{LogNormal}(\upsilon,\sigma)$ distribution is parameterized by the location parameter $\upsilon$ and the scale parameter $\sigma$, which are estimated based on information from the background writers (see Appendix~\ref{prior_estimation_normal_iw}) and the LKJ parameter $\eta$ is set equal to~1 for non-informative prior for the correlation matrix. The same approach for prior specification is applied to the other model parameters, as described in Section~\ref{bayesian_manova_2level}. 

For the Normal–LogNormal–LKJ prior setup, the marginal likelihood is not available in closed form, similarly to Section \ref{marginal_likelihood_estimators_summary}. 
Therefore, we estimate it using bridge sampling, as described in Section \ref{marginal_likelihood_estimators_summary}.
%For the extraction of posterior samples of the parameters, the MCMC algorithm No-U-Turn Sampler (NUTS) is implemented in {\it Stan} \citep[see][]{carpenter2017stan} version 2.26.1, via the {\it RStan} interface and {\it R} version 4.2.1. 

\subsection{Marginal Likelihood Estimators}
\label{marginal_likelihood_estimators_summary}

In this section, we describe the marginal likelihoods and their estimation for the two main cases considered in our data analysis. In the first case, we use either the full dataset $\bm{\mathcal{D}}$ and subsets of it corresponding to individual writers, denoted by $\bm{\mathcal{D}}_i$ for $i = 1, \dots, 13$, to evaluate and compare models $M_1-M_6$ described in Section \ref{modeling}; see Table \ref{overview_models}. 

In the second case, we focus on the primary comparison of interest by evaluating $H_1$ versus $H_2$ using the Bayes factor Eq. \eqref{eq_BF} and the associated marginal likelihoods. Here, three distinct datasets are used: $\bm{y}_1$, $\bm{y}_2$, and $\bm{X}$, as defined in the beginning of Section \ref{modeling}. These datasets are considered subsets of the full dataset $\bm{\mathcal{D}}$. Details about how  $\bm{y}_1$, $\bm{y}_2$, and $\bm{X}$ are constructed from $\bm{\mathcal{D}}$ depend on the comparison scenario, and it is described in detail in the experimental results in Section \ref{experimental_results}. 

Notably, the subset considered from the full dataset $\bm{\mathcal{D}}$ in model comparisons and evidence evaluation is grounded in the principle that the data should not be used twice (once in the prior and again in the likelihood), in order to avoid biasing in the results.

\begin{table}[!ht]
    \centering
    \footnotesize
    \begin{tabular}{|c|l|l|c|}
    \hline
         \textbf{Model} & \multicolumn{1}{|c|}{\textbf{Description (Model and Prior Specification)}} & \multicolumn{1}{|c|}{\textbf{Prior Abbreviation}} & \textbf{Section} \\ \hline
        $M_1$  & Normal model with Conjugate Normal-Inverse-Wishart & NIW Conjugate &  \ref{normal-inverse-wishart-conjugate} \\ \hline
         $M_2$  & Normal model with Hierarchical Normal-Inverse-Wishart & NIW Hierarchical &  \ref{normal-inverse-wishart-2level} \\\hline
          $M_3$  & Normal model with Normal-LogNormal-LKJ &  & \ref{normal-LogNormal-LKJ} \\ \hline
           $M_4$  & MANOVA model with Conjugate Normal-Inverse-Wishart & NIW Conjugate &  \ref{bayesian_manova_conjugate} \\\hline
            $M_5$  &   MANOVA model with Hierarchical Normal-Inverse-Wishart & NIW Hierarchical &  \ref{bayesian_manova_2level} \\ \hline
            $M_6$  &   MANOVA model with Normal-LogNormal-LKJ &  & \ref{MANOVA-normal-LogNormal-LKJ} \\ \hline
    \end{tabular}
    \caption{Overview of the models and prior specifications considered in the study.}
    \label{overview_models}
\end{table}

\subsubsection{Marginal likelihoods for Model Comparison using the Full Dataset}
\label{marginals_models}

In order to compare the proposed models, we calculate the Bayes factor for each writer as the ratio of the marginal likelihood for any given pair of models $M_{l}$ and $M_{\xi}$, for $l, \xi \in \{ 1,\dots,6\}$. The Bayes factor quantifies the relative evidence of the compared models based on the available data $\bm{\mathcal{D}}_i$ of each writer, $i=1,\dots,m$, the prior parameters are elicited using data from the remaining writers ($m-i$).  A higher Bayes factor indicates stronger support for one model over the other.
 
The marginal likelihood of each model using the data $\bm{\mathcal{D}}_i$ of each writer, $i=1,\dots,m$, is given by
\begin{equation}
    m(\bm{\mathcal{D}}_i|M_l) = \int
    f(\bm{\mathcal{D}}_i|\bm{\Theta},\bm{W},M_l)\pi(\bm{\Theta},\bm{W}|M_l) d(\bm{\Theta},\bm{W}).
\end{equation}
For any model $M_l$ (for $l=1,2,3,4,5,6$) and model parameters $\bm{\Theta}$ and $\bm{W}$. Models $M_1$ and $M_4$ are based on a conjugate prior setups, which means that the marginal likelihoods are available in an analytical form. Nevertheless, this is not the case for models ($M_2$,$M_3$) and ($M_5$,$M_6$), where the marginal likelihoods are not available in a closed-form expression. In such occasions, the marginal likelihoods of $M_2$, $M_3$, $M_5$, and $M_6$ were estimated using the bridge sampling approximation, a widely used Monte Carlo method for marginal likelihood estimation. This procedure was implemented via the {\it bridgesampling} package in {\it R} \citep{gronau2020bridgesampling}, which provides an efficient framework for computing normalizing constants for Bayesian modeling. Notably, the ``Warp-III'' variant of \cite{gronau2020computing}  is implemented within the {\it bridgesampling} package and refers to a transformation-based strategy to further enhance the efficiency and accuracy of the marginal likelihood estimates. By applying a nonlinear transformation to the posterior samples, ``Warp-III'' aims to better align the shapes of the posterior and proposal distributions. For a more detailed explanation of the bridge sampling approximation, see Appendix \ref{bridge_sampling_section}. Finally, in Section \ref{mce_per_writer}, we present the Monte Carlo errors of the marginal likelihood estimates for each considered model per writer.

\subsubsection{Marginal Likelihoods for Handwriting Evidence Evaluation}
\label{ml_propositions}

The marginal likelihoods (ML) needed to assess the Bayes factor in Eq. (\ref{eq_BF}) represent the measure of the evidence under the response data $\bm{y} = (\bm{y}_1, \bm{y}_2) = (\bm{y}_{w\ell j \bullet },w=1,2,\ell=1,..L,j=1,...,n_{w\ell})$ under the competing propositions $H_1$ and $H_2$, for a specific probabilistic model $M$. Here, by $M$ there are considered any of the six Bayesian models $\{M_1, M_2, M_3, M_4, M_5,$ $M_6\}$ introduced in~Section \ref{modeling}; see Table \ref{overview_models} for details. 

%The marginal likelihood is defined as the integral of the likelihood function over the prior distribution. 
For the numerator of Eq. (\ref{eq_BF}) the marginal likelihood  under the hypothesis $H_1$ is expressed:

\begin{equation}
    m(\bm{y}_1, \bm{y}_2|H_{1},M_l) = \int
    f(\bm{y}_1, \bm{y}_2|\bm{\Theta},\bm{W},H_{1},M_l)\pi(\bm{\Theta},\bm{W}|H_{1},M_l) d(\bm{\Theta},\bm{W}).
    \label{margi_lik_def_H1}
\end{equation}

For the denominator of Eq. (\ref{eq_BF}) under a hypothesis $H_2$ :

\begin{equation}
    m(\bm{y}_w|H_{2},M_l) = \int
    f(\bm{y}_w|\bm{\Theta},\bm{W},H_{2},M_l)\pi(\bm{\Theta},\bm{W}|H_{2},M_l) d(\bm{\Theta},\bm{W}), 
    \mbox{~for~} w\in\{1,2\},
    \label{margi_lik_def_H2_1}
\end{equation}
%
%\begin{equation}
%    m(\bm{y}_2|H_{k},M_l) = \int
%    f(\bm{y}_2|\bm{\Theta},\bm{W},H_{k},M_l)\pi(\bm{\Theta},\bm{W}|H_{k},M_l) d(\bm{\Theta},\bm{W}), 
%    \label{margi_lik_def_H2_2}
%\end{equation}
%
and any model $M_l$ (for $l=1,2,3,4,5,6$) and model parameters $\bm{\Theta}$ and $\bm{W}$. 
If hypothesis $H_1$ holds (i.e., the PoI is the writer of the questioned manuscript), model parameters are assumed equal, that is $\bm{\theta}_1=\bm{\theta}_2=\bm{\Theta}$ and $\bm{W}_1=\bm{W}_2=\bm W$. On the other hand, if hypothesis $H_2$ holds (i.e., the person of interest is not the writer of the questioned document), then the model parameters ($\bm{\theta}_1,\bm{\theta}_2$ and $\bm{W}_1,\bm{W}_2$) are assumed to be different across the two different sources $\bm{y}_1$ and $\bm{y}_2$.

In our model formulations, presented in Sections \ref{normal-inverse-wishart} and \ref{bayesian_manova}, the parameter vector $\bm{\Theta}$, under $H_2$, is denoted as $\bm{\Theta}= (\bm{\theta}_1, \bm{\theta}_2)$ for models $M_1$, $M_2$, $M_3$, and by $\bm{\Theta}=(\bm{\Theta}_1,\bm{\Theta}_2)$ for models $M_4$, $M_5$, $M_6$, where $\bm{\Theta}_w$ is specified as in~(\ref{eq_vartheta}), $w=1,2$. Furthermore, Models $M_1$ and $M_4$ are based on conjugate prior setups, which means their marginal likelihoods are available in analytical form. However, for models $M_2$, $M_3$, $M_5$, and $M_6$, the marginal likelihoods are not available in closed-form expressions. Therefore, we estimated them using the bridge sampling approximation with the ``Warp-III'' approach, implemented via the \textit{bridgesampling} package in \textit{R} \citep{gronau2020bridgesampling}.

%---------------------------------------------------------------
\section{Experimental Results}
\label{experimental_results}

Following the handwriting examination problem presented in Section~\ref{sn_methodologies}, a case scenario involving a document whose origin is contested is analysed in this section. These models arise from two likelihood structures and three different prior specifications (hence six models in total; see Table \ref{overview_models}) described in Section~\ref{modeling} were compared using the dataset described in Section~\ref{data}. The considered model approaches are the conjugate and non-conjugate versions of the Bayesian Normal with Normal-Inverse-Wishart or Normal-LogNormal-LKJ prior setup  ($M_1$, $M_2$, and $M_3$, respectively), and of the Bayesian MANOVA with Normal-Inverse-Wishart or Normal-LogNormal-LKJ prior setup models ($M_4$, $M_5$, $M_6$, respectively).

First, in Section~\ref{comparison_of_models}, we compare $M_{1}, \dots, M_{6}$ for each individual writer from the full available data $\bm{\mathcal{D}}_i$ of each writer $i=1,\dots, m$. For each writer $m$, this procedure enables pairwise comparisons among the models. Furthermore, when eliciting the prior parameters for the considered models, the data from all other writers (i.e., excluding the writer under evaluation) are used. Specifically, when comparing models for writer $i$, the data from the remaining $m-i$ writers are utilized for prior elicitation.

Subsequently, in Section~\ref{models_efficiency}, the performance of the considered models is evaluated within the Bayes factor framework, as outlined in Section~\ref{sn_methodologies}. Specifically, we conduct a series of experiments designed to simulate a real-world case study. For the same-writer experiments, each writer $i$ from the full dataset $\bm{\mathcal{D}}$ is selected, and their data are randomly divided into questioned and control sets. The Bayes factor in Eq.~\ref{eq_BF} is then assessed. The data from all other writers $(m - i)$ are used as background data $\bm{X}$ to elicit the prior parameters of the models under consideration. For the different-writer experiments, data are randomly selected from two different writers from the full dataset $\bm{\mathcal{D}}$, and the same procedure as described in the same-writer experiments is followed. 

%Firstly, in Section~\ref{comparison_of_models}, the Bayes factor for model comparison is assessed for each writer. Then, in Section~\ref{models_efficiency}, the performance of the considered models is evaluated based on the framework described in Section~\ref{sn_methodologies} by using the Bayes factor approach (Eq. \ref{eq_BF}).

%Finally, in Section~\ref{sensitivity}, a sensitivity analysis is performed in order to examine the impact of the choice of the degrees of freedom of the Inverse-Wishart distribution that models the within-writer variability. 

\subsection{Model Comparisons per Writer}
\label{comparison_of_models}

In this section, a model comparison is performed between the six models under consideration in Table \ref{overview_models}  described in Section \ref{modeling}. For each writer $i$, the model $M_l$ is fitted to the data $\bm{\mathcal{D}}_i$ characterizing this writer, and the prior parameters are fitted using data from the remaining writers (excluding the $i$ writer from the elicitation of the prior parameters).  The marginal likelihoods of all models under consideration for each writer $i$ are then estimated.  The Bayes factor for comparing two models  $M_l$ and $M_\xi$  for writer $i$ is given by

\begin{align*}
\textbf{BF}_{l,\xi}(\bm{\mathcal{D}}_i) = \frac{m(\bm{\mathcal{D}}_i\mid M_l)}{m(\bm{\mathcal{D}}_i\mid M_\xi)} \mbox{~for~} l , \xi \in \{1,2,3,4,5,6\}. 
\end{align*}

\textbf{Bayesian MANOVA vs. Normal Model.} 
First, the Bayesian MANOVA model is compared with the Bayesian Normal model, both in the conjugate ($M_4$ vs. $M_1$) and in the non-conjugate approaches ($M_5$ vs. $M_2$), ($M_6$ vs. $M_3$) versions. Hence, for each writer, the Bayes factors  $\rm{BF}_{4,1}$, $\rm{BF}_{5,2}$ and $\rm{BF}_{6,3}$ are calculated, respectively. 
Note that while the marginal likelihoods that are needed to calculate $\rm BF_{4,1}$ can be obtained analytically, bridge sampling described in Section~\ref{marginal_likelihood_estimators_summary} is implemented for the estimation of $\rm{BF}_{5,2}$ and $\rm{BF}_{6,3}$.
Table \ref{lr_of_models} presents the results (in log scale) for $\rm{BF}_{4,1}$ (second column), $\rm{BF}_{5,2}$ (third column) and $\rm{BF}_{6,3}$ (forth column).

\begin{table}[ht!]
\begin{center}
\small
\begin{tabular}{|c|c|c|c|} 
 \hline
 & \multicolumn{2}{c|}{\textbf{Normal-Inverse-Wishart}}  & \multicolumn{1}
  {c|}{\textbf{Normal-LogNormal-LKJ}}\\
 \hhline{~---}
  \textbf{Writer} & \multicolumn{1}{c|}{\textbf{Conjugate Approach}} & \multicolumn{1}
  {c|}{\textbf{Hierarchical Approach}} & \multicolumn{1}
  {c|}{\textbf{Covariance Decomposition}}\\
 & \multicolumn{1}{c|}{$\log BF_{4,1}^*$ } & \multicolumn{1}{c|}{$\log BF_{5,2}^{**}$} & \multicolumn{1}{c|}{$\log BF_{6,3}^{**}$}\\
 \hline
        1 &~\,20.6  &~\,19.1 &~\,16.6\\ \hline
        2 &  174.7 &   173.4 & 164.5\\ \hline
        3 &  113.5 &   115.7 & ~\,89.7\\ \hline
        4 &~\,92.8 & ~\,93.1 & ~\,78.8\\ \hline
        5 &  205.9 &  203.8 & 202.9 \\ \hline
        6 &  116.3 &  111.7 & 114.2\\ \hline
        7 &  156.8 &  157.3 & 152.1\\ \hline
        8 &  132.8 &  131.8 & 128.8\\ \hline
        9 &  157.4 &  152.6 & 144.9\\ \hline
        10&~\,53.6  &~\,45.3 &~\,42.3\\ \hline
        11&  215.6 & 184.9 & 181.6\\ \hline
        12&  196.7 &   193.0 & 198.3\\ \hline
        13&  155.4 &   159.0 & 154.6\\ 
          \hline
        \multicolumn{3}{l}{\footnotesize 
        \it $^*$Closed form expression;  $^{**}$Bridge Sampling estimate}
\end{tabular}
\end{center}
\vspace{0.05cm}
\caption{Logarithmic Bayes factors of MANOVA models ($M_3$ or $M_4$ or $M_6$) vs. the Normal models ($M_1$ or $M_2$ or $M_5$).
%for each writer $i=1,\dots,13$. 
}
\label{lr_of_models}
\end{table}

From these results, it can be observed that  $\log\rm{BF}_{4,1}$, $\log\rm{BF}_{5,2}$ and $\log\rm{BF}_{6,3}$ are markedly higher than the value of five. These high BF values suggest very strong evidence  \citep{kass1995bayes} in favor of the MANOVA formulation. Thus, with reference to the available dataset, it can be claimed that the Bayesian MANOVA with characters as predictors fits the data per writer better for any of the prior setups.

{\bf Conjugate vs. non-Conjugate.}
Next, attention is given to the comparison between the conjugate and non-conjugate versions of each model, specifically considering Normal-Inverse-Wishart hierarchical priors and the Normal-LogNormal-LKJ prior vs. conjugate versions. Bayes factors $\rm{BF}_{1,2}$, $\rm{BF}_{1,3}$, $\rm{BF}_{2,3}$, $\rm{BF}_{4,5}$, $\rm{BF}_{4,6}$ and $\rm{BF}_{5,6}$ are then calculated to compare the two-level random effects model and the MANOVA model with different prior setups (conjugate, Normal-Inverse-Wishart hierarchical prior, Normal-LogNormal-LKJ prior). As already pointed out in Section~\ref{modeling}, handwriting literature emphasises the importance of modeling both within and between writers' variability. The available dataset will therefore be used to assess whether the results are consistent with this theory. This is achieved by considering the comparison between the conjugate and non-conjugate versions of the Normal and MANOVA models with a Normal-Inverse-Wishart prior. Table \ref{lr_of_conjugate_models2} presents the assessed Bayes factors per writer. According to these results, for the Normal model, the Bayes factors provide evidence supporting the non-conjugate prior (and thus, the need to also account for the between-writers variability) for 9 out of 13 writers. On the other hand, for the MANOVA model with Normal-Inverse-Wishart of conjugate and non-conjugate prior setup, the Bayes factors do not show a clear preference for the non-conjugate prior, which is only supported for about half of the writers (seven out of 13). Thus, this issue might require further investigation.

{\bf Inverse-Wishart vs. LogNormal-LKJ.}
Finally,  we compare models $M_1$, $M_2$, $M_4$, $M_5$ (based on the Normal-Inverse-Wishart prior) with models $M_3$ and $M_6$ (their corresponding Normal-LogNormal-LKJ prior counterparts). Table \ref{lr_of_iw_vs_lkj_models} presents the assessed Bayes factors $\rm{BF}_{2,3}$ and $\rm{BF}_{5,6} $ for each writer. Based on the results of this table, the Bayes factors support the Normal-LogNormal-LKJ prior over the corresponding Normal-Inverse-Wishart prior for 9 out of 13 writers under the Bayesian MANOVA. The evidence is even more systematic for the Bayesian Normal hierarchical models, where the Normal-LogNormal-LKJ prior is favored for almost all writers (12 out of 13). Similar results are presented for conjugate Normal-Inverse-Wishart and Normal-LogNormal-LKJ prior in Table \ref{lr_of_conjugate_lkj_models} at Appendix \ref{model_comparisons_appendix} for the estimated Bayes factors $\rm{BF}_{1,3}$ and $\rm{BF}_{4,6}$ per writer. 
%In the subsequent experiments on model accuracy, we include all models under consideration to ensure a comprehensive evaluation of them.

\begin{table}[!ht]
\begin{center}
 \small
\begin{tabular}{|c|c|c|c||c|c|c|} 
 \hline
 & \multicolumn{6}{c|}{\textbf{Normal-Inverse-Wishart}}\\
 \hhline{~------}
 & \multicolumn{3}{c||}{\textbf{Normal}} & \multicolumn{3}{c|}{\textbf{MANOVA}}\\
  \textbf{Writer}  & \multicolumn{3}{c||}{$\log BF_{1,2}^*$ } & \multicolumn{3}{c|}{$\log BF_{4,5}^*$}\\
\hhline{~------}

 & \multicolumn{1}{c|}{\textbf{Value}} & \multicolumn{1}{c|}{\textbf{Sign}} & \multicolumn{1}{c||}{\textbf{Interpretation}} &
 \multicolumn{1}{c|}{\textbf{Value}} & \multicolumn{1}{c|}{\textbf{Sign}} & \multicolumn{1}{c|}{\textbf{Interpretation}}\\
 \hline
    1 & ~\,~\,0.14 & + & Bare Mention & ~\,~\,1.60 & + & Substantial \\ \hline
        2 & ~\,-4.92 & - & Strong & ~\,-3.58 & - & Strong \\ \hline
        3 & ~\,-3.77 & - & Strong &  ~\,-6.02 & - & Extreme \\ \hline
        4 & ~\,-2.44 & - & Substantial & ~\,-2.72 & - & Substantial\\ \hline
        5 & ~\,~\,1.85 & + & Substantial & ~\,~\,3.99 & + & Strong \\ \hline
        6 & ~\,-2.13 & - & Substantial & ~\,~\,2.44 & + & Substantial \\ \hline
        7 & ~\,-2.52 & - & Substantial & ~\,-3.05 & - & Strong \\ \hline
        8 & ~\,-3.68  & - & Strong & ~\,-2.66 & - & Substantial \\ \hline
        9 & ~\,-0.17 & - & Bare Mention & ~\,~\,4.69 & + & Strong\\ \hline
        10 & ~\,~\,6.30 & + & Extreme & ~\,14.62 & + &  Extreme\\ \hline
        11 & ~\,~\,3.33 & + & Strong & ~\,34.05 & + & Extreme \\ \hline
        12 & ~\,-5.45 & - & Extreme & ~\,-1.72 & - & Substantial\\ \hline
        13 & -12.54 & - & Extreme & -16.09 & - & Extreme \\ 
  \hline
  \multicolumn{7}{l}{\footnotesize 
        \it $^{*}$Bridge Sampling estimate}
\end{tabular}
\end{center}
\caption{Logarithmic Bayes factors (per writer) comparing the conjugate and non-conjugate approaches of the Normal-Inverse-Wishart prior for the Bayesian Normal and  MANOVA models.}
%per writer
\label{lr_of_conjugate_models2}
\end{table}

\begin{table}[!ht]
\begin{center}
 \small
\begin{tabular}{|c|c|c|c||c|c|c|} 
 \hline
  & \multicolumn{3}{c||}{\textbf{Normal}} & \multicolumn{3}{c|}{\textbf{MANOVA}}\\
  & \multicolumn{3}{c||}{\textbf{Inverse-Wishart vs LogNormal-LKJ}} & \multicolumn{3}{c|}{\textbf{Inverse-Wishart vs LogNormal-LKJ}}\\
\textbf{Writer}  & \multicolumn{3}{c||}{$\log BF_{2,3}^*$ } & \multicolumn{3}{c|}{$\log BF_{5,6}^*$}\\
\hhline{~------}

 & \multicolumn{1}{c|}{\textbf{Value}} & \multicolumn{1}{c|}{\textbf{Sign}} & \multicolumn{1}{c||}{\textbf{Interpretation}} &
 \multicolumn{1}{c|}{\textbf{Value}} & \multicolumn{1}{c|}{\textbf{Sign}} & \multicolumn{1}{c|}{\textbf{Interpretation}}\\
 \hline
    1 & -17.65 & - & Extreme & -15.10 & - & Extreme \\ \hline
        2 & -28.13 & - & Extreme & -19.47 & - & Extreme \\ \hline
        3 & -34.46 & - & Extreme & -31.75 & - & Extreme \\ \hline
        4 & -48.50 & - & Extreme & -45.21 & - & Extreme\\ \hline
        5 & ~\,~\,5.39 & + & Extreme & ~\,~\,6.34 & + & Extreme \\ \hline
        6 & -17.67 & - & Extreme &  -20.14 & - & Extreme \\ \hline
        7 & ~\,-3.80 & - & Strong & ~\,~\,1.47 & + & Substantial \\ \hline
        8 & ~\,-2.00  & - & Substantial & ~\,~\,1.00 & + & Bare mention \\ \hline
        9 & ~\,-2.94 & - & Substantial & ~\,~\,4.69 & + & Strong\\ \hline
        10 & -23.91 & - & Extreme & -20.98 & - &  Extreme\\ \hline
        11 & -37.50 & - & Extreme & -34.19 & - & Extreme \\ \hline
        12 & -13.76 & - & Extreme & -19.12 & - & Extreme\\ \hline
        13 & -15.49 & - & Extreme & -11.08 & - & Extreme \\ 
  \hline
  \multicolumn{7}{l}{\footnotesize 
        \it $^{*}$Bridge Sampling estimate}
\end{tabular}
\end{center}
\caption{Logarithmic Bayes factors (per writer)
comparing the within-writer covariance prior setups (a) Inverse-Wishart or (b) LogNormal-LKJ prior of Bayesian Normal and MANOVA models.}
%per writer
\label{lr_of_iw_vs_lkj_models}
\end{table}

\newpage
\subsection{Accuracy of Models}
\label{models_efficiency}

In this section, the performance of the models under consideration is measured by means of simulation studies where the propositions of interest are whether a given individual is the writer of a questioned manuscript versus the alternative proposition that the writer is an unknown individual  (see Section \ref{sn_methodologies}). To this end, the models will be compared to study their ability to deliver Bayes factor values that support the correct proposition.

To assess the performance of the proposed models whenever proposition $H_1$  (same writer case) holds, a variety of simulated case studies were generated considering character measures from the same writer as questioned $\bm{y}_1$ and control $\bm{y}_2$ data, respectively. Therefore, the data of each writer from the available database was divided into two parts, one used as the questioned data and the other as control data. The proportion $\pi_{split}$ of data used as questioned material was randomly taken between 0.35 and 0.65, $\pi_{split}\in (0.35, 0.65)$, while the remaining data was taken as control material. This process was repeated 100 times (using different random splits) for each of the 13 available writers,  resulting in a total number of 1300 same-writer comparisons. The remaining writers serve as background data $\bm{X}$ to elicit the prior parameters for the models, namely, we use data from all writers except the one being analyzed. This made it possible to assess the false negative rate, that is, the percentage of cases giving rise to a Bayes factor less than~1 when it should be greater than~1. %Only one BF (out of a total number of 1300 comparisons) falsely supporting the proposition $H_2$ was obtained with the NIW model when only the character o was analysed. However, it is worth emphasising that this value is of limited magnitude and, based on the verbal scale proposed by \citet{kass1995bayes}, provides very weak support against the $H_1$ hypothesis. 

%The same procedure is followed when comparing material from different writers, i.e., $H_2$ is true. For this purpose, the data for all possible pairs of writers were randomly selected and divided into questioned and control data, following the same procedure as before. A part of the measurements originating from the first writer in the couple was used to act as questioned data, while a part of the measurements originating from the second writer was used to act as control data. This process was repeated 100 times for each of the 78 writer pairs, resulting in a total of 7800 comparisons between different writers. The same procedure is used to elicit the prior parameters for the models; specifically, data from all writers except the two being analyzed are utilized. This made it possible to assess the false positive rate, defined as the percentage of cases in which the Bayes factor is greater than one when one would expect it to be less than one.

A similar procedure is followed when comparing material from different writers, i.e., $H_2$ is true. For each pair of writers, case studies were generated by treating the measurements of the first writer as questioned data $\bm{y}_1$ and those of the second writer as control data $\bm{y}_2$. To account for the effect of sample size,  the same approach as in the simulated procedure for $H_1$ was followed. Hence, only a random subset of the data from each writer, in each pair, was considered. Specifically, a proportion $\pi_{split} \in (0.35, 0.65)$ of the measurements was randomly selected from the first individual of each pair (i.e., the writer of the questioned document), while the proportion of $(1-\pi_{split})$ of the measurements of the second individual of the pair was considered as the control data. This process was repeated 100 times for each of the 78 pairs of writers, resulting in a total number of 7800 different comparisons. The remaining writers serve as background data $\bm{X}$ to elicit the prior parameters for the models, namely, we use data from all writers except the two that are being analyzed. In these comparisons, the objective was to assess the false positive rate, defined as the percentage of cases in which the Bayes factor was greater than~1 when it should have been less than~1.

\begin{figure}[!ht]
\centering

\begin{subfigure}{\textwidth}
 \centering
\caption{Same Writer Comparisons}
    \includegraphics[height=6.9cm, width=1\textwidth]{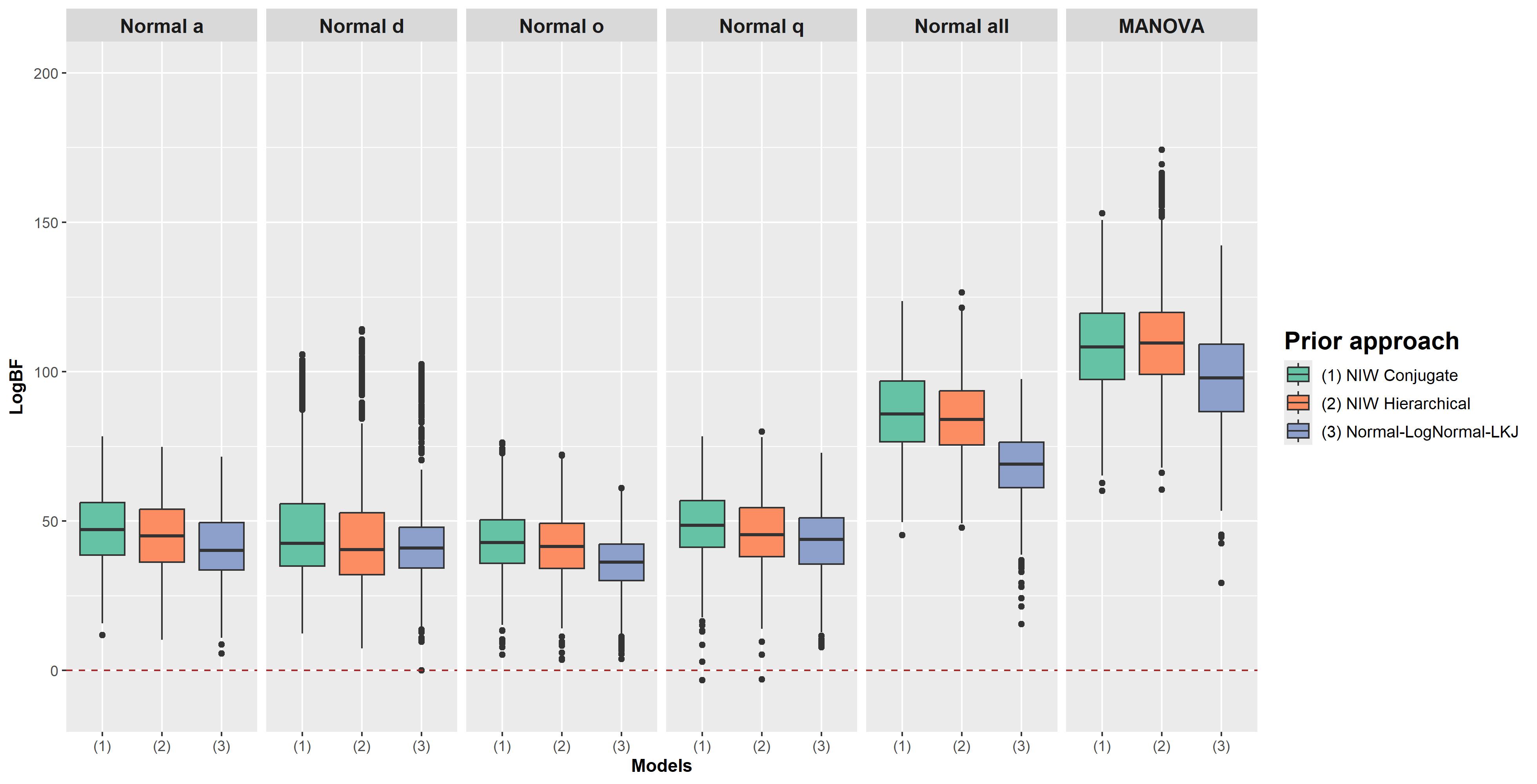}
\label{fig1:ss_logbf}
\end{subfigure}
\hfill
\hfill
\begin{subfigure}{\textwidth}
\centering
\caption{Different Writers Comparisons}
    \includegraphics[height=6.9cm, width=1\textwidth]{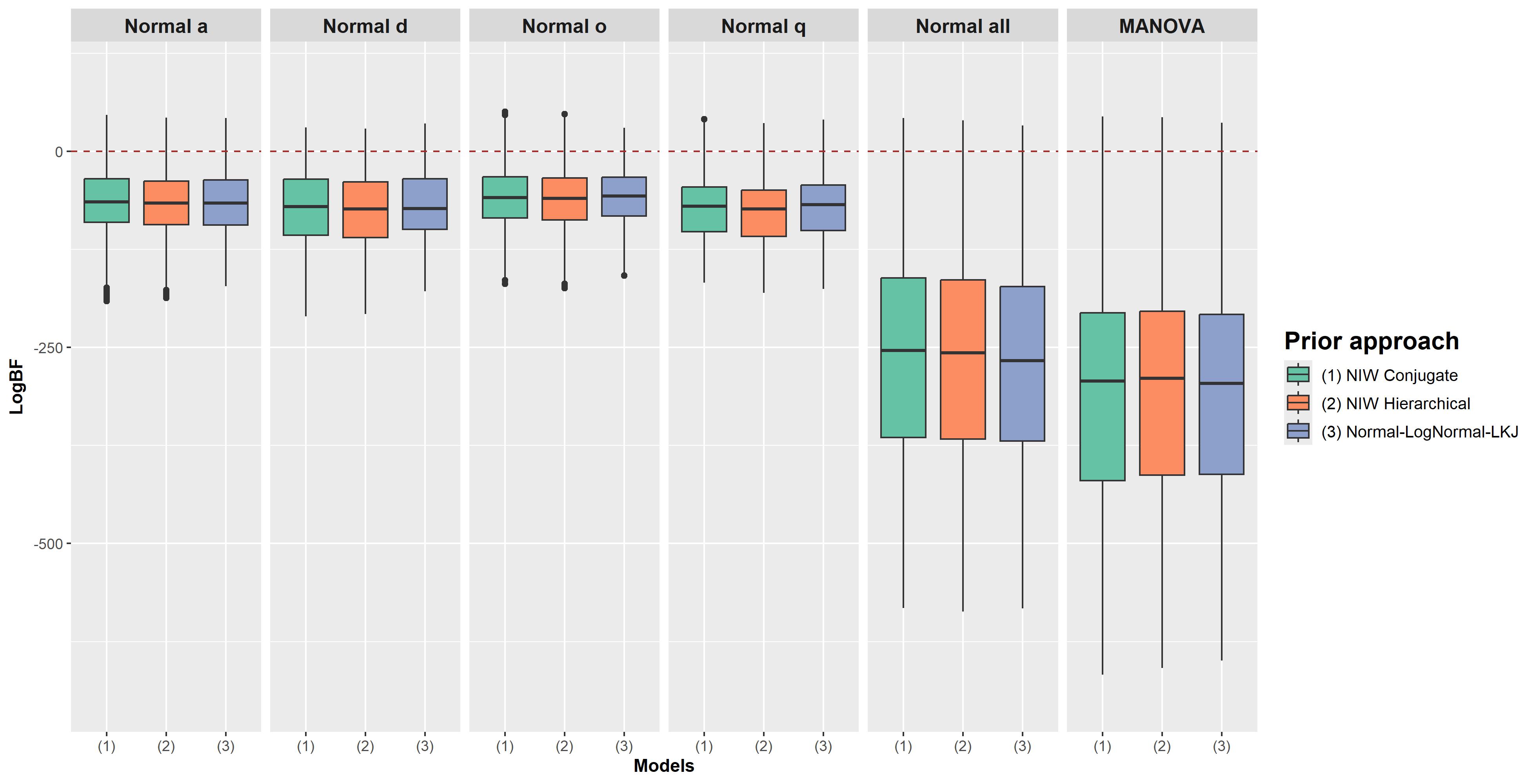}
\label{fig1:ds_logbf}
\end{subfigure}
\caption{Logarithmic Bayes factors for handwriting %authentication 
evaluation ($\log\rm{BF}$) for %the 
same writer (a) and different writers comparisons (b), %with 
using the data modeling approaches %of 
described in Sections~\ref{normal-inverse-wishart} and~\ref{bayesian_manova}.}
\label{fig1:logbf_experiments}
\end{figure}

The logarithm of the Bayes factors \eqref{eq_BF}, over all simulated case-scenarios,  obtained for testing competing hypotheses $H_1$ and $H_2$ (see Section \ref{ml_propositions}) is provided in Figures  \ref{fig1:ss_logbf} and \ref{fig1:ds_logbf} for same-writer and different-writers comparisons, respectively.  In these figures, the first five panels (from the left) provide the Bayes factors for the Bayesian Normal models ($M_1$, $M_2$, $M_3$) for characters {\em a}, {\em d}, {\em o}, and {\em q}  and for all characters together. The last panel provides the Bayes factor for the MANOVA models ($M_4$, $M_5$, $M_6$) using all characters together. In each panel, the BFs obtained analytically using the conjugate prior are depicted in the first boxplot (from left), while Boxplots 2 and 3 represent the BFs obtained using the hierarchical Normal-Inverse-Wishart and Normal-LogNormal-LKJ prior, where the marginal likelihoods are estimated by the bridge sampling described in Appendix~\ref {bridge_sampling_section}.

Looking at these box-plots, it is evident that almost all Bayes factors, regardless of the statistical modeling approach, support the correct hypothesis. 
Specifically, (almost) all the Bayes factors (in logarithmic scale) are well placed above zero for the first case-scenario (same writer) and below zero in most comparisons for the second case-scenario (different writers). 
In particular, in the case of the same writers (i.e., under $H_1$), we observe only one false negative case for 
character $q$ under NIW conjugate and the  NIW hierarchical approaches, and for character {\em d} under the Normal-LogNormal-LKJ prior, while no false negatives occurred for the remaining characters and approaches. 

As far as the assessment of the performance of the different methods under hypothesis $H_2$, the number of false positives %(i.e., cases giving rise to Bayes factors, in logarithmic scale, greater than~0)
over the total number of 7800 different writers comparisons is presented in Table \ref{fp_random}, for both modeling approaches and different prior setups. 
Specifically, the total number of false positives ranges from a minimum of~47 to a maximum of~589  over a total number of 7800 different writer comparisons (i.e., 0.6\% -- 7.6\%).  

For individual characters, the highest false positive rates (up to 7.6\% for NIW Conjugate approach) were observed for character {\em a}, while character {\em q} consistently produces the lowest rates. When considering all characters, the false positive rate drops further from $2.2\%$, which is the lowest false positive rate for the character {\em q} (the best performing character), to $0.9\%$ under the Normal-LogNormal-LKJ prior setup.

Concerning the prior setups, the NIW Conjugate and NIW Hierarchical models produced similar false positive rates for most characters, although the Hierarchical version generally performs slightly better. 
The Normal-LogNormal-LKJ setup resulted in lower false positive rates for characters {\em a} and {\em o}, but higher rates for characters {\em d} and {\em q} compared to the NIW approaches. 

The MANOVA model exhibits the smallest proportion of false positives among all methods, with rates consistently below 0.7\%. 
This indicates superior performance for MANOVA in cross-writer discrimination when all characters are considered together. 
Furthermore, only minor differences were observed among the different prior setups within the MANOVA formulation.

%
% *** put in only in the phd **** 
%Finally, it should be noted that all models and approaches exhibited reduced accuracy for the pair of writers 6 and 8. In this case, even the best-performing model, MANOVA, shows a very high false positive rate (around 50) of producing a Bayes factor supporting the incorrect hypothesis ($H_1$ while $H_2$ is true. This result is not surprising given the strong similarity between these writers, as also highlighted by the Mahalanobis distance (Figure~\ref{mahalanobis})."

\begin{table}[!ht]
\begin{center}
\begin{tabular}{|c|c|c|c|c|} 
 \hline
&  &\multicolumn{2}{c|}{\textbf{Normal-Inverse-Wishart}}  & \multicolumn{1}{c|}{\textbf{Normal-LogNormal-LKJ}}\\
 \hhline{~~--~}
  \textbf{Model} & \textbf{Characters} & \multicolumn{1}{c|}{\textbf{Conjugate}} & \multicolumn{1}
  {c|}{\textbf{Hierarchical}}& \\ 
  %\multicolumn{1}{c|}{\textbf{Covariance Decomposition}}\\
 \hline
        &{\em a} &589 (7.6\%) &560 (7.2\%) &545 (7.0\%)\\ 
        \hhline{~----}
        &{\em d} &  491 (6.3\%) &  308  (3.9\%) & 589 (7.6\%) \\ 
        \hhline{~----}
    Normal &{\em o} & 410 (5.3\%) &   363 (4.7\%)& 301 (3.9\%) \\ 
        \hhline{~----}
        &{\em q} & 213 (2.7\%) & 168  (2.2\%) & 265 (3.4\%)\\ 
        \hhline{~----}
       & all &  113 (1.4\%) &   \; 86 (1.1\%) &  \; 68 (0.9\%)\\ \hline
       MANOVA &all &  \; 53 (0.7\%) &  \; 51 (0.7\%) &  \; 47 (0.6\%)\\ 
       \hline

\end{tabular}
\end{center}
\caption{Number of false positives over %a total of 7800 generated comparison datasets; For each comparison of writers (out of 78 combinations), we have considered 100 random splits with the split percentage taken randomly from the  uniform (0.35,0.65).
7800 different writer comparisons (percentages in brackets).}
\label{fp_random}
\end{table}

\newpage
\section{Sensitivity Analysis of Prior Elicitation}
\label{sensitivity_analysis_prior}

%The prior elicitation is particularly significant in hierarchical models, where priors specified at different levels can shape the resulting posterior distribution. 

In Bayesian hierarchical modeling, prior elicitation is critical, as it can have a substantial impact on the resulting Bayes factor values. In this study, we implement a subjective prior methodology in which prior distributions are informed by empirical data derived from handwriting samples of a group of writers, different from the ones being compared. This approach explicitly incorporates prior knowledge regarding the characteristics of the handwriting features based on Fourier analysis used in the study. Such a strategy is particularly advantageous in our case, where we wish to determine whether the available handwriting evidence originates from the same writer or from a different writer.

To address the influence of this subjective prior framework, in Section \ref{bootstrap_section}, we implement a sensitivity analysis of the prior elicitation framework. Following the evaluation procedure described in Section \ref{models_efficiency}, namely the data of a writer or pair of writers from the full dataset $\bm{\mathcal{D}}$ are randomly selected and serve as the questioned data $\bm{y}_1$ and control data $\bm{y}_2$, while the remaining writers are considered as the background data $\bm{X}$. The sensitivity analysis is conducted by constructing the prior distributions from a random subsample of each background writer’s data rather than from the entire background dataset $\bm{X}$.  Specifically, for each background writer, a random subsample with replacement comprising 50\% of their data is considered.

%As the subsample percentage increases, the results become more closely aligned with those obtained when considering the entire background dataset. 
%Therefore, we report this percentage, as it is worthy in the context of our study.

Furthermore, for the degrees of freedom of the Wishart prior in $M_{1-2}$ and $M_{4-5}$, we set $\nu = p+2$ as the default low-information value, while for the LKJ prior in $M_3$ and $M_6$, we use $\eta = 1$. Hence, in the second part of the sensitivity analysis in Section \ref{section_specification_summary}, we examine how different prior values for these variance–covariance parameters affect the posterior Bayesian evidence, as measured by the Bayes factor.

\subsection{Subsampling of Background Data}
\label{bootstrap_section}

In this section, we investigate the effect of the prior specification by applying subsampling to the background data used for prior elicitation. 
The procedure of the randomly selected cases that are used for the evaluation of the two hypotheses ($H_1$ vs. $H_2$) is the same as the one described in Section \ref{models_efficiency}. 
Specifically, to isolate the effect of prior elicitation, we utilize a single random data split for both the same-writer and different-writer experiments, thereby minimizing potential confounding effects arising from data partitioning. This analysis focuses on the Bayesian MANOVA model, which has been identified as the most effective and recommended approach in Section \ref{models_efficiency}. 
To assess the robustness and stability of the Bayes factor and the resulting decision, we performed 30 iterations of random subsampling with replacement, using 50\% of the background data for each writer in each evaluation case. To clarify further, three distinct datasets are considered in our analysis: (i) the questioned data $\bm{y}_1$, (ii) the control data $\bm{y}_2$, and (iii) the background data $\bm{X}$, our objective in this section is to evaluate the effect of varying the third dataset.

\begin{figure}[!ht]
    \centering
    \includegraphics[width=1\textwidth]{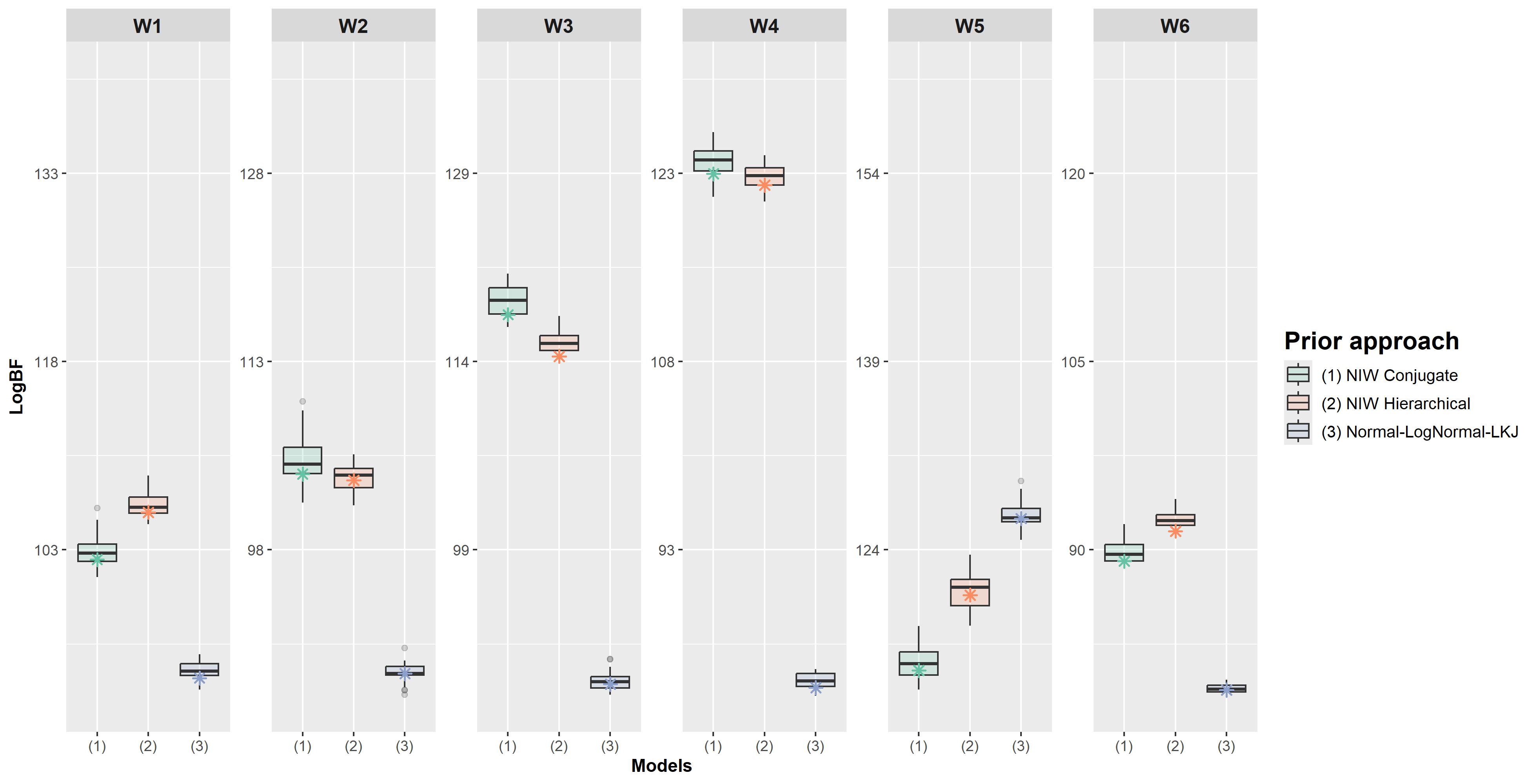} 
    \par\vspace{0.2cm} % optional spacing
    \parbox{\textwidth}{%
        \footnotesize\it
        \hspace{0.5cm} $^*$indicates the $\log\mathrm{BF}$ using the complete background dataset for each case.
    }
    \caption{
        Boxplots of Logarithmic Bayes factors ($\log\mathrm{BF}$) for handwriting evaluation for the same writer scenarios over different subsamples of background data for the Bayesian MANOVA approach.
    }
    \label{same_source_bootstrap}
\end{figure}

\begin{figure}[!ht]
    \centering
    \includegraphics[width=1\textwidth]{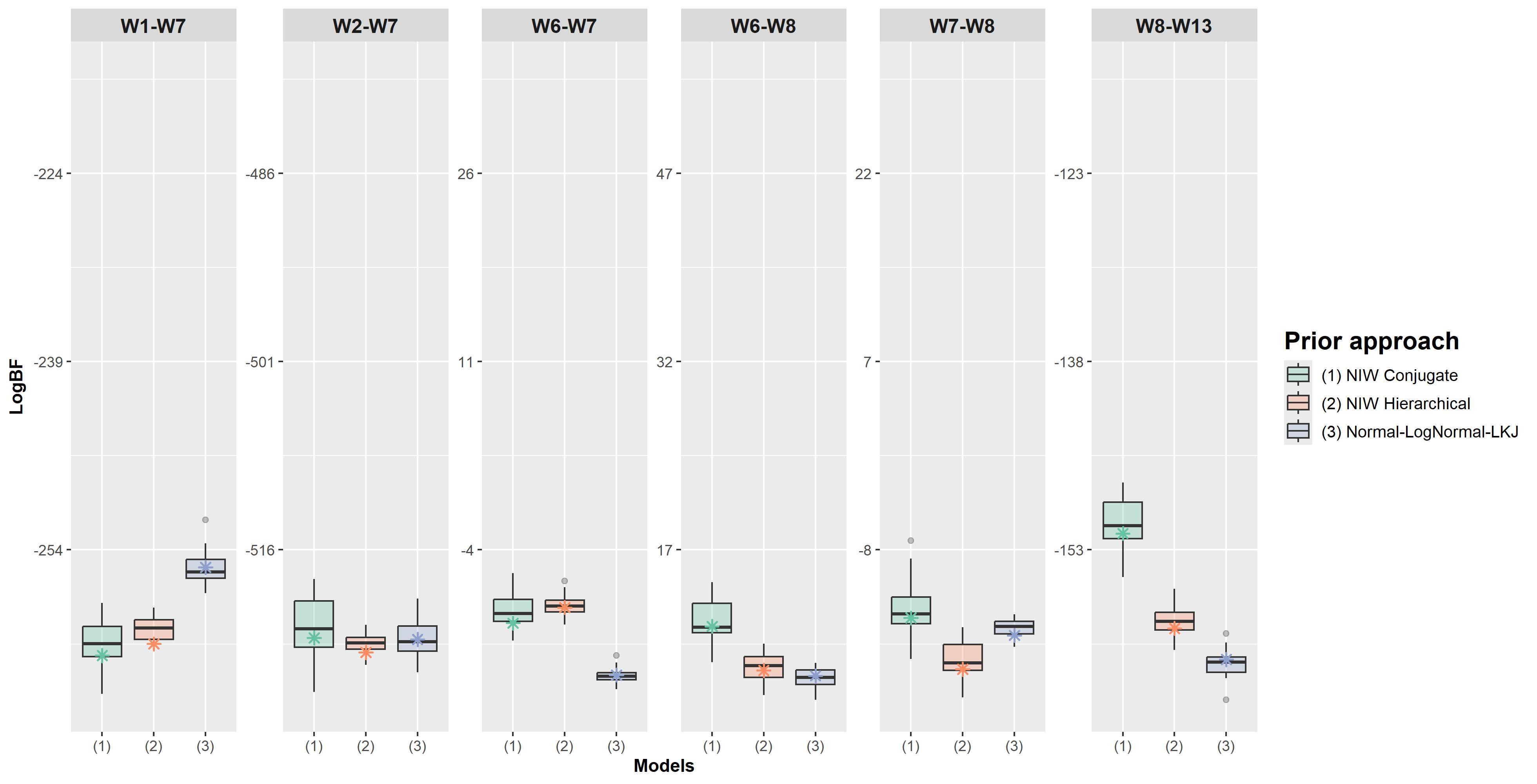}
    \par\vspace{0.2cm} % optional spacing
    \parbox{\textwidth}{%
        \footnotesize\it
        \hspace{0.5cm} $^*$indicates the $\log\mathrm{BF}$ using the complete background dataset for each case.
    }
    \caption{
    Boxplots of Logarithmic Bayes factors ($\log\mathrm{BF}$) for handwriting evaluation for the different writers scenarios over different subsamples of background data for the Bayesian MANOVA approach. }
    \label{different_source_bootstrap}
\end{figure}

Figures \ref{same_source_bootstrap} and \ref{different_source_bootstrap} present the logarithmic Bayes factors (LogBF) values for handwriting evaluation across 30 different elicited priors, comparing the performance of 
 Bayesian MANOVA models under three prior setups (conjugate, hierarchical Normal-Inverse-Wishart, Normal-LogNormal-LKJ). Figure \ref{same_source_bootstrap} presents the results for same-writer comparisons, while Figure \ref{different_source_bootstrap} shows comparisons between different writers. The asterisks in  Figure \ref{different_source_bootstrap} indicate the LogBF values computed using a prior obtained from the complete background data, serving as a reference point for the subsampled distributions. 
 These reference points are generally located near the boxplots, indicating that our proposed approach is relatively robust (both in terms of the Bayes factor values and the resulting decision) to variations in the background data used for prior elicitation, and remain close to the Bayes factor obtained from the full dataset.

% TO BE CHECKED IF THEY ASKED
%It can be observed that the LogBF resulting from the conjugate prior approach using the entire background dataset tends to smaller values compared to LogBF values obtained from subsamples. This occurs because the estimated scale matrix of the Inverse-Wishart prior, calculated from the whole background data, is typically larger than that estimated from subsamples (see Eq. \eqref{within_writer_u} in Appendix \ref{prior_estimation_normal_iw}). As a result, the prior becomes more diffuse in this cases, leading to a wider posterior and consequently smaller marginal likelihoods. In contrast, the Normal-LogNormal-LKJ prior approach is more flexible due to its covariance decomposition, which allows it to avoid this effect.

Overall, the results indicate that the support provided by LogBF is not affected by the constructed prior elicitation and correctly identifies the true hypothesis in nearly all cases. Nevertheless, a certain degree of variability is observed, which is expected. For example, comparisons involving different writers (e.g., W8–W13, W7–W8) display notable variability in LogBF values of the conjugate approach, although these do not alter the overall support in favor of \(H_2\). This pattern indicates a certain variability of the Bayes factor to the choice of background data, whereas same-writer comparisons (e.g., for W3, W4) demonstrate substantially greater stability. Further examples can be found in Appendix \ref{section_sensitivity_analysis_appendix}.

%The choice of prior can influence both the central tendency and the variability of the LogBF, with some priors (notably the conjugate prior) occasionally producing greater variability. %These results highlight the sensitivity of the approach to the specific prior used. 

To assess the impact of this variability, we examine four writer pairs (W5–W6, W6–W7, W6–W8, and W7–W8) that exhibit the highest similarity and were selected based on having cases with $ |\log BF| < 10$. For the remaining writer pairs, the observed variability does not appear to affect the overall level of support. 
%In some cases, different prior elicitation can result in differences of up to 5 LogBF units, which may substantially affect the support for competing hypotheses. To quantify this observation, we conducted extensive experiments on writers whose LogBF values are closer to zero. As observed in Section \ref{models_efficiency}, this occurs only in different writers comparisons, specifically, we considered comparisons between writers W5–W6, W6–W7, W6–W8, and W7–W8.
To elaborate further, we performed 30 random data selections from each pair of writers, following the procedure described in Section \ref{models_efficiency}, namely changing the questioned and control datasets. Then, for each pair of writers, subsamples with replacement were drawn, comprising 50\% of the data from the remaining writers, which served as the background dataset. In total, this procedure resulted in 3600 comparisons (case studies) for analysis.

Table \ref{different_support_table} summarizes the sensitivity of support for competing hypotheses under each MANOVA prior specification. The NIW conjugate prior approach presents the highest number of cases with inconsistent support (i.e., support shifts across background data subsamples) among the 30 priors elicited from different background data subsets (9 cases, or 7.5\% of 120 cases). For these nine cases, the average LogBF range is found to be 8.51 units, while the widest LogBF subsampling interval (worst case scenario) takes values in the interval $(-6.24, 4.79)$. Conversely, the Normal-LogNormal-LKJ prior setup exhibits fewer cases (2 cases, 1.67\%) and smaller average differences, indicating a relatively more stable inference under prior variation.  The NIW hierarchical prior shows intermediate behavior between these two setups.

%This suggests greater sensitivity to prior elicitation under this setup, because we observe a wide range of outcomes—from strong support for one hypothesis to strong support for its competing alternative. 
%These findings reinforce the importance of carefully selecting and validating prior formulations in Bayesian models, as prior choice can meaningfully influence the strength and direction of evidence, especially when LogBF is near zero and hypotheses are competitively supported. We suggest that this approach becomes more robust as the number of background writers is sufficient. 

\begin{table}[!ht]
\centering
\renewcommand{\arraystretch}{1.15}
\begin{tabular}{l c cc}
\toprule
 \multicolumn{4}{c}{\textsc{Inconsistent Cases$^{*}$}} \\ 
 \cmidrule(lr){1-4}
 & & \multicolumn{2}{c}{\textsc{LogBF Summaries}} \\
\cmidrule(lr){3-4}
\textbf{Prior setup} & \textbf{Support shifts$^{**}$ (\%)} & \multicolumn{1}{c}{\textbf{Avg.\ range$^\dagger$}} & \multicolumn{1}{c}{\textbf{Worst case interval$^\ddagger$}} \\
\midrule
NIW Conjugate           & 9 (7.50) & 8.51 & $(-6.24,\;4.79)$ \\
NIW Hierarchical        & 5 (4.17) & 3.83 & $(-2.69,\;1.48)$ \\
Normal–LogNormal–LKJ    & 2 (1.67) & 3.42 & $(-1.49,\;3.05)$ \\
\bottomrule
\end{tabular}

\vspace{0.5em}
\begin{minipage}{0.9\linewidth}
\footnotesize\itshape
$^*$ Results are based on 30 random subsets of four compared writer pairs and 30 random 50\% subsamples of background writers’ data ($4 \times 30 \times 30 = 3600$).\\
$^{**}$ Number of inconsistent-support cases across 30 random subsets of compared writers ($4\times30$=120). \\
%$^{**}$ Summaries over $4\times30=120$ writer pairs. \\
$^\dagger$ Average logBF range across inconsistent cases; range is computed over 30 background-data subsamples. \\
$^\ddagger$ LogBF interval $(\min,\max)$ of the case with highest range of logBFs.
\end{minipage}

\vspace{0.5em}
\caption{Sensitivity analysis for the most similar writer pairs {\scriptsize(W5–W6, W6–W7, W6–W8, and W7–W8)}
under different prior setups. }
\label{different_support_table}
\end{table}

\newpage
\subsection{Specification of the Inverse-Wishart's Degrees of Freedom and the parameter of the LKJ distribution }
\label{section_specification_summary}

In Bayesian hierarchical modeling, the choice of priors for covariance matrices is critical, since it has a direct impact on the resulting Bayes factor values,  as observed in  Section~\ref{experimental_results}. 

The Inverse-Wishart prior is a traditional choice for modeling covariance matrices. It is a conjugate prior for the multivariate Normal distribution, which simplifies the computational aspects of Bayesian inference. The Inverse-Wishart prior is parameterized by a scale matrix and by the degrees of freedom, which control the prior's concentration around the scale matrix. 

On the other hand, the covariance decomposition with the LKJ approach provides a more flexible alternative for modeling correlation matrices. 
The LKJ prior is parameterized by a shape parameter, $\eta$, which controls the concentration of the prior around the identity matrix. 
When $\eta = 1$, the LKJ prior is uniform over the space of correlation matrices, making it a non-informative prior. 
Higher values of $\eta$ result in stronger concentration around the identity matrix, reflecting a prior belief in weaker correlations. 
In this section, we present a sensitivity analysis for different values of the degrees of freedom in the approach using the Inverse-Wishart prior and for different values of $\eta$ in the LKJ prior approach. 

In the experiments using the Inverse-Wishart prior, the degrees of freedom $\nu$ vary from the minimum admissible value of 11\footnote[1]{That is, $p+2=11$  for $p=9$.} and from 20 to 50, with increasing in steps of 10, that is, $\nu \in \{ 11,20,30,40,50\}$. 
The Bayes factor has therefore been calculated for each value of $\nu$, for all comparisons between characters from different writers,  considering the Bayesian Normal and MANOVA models with Normal-Inverse-Wishart prior setups (conjugate and non-conjugate). 

The results show that the Bayes factor is quite sensitive to the choice of degrees of freedom, which has an incremental effect on its value; see Appendix \ref{dof_iw_appendix}. 
This occurs because higher degrees of freedom make the prior of $\bm{W}$ more concentrated within a smaller region of $\mathbb{R}^p$ \citep{gaborini2021bayesian}.
Therefore, for larger values of $\nu$, the variability of the variance-covariance matrix decreases. Consequently, we expect that as $\nu$ becomes large, the resulting log-Bayes factors will approach an upper bound which will always support the hypothesis that measurements originate from the same writer (even if the measurements originate from different writers); see Appendix \ref{dof_iw_appendix} for more details.

Let us now consider the experiments using the LKJ prior approach; the parameter $\eta$ was set to values ranging from the non-informative value of one (1) to increasingly informative values of 2, 5, 10, and 20.  
Higher values of $\eta$ result in a greater concentration of the prior around the identity matrix, reflecting the assumption that Fourier coefficients are uncorrelated (an assumption motivated by the Fourier analysis). 
For each value of $\eta$, the Bayes factor was computed for all pairwise comparisons between different writers, employing the two considered models: the Bayesian Normal and MANOVA models with Normal-LogNormal-LKJ prior setup. Results show that the Bayes factor is quite sensitive to the choice of $\eta$, particularly for values above five (5), where its effect becomes increasingly more marked; see Appendix \ref{eta_lkj_appendix} for a related discussion. 
This sensitivity arises because higher values of $\eta$ imply stronger prior beliefs in the absence of correlation among variables. %Further details can be found in Appendix \ref{eta_lkj_appendix}

The empirical comparison of the two approaches for modeling the variance-covariance matrix (the Inverse-Wishart prior and the LKJ prior) suggests that the LogNormal-LKJ model generally provides more robust results, especially for the Bayesian MANOVA model.  
This reflects the model's flexibility arising from the use of the covariance decomposition. 
The log-normal prior provides the appropriate information for the standard deviations that lie on the diagonal of the covariance matrices, while the LKJ prior controls the information about the correlation structure. 
Based on our analysis, we recommend using the LogNormal-LKJ prior with the use of an informative LogNormal prior derived from background data measurements. 
This approach provides a solid foundation for the implementation of the proposed models using the LogNormal-LKJ prior setup.

%\clearpage
%\newpage
\section{Conclusions and Future Work}
\label{conclusions}

This study deals with the challenge of managing the uncertainty in forensic handwriting examination, which affects both inferential and decision-making processes in legal contexts. An experimental study has been conducted to facilitate (a) the management of complex multivariate data, (b) the combination of multiple sources of information (in this specific context, characters of different types) for the assessment of a joint probabilistic value, and (c) the identification of optimal modeling approaches. 
The proposed Bayesian statistical framework provides a comprehensive quantitative tool for uncovering valid and informative patterns in handwriting data, with the potential to substantially impact several areas of handwritten document analysis.

The proposed Bayesian probabilistic framework has been applied to a series of forensic case studies generated from real handwriting data. 
In particular, the performance of a two-level random effects model proposed for handwriting examination by \cite{bozza2008probabilistic} was compared with that of a Bayesian hierarchical MANOVA model, whose implementation is novel in this context. The study addressed the following five research questions:

\begin{enumerate}[label=(\roman*)]
\item Which model fits the data better?
\item Which model has better discrimination performance (lower false positive and false negative rates)?
\item Does modeling the variability between writers have an impact on the final results?
\item Which character is most informative with respect to the ultimate goal of identifying the writer of a questioned handwritten manuscript?
\item Do the results of the implemented models remain robust across different background data and prior specifications?
\end{enumerate}

To address  the first research question, six alternative Bayesian models have been considered and compared. These models arise from two likelihood structures: (a) a multivariate Normal model (See Section \ref{normal-inverse-wishart}), and (b) a MANOVA model that accounts for character-level variability (See Section \ref{bayesian_manova}). For each likelihood, three different prior formulations are examined, resulting in distinct Bayesian models: (i) a conjugate Normal-Inverse-Wishart prior, (ii) a hierarchical Normal-Inverse-Wishart prior, and (iii) a Normal-LogNormal-LKJ prior specification. The results, presented in Table \ref{lr_of_models}, showed strong support in favor of the Bayesian MANOVA model. The incorporation of the between-writers variability into the Normal-Inverse-Wishart (NIW) prior formulation appears to have a positive effect, with Bayes factors supporting this prior setup (and thus against the conjugate model) for~9 out of~13 writers, as shown in Table~\ref{lr_of_conjugate_models2}. For the MANOVA model with Normal-Inverse-Wishart prior, no clear advantage emerges between the conjugate and the hierarchical prior setups, as the results support both approaches (see Table~\ref{lr_of_conjugate_models2}). Finally, we compared the two prior setups Normal-LogNormal-LKJ prior and the Normal-Inverse-Wishart. The Bayes factors, presented in Table \ref{lr_of_iw_vs_lkj_models}, support the Normal-LogNormal-LKJ prior over the Normal-Inverse-Wishart prior for most writers in both statistical models.

In order to address the research questions (ii)--(iv), a series of forensic case scenarios has been simulated by randomly selecting measurements from the available data to act as questioned and control material, respectively. Regarding the second research question, the Bayesian MANOVA model performed better than the Bayesian Normal model, with a false positive rate of around $0.6\%$ compared to the $0.9\%$ for the Normal model when all characters are analysed jointly. This corresponds to a reduction in the false positive rate of approximately $33.3\%$ compared to the Bayesian Normal model. This is not surprising, as the latter model does not take character-level variability into account. Furthermore, it was observed that employing the Normal-LogNormal-LKJ prior in the Bayesian MANOVA framework resulted in a 0.1\% reduction in false positives compared to the Normal-Inverse-Wishart prior. This finding aligns with the results obtained from model fitting. However, when each character was analyzed separately using the Bayesian Normal model, the Normal-Inverse-Wishart prior demonstrated superior performance for two of the four characters considered. With respect to supporting the correct hypothesis when the handwritten material originated from the same writer (i.e., under $H_1$), both models consistently classified all relevant cases correctly, with the sole exception of the NIW model for character {\em q}, which exhibited an exceedingly small proportion of misclassifications ($<0.1\%$; one over 1200 generated scenarios).

Concerning research question (iii), the results presented in Section \ref{models_efficiency} show that the hierarchical prior setup, which accounts for between-writer variability, achieves better overall performance with either Normal-LogNormal-LKJ or Normal-Inverse-Wishart prior compared to the conjugate approach (see Table~\ref{fp_random}). In contrast, the MANOVA model shows better results for the Normal-LogNormal-LKJ prior setup. This finding, along with the other experiments discussed in Section \ref{models_efficiency}, where the accuracy of proposed models for handwriting discrimination has been tested, is in agreement with the results of the model comparison for the first research question (see Section \ref{comparison_of_models}). Similarly, a consistent reduction is observed between the conjugate and the non-conjugate version of the Bayesian Normal model with Normal-Inverse-Wishart prior. However, these results require further investigation and validation by ideally implementing the proposed methodology in a variety of different handwriting datasets by also involving a larger number of writers.

The analyses conducted to address research question (iv) reveal which character, among the four analyzed, is the most informative in terms of discrimination between the two hypotheses under investigation. Character {\em q} has consistently been found to produce fewer instances of misleading evidence, with a false positive rate of 2.2\%. On the other hand, the character {\em a} is the least informative in terms of discriminatory power, with a false positive rate reaching 7.2\% with the hierarchical NIW prior. Finally, a false positive rate of~$3.9$ and~$4.7$, respectively, is observed for the characters {\em d} and {\em o}.

 A sensitivity analysis was performed to evaluate the robustness of the results to prior elicitation and specification, addressing research question (v). By conducting subsampling on 50\% of background writers’ data of each evaluation case, we observed that the Bayes factor did not change to a degree that would alter the model support decisively. In particular, under the MANOVA model with the Normal-LogNormal-LKJ prior, the support for the tested hypotheses was reversed in only two of the 120 comparisons.  These two cases involved the writers with the most similar handwriting text. This finding highlights the model’s robustness in providing consistent support for or against a given hypothesis (see Section~\ref{sensitivity_analysis_prior}). Furthermore, the degrees of freedom parameter of the Inverse-Wishart distribution and the $\eta$ parameter of the LKJ distribution play a crucial role in the behavior of the resulting Bayes factors. The Bayes factors are notably sensitive to the specification of these prior parameters. Specifically, higher values of the degrees of freedom and $\eta$ lead to increasingly informative priors, which produce larger Bayes factors and thus stronger support for (the false) hypothesis $H_1$. Given this sensitivity, careful consideration is required when specifying the prior degrees of freedom (in the Inverse-Wishart distribution) and the prior parameter $\eta$ (in the LKJ distribution). A recommended approach is to adopt weakly informative priors. This can be achieved by setting the degrees of freedom of the Inverse-Wishart distribution to its minimum value $p+2$, and by specifying $\eta =1$ for the LKJ distribution. 

Finally, there is the open issue concerning the scale of the resulting Bayes factors. 
Their magnitude in this study is in accordance with the ones from previous related work \citep{bozza2008probabilistic}, but are open to critical remarks on robustness that cannot be demonstrated empirically. Similar situations have been discussed in the field of genetic (DNA) evidence (see e.g. \citet{Hopwoodetal2012}). 
However, values of this magnitude can be difficult to justify, especially with such a limited database.

For future work, we suggest including neighbouring characters of loop characters in the analysis, which may provide more information and improve the accuracy of results. Such data can be modelled by a two-way MANOVA model with neighbouring characters and loop characters as dummy variables.

\section{Acknowledgements}
We are grateful to the Swiss National Science Foundation for its generous support through grant number 100011$\underline{\ }$204554/1 (The anatomy of forensic inference and decision), which enabled us to conduct this research. We also thank Dr. Raymond Marquis (University of Lausanne) for providing us with the data and supporting and consulting us in this extensive handwriting analysis. 
%Furthermore, we appreciate the constructive feedback and suggestions from the anonymous reviewers, which helped us to improve the quality of our paper.

\newpage
\appendix

\renewcommand\thefigure{\thesection.\arabic{figure}}    
\renewcommand\thetable{\thesection.\arabic{table}}    

\setcounter{figure}{0}
\setcounter{table}{0}

\section*{APPENDIX}

\section{Descriptive Analytics of Fourier Coefficients and Surface size}
\label{descriptive_analysis}

\begin{figure}[!ht]
    \centering
\includegraphics[width=0.99\textwidth]{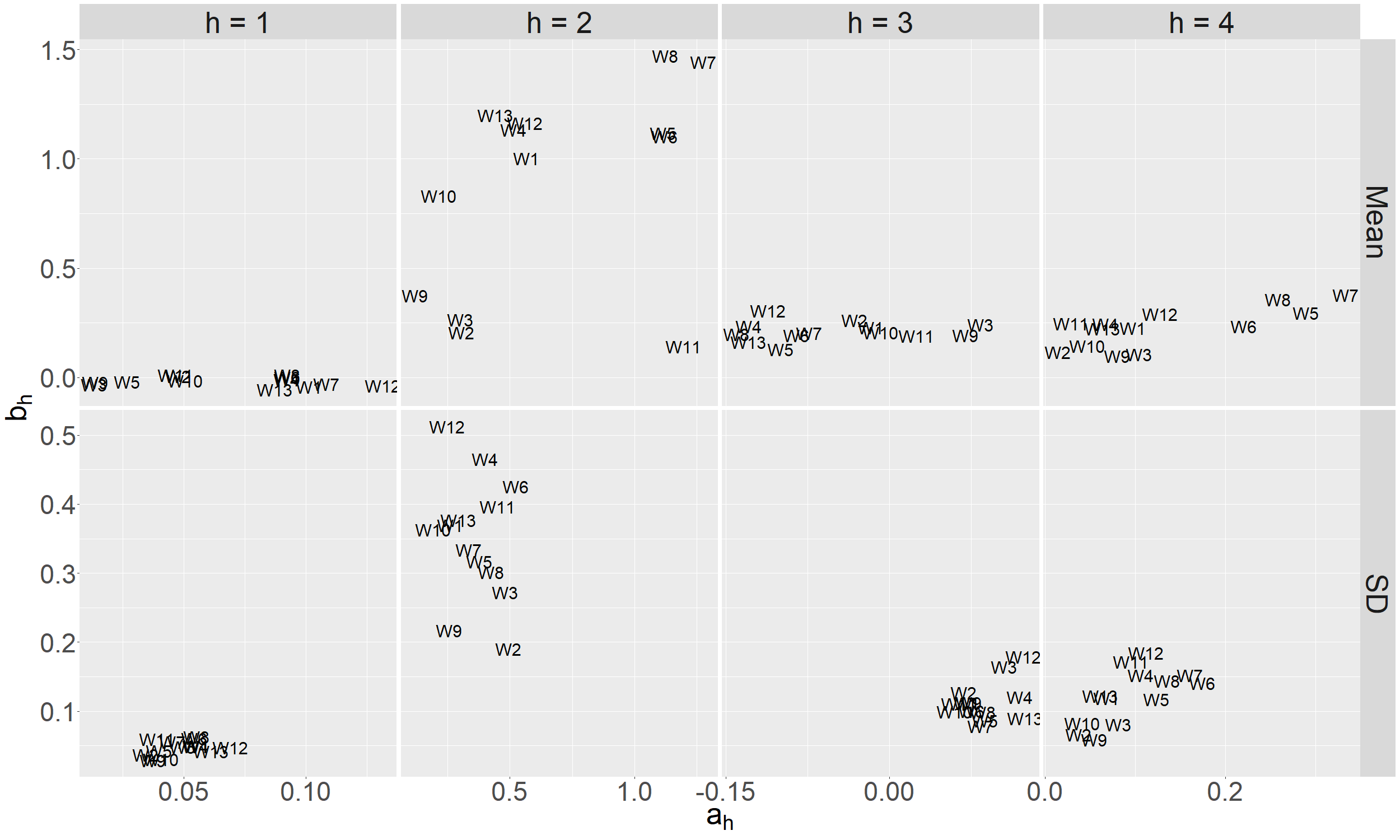}
\caption{Mean and standard deviation of every pair of Fourier coefficients $(a_h,b_h)$, for each harmonic, each writer, and across all characters.}
\label{coefficient_representation}
\end{figure}

In this section, we present a basic descriptive analysis of the main characteristics of the sample used for the implementation of the proposed methodology. The Fourier coefficients can be represented in Cartesian coordinates as in Figure \ref{coefficient_representation}, where there are depicted, for each harmonic, each writer and across all characters, the mean and the standard deviation of the pair of coefficients ($a_h$,$b_h$). This graphical representation can be informative of the discriminating power of each harmonic. In particular, the second harmonic measuring ellipticity seems the most informative for discriminating purposes. For an analytical description of the available characters' features, the reader can refer to graphical illustrations in Figure \ref{fig:figures}.

\begin{figure}%[!ht]
    \centering
\includegraphics[width=0.99\textwidth]{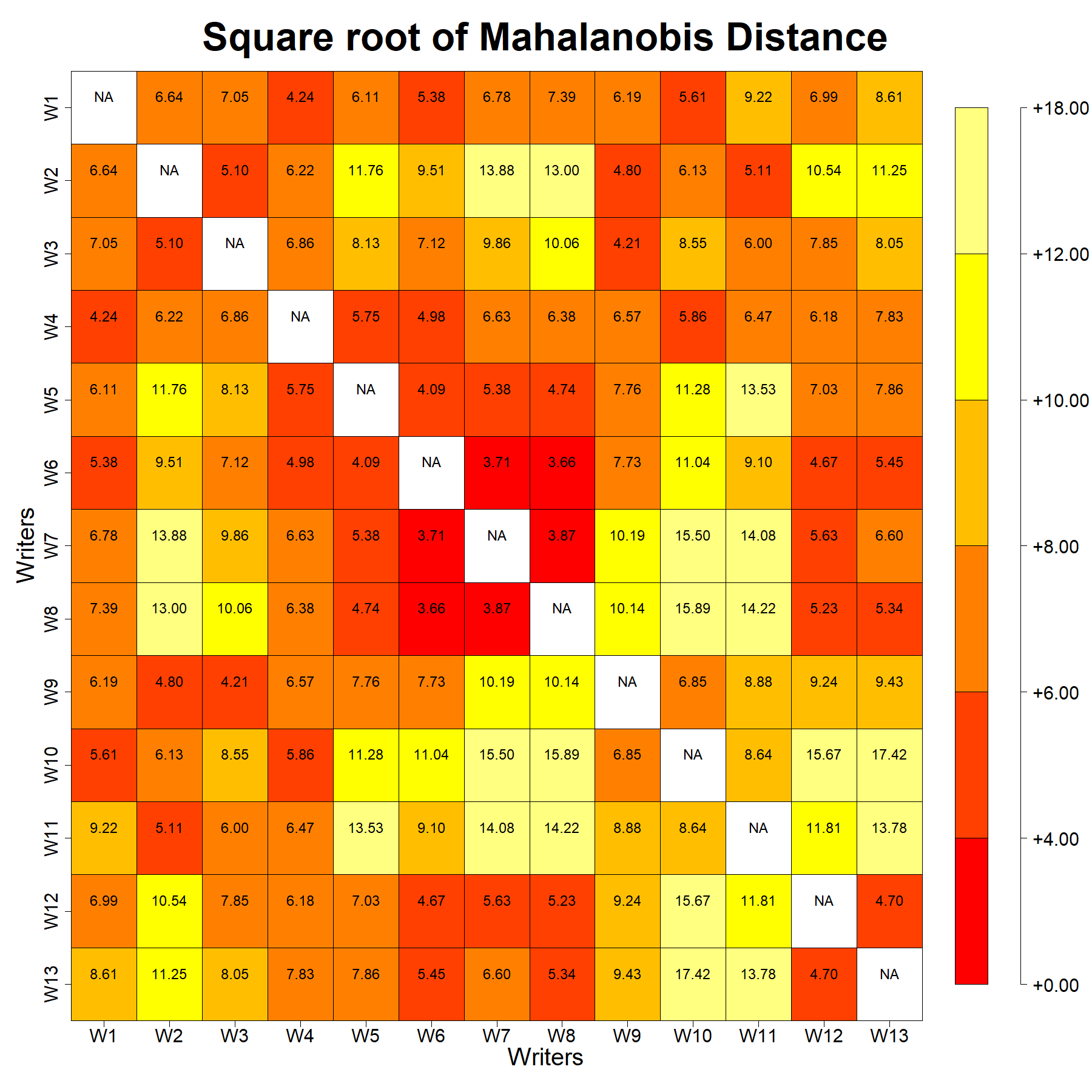}
\caption{Square root of the Mahalanobis distance between writers of Section \ref{data} using the Fourier coefficients and the surface size across all characters as described in Section \ref{fourie_harmonics}.}
\label{mahalanobis}
\end{figure}

Finally, the Mahalanobis distance has been quantified to measure the variability between pairs of writers across all characters. The Mahalanobis distance between two writers is calculated by measuring how far the feature vectors of one writer's samples are from the mean and covariance structure of another writer's samples. From Figure~\ref{mahalanobis}, where the square roots of Mahalanobis distances are represented, it can clearly be observed that while some pairs of writers present very small distances (e.g., writers 6, 7 and 8), other pairs of writers are characterized by greater distances (e.g., writers 10 and 13), confirming more pronounced peculiarities that should make them easier to discriminate.

\clearpage
\newpage
%\subsection{Data Plots}
%\label{data_plots}

\begin{figure}[!ht]
    \centering 
\subfloat[a][Box-plots of the surface size(in $cm^2$) of each loop character and per writer.]{
    \includegraphics[width=1\textwidth]{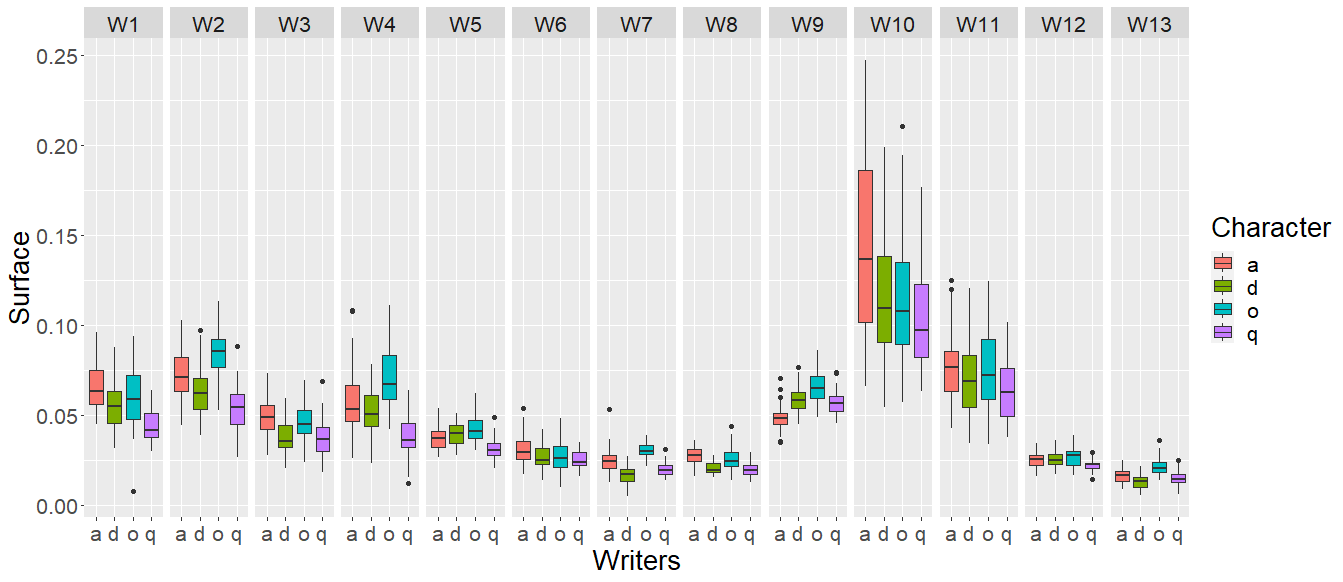}}

\subfloat[b][Box-plots of   Fourier coefficient  $a_1$ for each loop character and writer.]{
    \includegraphics[width=1\textwidth]{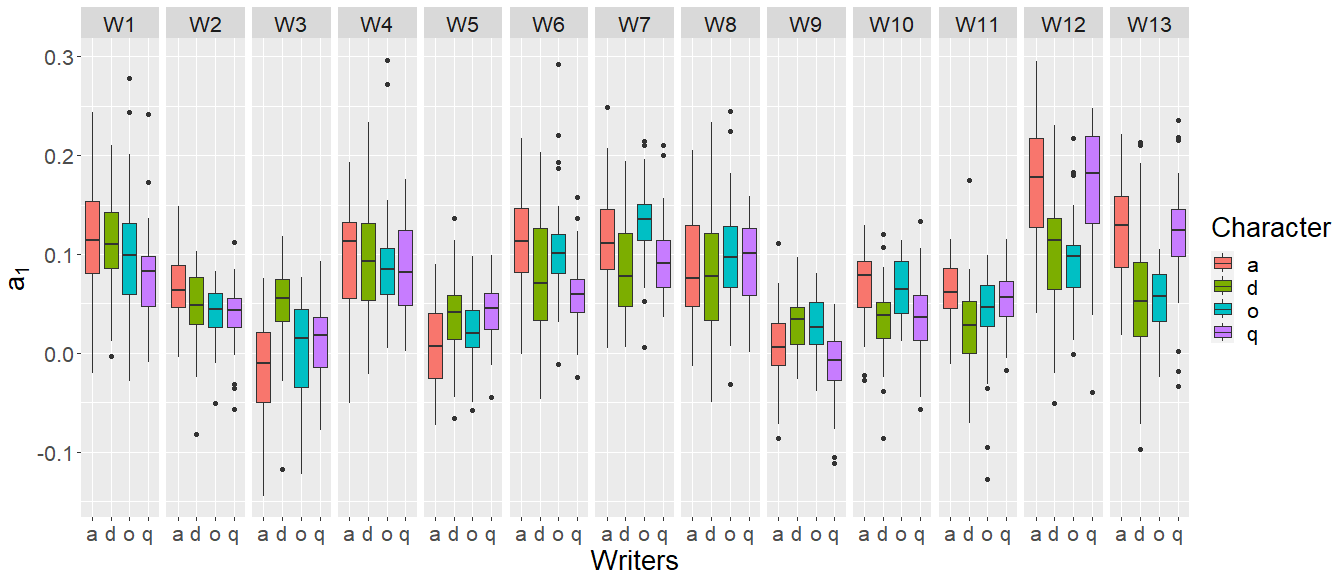}}

\subfloat[c][Box-plots of Fourier coefficient $b_1$ for each loop character and writer.]{
    \includegraphics[width=1\textwidth]{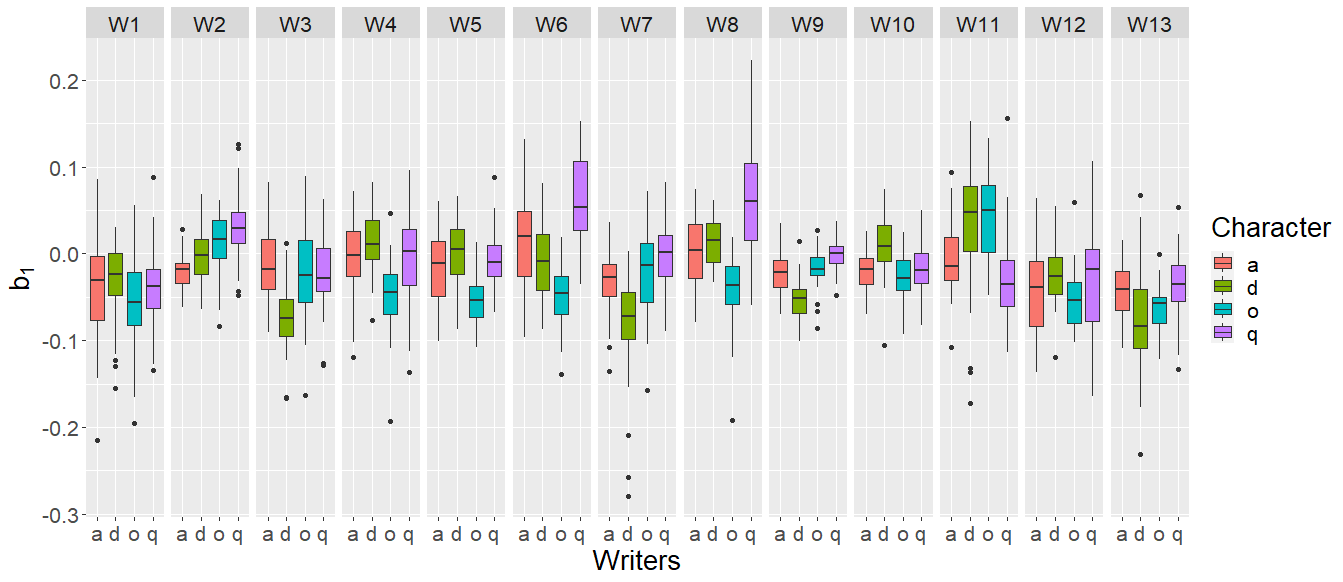}}

\end{figure}
\begin{figure}[!ht]
\centering 
\ContinuedFloat 
\subfloat[d][Box-plots of   Fourier coefficient $a_2$ for each loop character and writer.]{
    \includegraphics[width=1\textwidth]{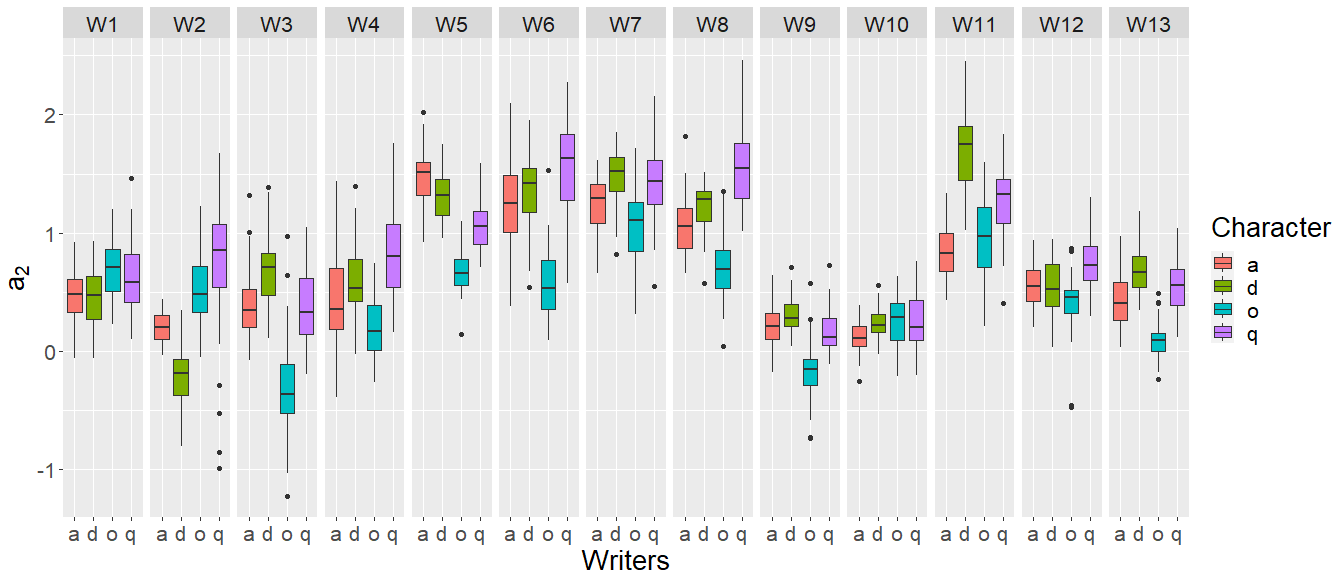}}

\subfloat[e][Box-plots of Fourier coefficient $b_2$ for each loop character and writer.]{
    \includegraphics[width=1\textwidth]{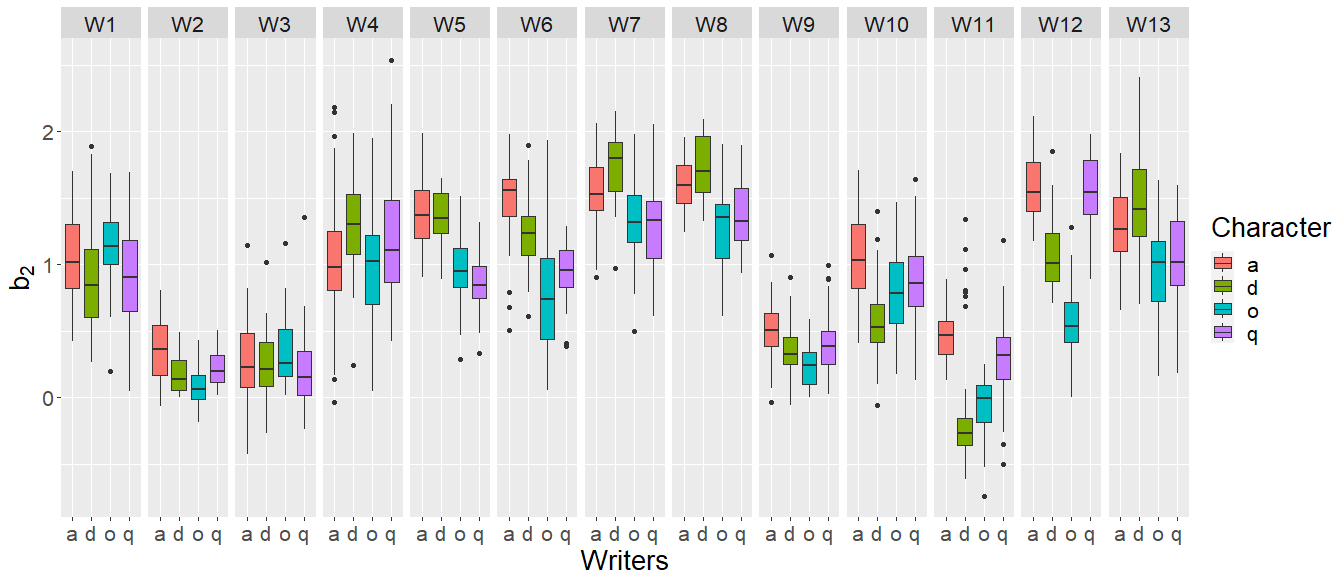}}

\subfloat[f][Box-plots of Fourier coefficient $a_3$ for each loop character and writer.]{
    \includegraphics[width=1\textwidth]{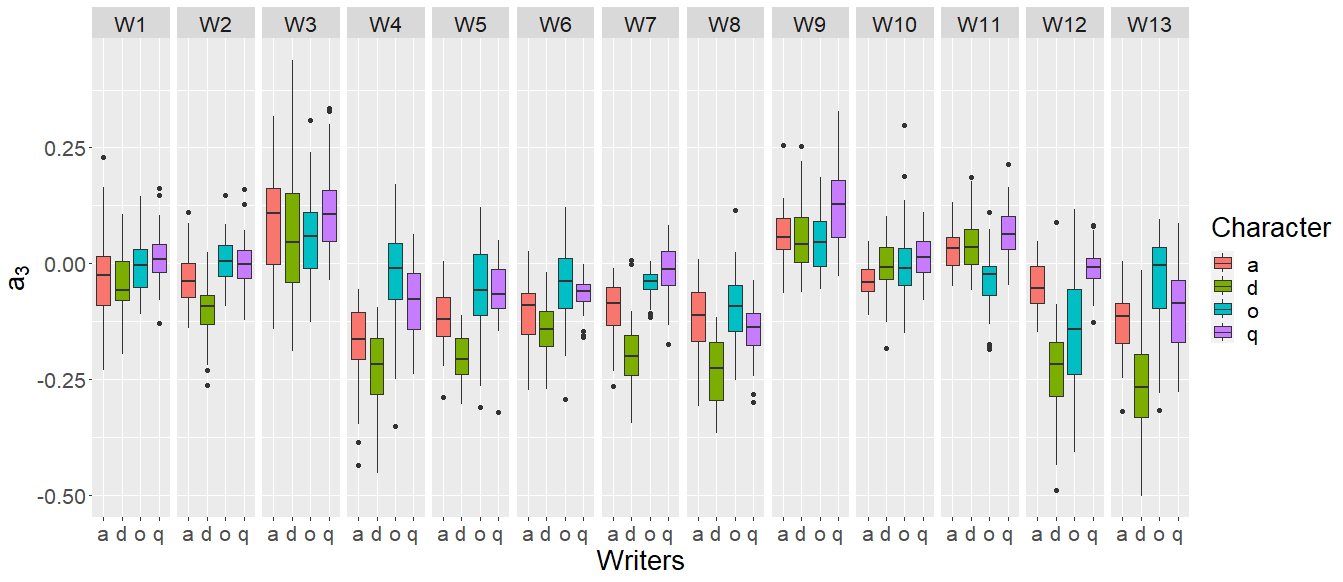}}

\end{figure}
\begin{figure}[!ht]
\centering 
\ContinuedFloat 

\subfloat[g][Grouped box-plot of the $b_3$  Fourier coefficient per loop character and per writer.]{
    \includegraphics[width=1\textwidth]{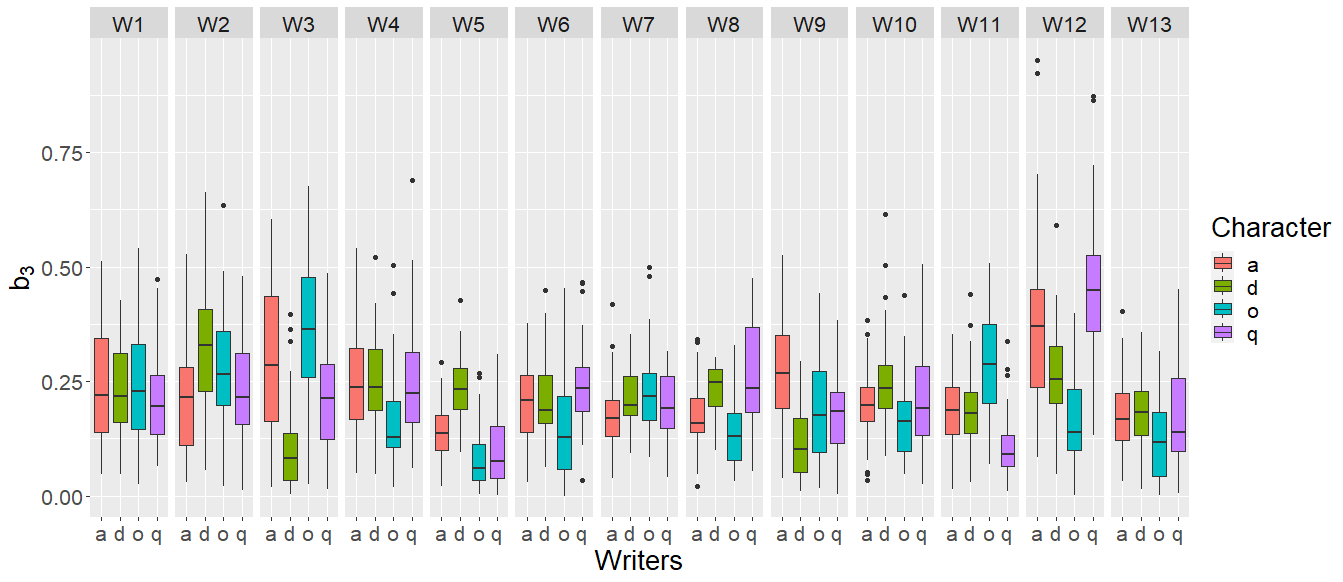}}
    
\subfloat[i][Box-plots of Fourier coefficient $a_4$ for each loop character and writer.]{
    \includegraphics[width=1\textwidth]{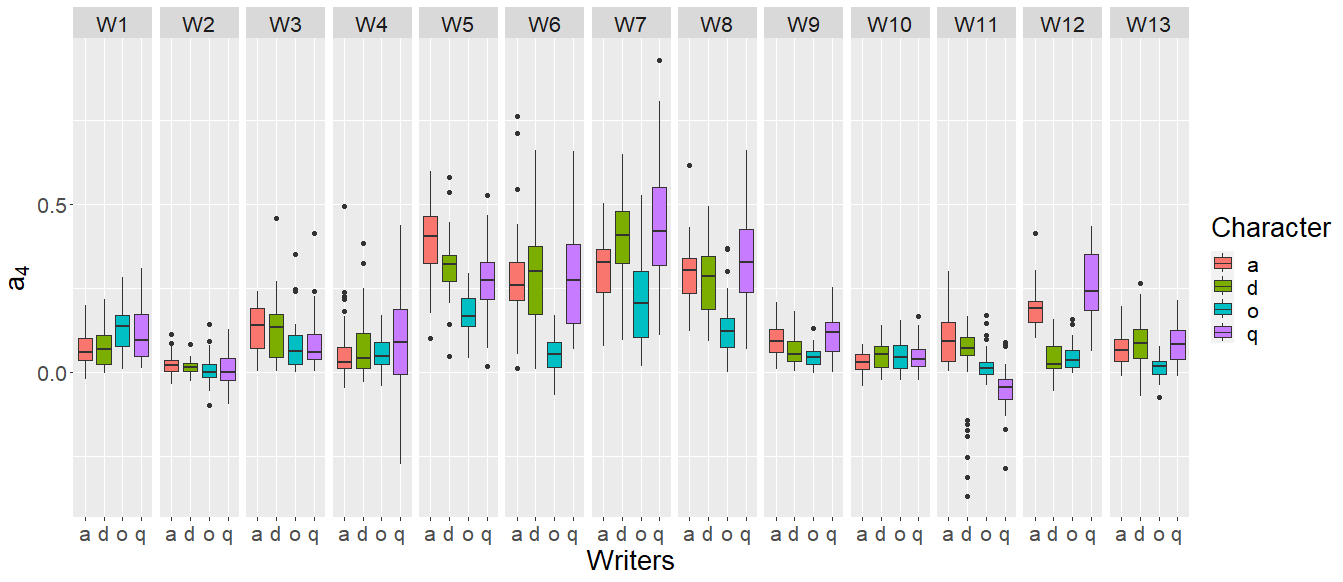}}

\end{figure}
\newpage

\begin{figure}[!ht]
\centering 
\ContinuedFloat
\subfloat[h][Box-plots of Fourier coefficient $b_4$ for each loop character and writer.]{
    \includegraphics[width=1\textwidth]{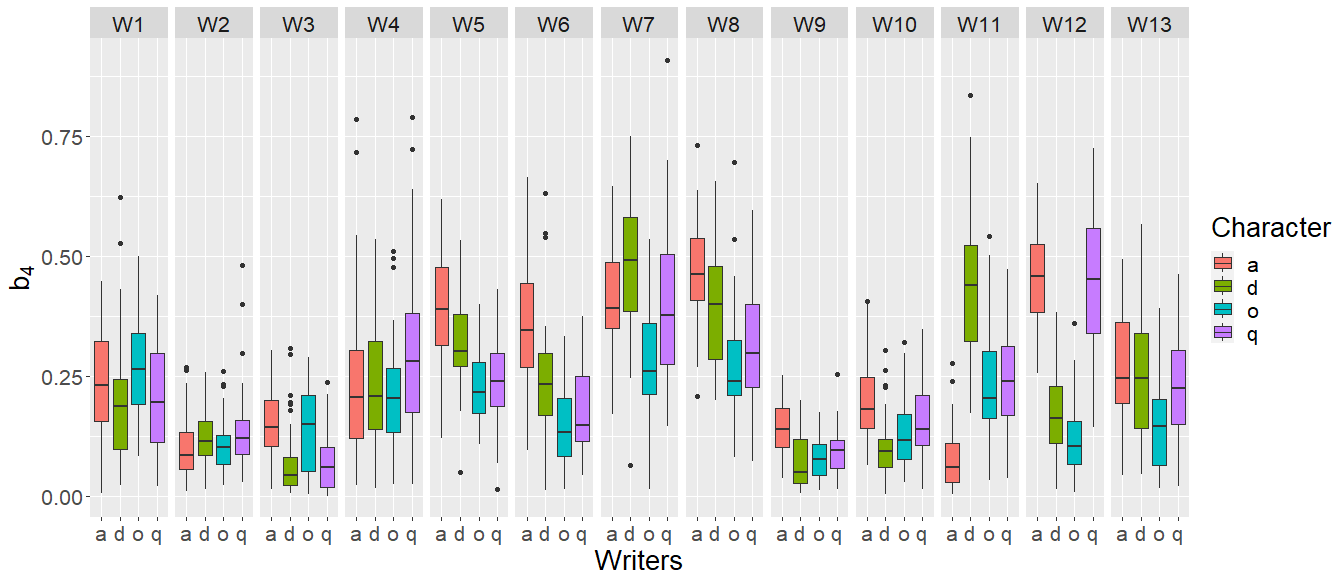}}

\caption{Grouped box-plot of the surface size and the first four pairs of Fourier coefficients per loop character and per writer. The surface size is at $cm^2$}
\label{fig:figures}
\end{figure}

\clearpage
\newpage
\section{Assumptions Underlying the Bayes Factor}
\label{assumption_BF}

The Bayes Factor (BF) employed for handwriting evaluation in this study is defined as follows:

\begin{equation}
BF = \frac{m(\bm{y}_1, \bm{y}_2 \mid H_1)}{m(\bm{y}_1, \bm{y}_2 \mid H_2)} = \frac{m(\bm{y}_1 \mid \bm{y}_2, H_1)  m(\bm{y}_2 \mid H_1)}{m(\bm{y}_1 \mid \bm{y}_2, H_2) m(\bm{y}_2 \mid H_2)}
\end{equation}

This expression can be further decomposed as:
\begin{equation}
BF = \frac{m(\bm{y}_1 \mid \bm{y}_2, H_1)}{m(\bm{y}_1 \mid H_2)} \times \frac{m(\bm{y}_2 \mid H_1)}{m(\bm{y}_2 \mid H_2)}
\end{equation}

Under the assumption that the likelihood of $\bm{y}_1$ does not depend on $\bm{y}_2$ when $H_2$ holds, the Bayes Factor simplifies to:
\begin{equation}
BF = \frac{m(\bm{y}_1 \mid \bm{y}_2, H_1)}{m(\bm{y}_1 \mid H_2)}
\end{equation}

Furthermore, if the likelihood of $\bm{y}_2$ (i.e., the control measurements) is independent of whether $H_1$ or $H_2$ is true, it follows that:
\begin{equation}
m(\bm{y}_2 \mid H_1) = m(\bm{y}_2 \mid H_2)
\end{equation}

The marginal likelihood $m(\bm{y}_1 \mid \bm{y}_2, H_1)$ can be expressed as:
\begin{equation}
m(\bm{y}_1 \mid \bm{y}_2, H_1) = \int_\theta f(\bm{y}_1 \mid \theta) f(\theta \mid \bm{y}_2, H_1) d\theta
\end{equation}
where $f(\theta \mid \bm{y}_2, H_1)$ denotes the posterior distribution of $\theta$ given $\bm{y}_2$ under $H_1$ (we denote model parameters equal to $\theta$ for simplicity only in this section for all considered model parameters of this study see Section \ref{modeling}).

Alternatively, using Bayes’ theorem, this can be rewritten as:
\begin{equation}
m(\bm{y}_1 \mid \bm{y}_2, H_1) = \frac{\int_\theta f(\bm{y}_1 \mid \theta) f(\bm{y}_2 \mid \theta) f(\theta \mid H_1) d\theta}{m(\bm{y}_2 \mid H_1)}
\end{equation}

Similarly, the marginal likelihood under $H_2$ is given by:
\begin{equation}
m(\bm{y}_1 \mid H_2) = \int_\theta f(\bm{y}_1 \mid \theta) f(\theta \mid H_2) d\theta
\end{equation}

Combining these results, and noting that $m(\bm{y}_2 \mid H_1) = m(\bm{y}_2 \mid H_2)$, the Bayes Factor can be expressed as:
\begin{equation}
BF = \frac{m(\bm{y}_1 \mid \bm{y}_2, H_1)}{m(\bm{y}_1 \mid H_2)} = \frac{\int_\theta f(\bm{y}_1 \mid \theta) f(\bm{y}_2 \mid \theta) f(\theta \mid H_1) d\theta}{\int_\theta f(\bm{y}_1 \mid \theta) f(\theta \mid H_2) d\theta \int_\theta f(\bm{y}_2 \mid \theta) f(\theta \mid H_2) d\theta}
\end{equation}

Note that the implemented Bayes Factor assumes independence between the questioned and control material under $H_2$. This assumption implies that possible disguised behavior is not considered. This formulation is presented in the main text of the paper.

\section{Dummy variables}
\label{dummy_variables}

For each writer, the character indicator is transformed into a set of dummy variables.  
It is common practice to omit one of the dummy variables from the equation for identifiability reasons in regression models, as first introduced by \cite{suits1957use}. Moreover, the category corresponding to the omitted dummy variable serves as the reference group against which all other levels are compared.
In our case, the character {\em a} is selected as the reference category; 
see Tables \ref{b1} and \ref{b2}.

\begin{table}[ht]
        \large
        \begin{minipage}{0.5\linewidth}
            \centering
            \tikzmarknode{A}{%contents
            \begin{tabular}{|c|c|}
            \hline
            \textbf{Indicator}&
                 \makecell{\textbf{Fourier}\\ \textbf{Coefficients}}\\
                 \hline
                 a & \dots \\
                 \hline
                 d & \dots  \\
                 \hline
                 o & \dots  \\
                 \hline
                 q & \dots \\
                 \hline
            \end{tabular}}   
            \caption{Processed Data}\label{b1}
        \end{minipage}\hfill
        \begin{minipage}{.5\linewidth}
            \centering
            \tikzmarknode{B}{%contents
            \begin{tabular}{|c|c|c|c|c|}
            \hline
                \textbf{a} &
                \textbf{d} &
                \textbf{o} &
                \textbf{q} & \makecell{\textbf{Fourier}\\ \textbf{Coefficients}}\\
                \hline
                  1 & 0 & 0 &0 &\dots \\
                 \hline
                  1 & 1 & 0 &0 & \dots  \\
                 \hline
                  1 & 0 & 1 &0 & \dots  \\
                 \hline
                  1 & 0 & 0 &1 & \dots \\
                 \hline
            \end{tabular}}  
            \caption{Regression Data} \label{b2}
        \end{minipage}
    \begin{tikzpicture}[remember picture, overlay]
    \path (A) -- node[draw, text width=2em, single arrow, thick, black]{} (B);
    \end{tikzpicture}   
    \end{table}

\newpage
\section{Prior Elicitation}
\subsection{Prior Parameters Elicitation for Bayesian Normal Model}
\label{prior_estimation_normal_iw}

For the Normal-Inverse-Wishart model, the prior parameters can be elicited from the background dataset. The prior mean of writer $i$ for character $\ell$, denoted by $\bm{\theta}_{i\ell}$, is given by the sample mean of the corresponding background data over all repetitions, that is
\begin{equation}
    \widehat{\bm{\theta}}_{i\ell} = \frac{\sum_{j=1}^{n_{i\ell}} \bm{X}_{i\ell j}}{n_{i\ell}},
    \label{partial_mean}
\end{equation}
where $i =  1,\dots,m$, 
$\ell = 1,\dots, L$, 
$\widehat{\bm{\theta}}_{i\ell}$ and $\bm{X}_{i\ell j}$ are vectors of length $p$; the elements of the latter vector are the background Fourier coefficients and the surface size of writer $i$ for character $\ell$ over all repetitions $n_{i\ell}$.
The overall mean of loop character $\ell$ is denoted by $\bm{\mu}_\ell$ is estimated as
\begin{equation}
    \widehat{\bm{\mu}}_\ell = \sum_{i=1}^{m} \widehat{\bm{\theta}}_{i\ell}\frac{n_{i\ell}}{n_\ell}
    \label{overall_mean}
\end{equation}
where $n_\ell=\sum_{i=1}^mn_{i\ell}$. Equivalently, the overall mean across all characters and writers is simply given by 
$\widehat{\bm{\mu}} = \tfrac{1}{n} \sum_{i=1}^{m}\sum_{\ell=1}^{L} n_{i\ell}\widehat{\bm{\theta}}_{i\ell} $; where $n=\sum_{\ell=1}^Ln_\ell$. 

The between-writer covariance matrix $\bm{B}_\ell$ for loop-character $\ell$ is elicited by setting it equal to the sample covariance of the corresponding background data, that is 
%\begin{equation}
%    \hat{\bm{B}} = \frac{1}{m-1} \sum_{i=1}^{m} n_i(\hat{\bm{\theta}}_i-\hat{\bm{\mu}})(\hat{\bm{\theta}}_i-\hat{\bm{\mu}})'
%    \label{between_writer_cov}
%\end{equation}
\begin{equation}
    \widehat{\bm{B}}_\ell = \frac{1}{n-1} \sum_{i=1}^{n} (\bm{X}_{ij\ell}-\widehat{\bm{\mu}}_\ell)(\bm{X}_{ij\ell}-\widehat{\bm{\mu}}_\ell)'.
    \label{between_writer_cov}
\end{equation}

Following the parametrization of the Inverse-Wishart distribution in R and Stan, its mean is given by $E[\bm{W}_{i\ell}]=\bm{U}_\ell/(\nu-p-1)$, where $p$ is the number of Fourier coefficients and the surface size retained for each loop character representing our response data. 
Hence, parameters $\bm{U}_\ell$ (that describe the within-writer variation) are elicited by setting $\bm{U}_\ell$ equal to
\begin{equation}
    \widehat{\bm{U}}_\ell =  \widehat{\bm{W}} _{i\ell}(\nu-p-1), 
    \label{within_writer_u}
\end{equation}
where $\widehat{\bm{W}} _{i\ell}$ can be estimated by 
\begin{equation}
    %E[\bm{W}_{i\ell}]=
    \widehat{\bm{W}} _{i\ell}= \frac{1}{n-m} \sum_{i=1}^{m}\sum_{j=1}^{n_i} (\bm{X}_{ij\ell}-\widehat{\bm{\theta}}_{i\ell})(\bm{X}_{ij}-\widehat{\bm{\theta}}_{i\ell})'~.
    \label{within_writer_cov}
\end{equation}

We consider values of $\nu$ such that $\nu \geq p+2$ so that the prior mean is well-defined; the reader can refer to \cite{press2005applied} for further details.
Finally, the prior parameter $k_0$  is selected based on a grid search in the (0,1) interval for all writers using background data. Specifically, we choose the value of $k_0$ which maximizes the marginal likelihood in the background data.

For the Normal-LogNormal-LKJ prior approach, the parameters of the Normal prior are estimated consistently using the formulations presented in Equations \eqref{overall_mean} and \eqref{between_writer_cov}. Subsequently, by extracting the diagonal elements of the estimated covariance matrix $\widehat{\bm{W}} _{i\ell}$, we derive the prior parameters of the LogNormal distribution as follows:

\begin{equation}
    \hat{\upsilon}_\ell = \frac{1}{p} \sum_{\kappa=1}^p \log(\widehat{\bm{W}} _{i\ell\kappa\kappa})
    \label{location_lognormal}
\end{equation}

\begin{equation}
    \hat{\sigma}_\ell = \frac{1}{p-1} \sum_{\kappa=1}^p \left( \log(\widehat{\bm{W}} _{i\ell\kappa\kappa}) - \hat{\upsilon}_\ell\right)
    \label{scale_lognormal}
\end{equation}

 For $\eta$ parameter of LKJ distribution is set equal to one.

\subsection{Prior Parameters Estimation for Bayesian MANOVA}
\label{prior_estimation_bayes_regression}

Similarly, for the  Bayesian MANOVA model, the prior parameters are also elicited using the available background data. The elicitation of prior parameters 
for reference character `{\em a}'
$\bm{\mu}_1$ and $\bm{B}_1$ of $\bm{\theta}_1$ 
is performed by Equations \ref{overall_mean}--\ref{between_writer_cov} for $\ell=1$. 
For the prior parameters  $\bm{\mu}_\ell$ and $\bm{B}_\ell$, i.e. for $\ell=2,3,4$, we use the same equation but instead of the original data $\bm{X}_{i\ell j}$ we consider the differences from character $a$, i.e. $d_{i\ell j}=\bm{X}_{i\ell j}-\widehat{\bm{\mu}}_1$.

The trace matrix of the Inverse-Wishart distribution $\bm{U}$ for the within-writer variation is elicited from Equations \ref{within_writer_u} and \ref{within_writer_cov}
in the same way as in the Normal model and by considering all characters together; for more details see \cite{press1980bayesian}. 

Finally, the $\bm{K}_0$ of the conjugate approach is elicited based on a grid search in $(0,1)^{L}$ for all characters of background writers: the value that maximizes the marginal likelihood is selected.

For the Normal-LogNormal-LKJ prior approach, the parameters of the Normal prior are estimated consistently using the formulations presented in Equations \eqref{overall_mean} and \eqref{between_writer_cov} by calculating the differences as described in he beginning of this section. Furthermore, by extracting the diagonal elements of the estimated covariance matrix $\widehat{\bm{W}}$, we derive the parameters of the LogNormal distribution based on Equations \eqref{location_lognormal}, \eqref{scale_lognormal}, and the $\eta$ parameter of the LKJ distribution is set equal to one.

\section{Bridge Sampling}
\label{bridge_sampling_section}
\setcounter{table}{0}
\renewcommand{\thetable}{\Alph{section}\arabic{table}}

The most known and effective Monte Carlo estimator of the marginal likelihood was introduced by  \cite{meng1996simulating}. 
This method is based on the following simple identity:

\begin{eqnarray}  
    m(\bm{y}) = \frac{\int h(\bm{\Theta},\bm{W})g(\bm {\Theta},\bm{W})f(\bm{y}|\bm {\Theta},\bm{W})\pi(\bm {\Theta},\bm{W})d(\bm {\Theta},\bm{W})}{\int h({\bm \Theta},\bm{W})g({\bm \Theta},\bm{W})\frac{f(\bm{y}|{\bm \Theta},\bm{W})\pi({\bm \Theta},\bm{W})}{m(\bm{y})}d({\bm \Theta},\bm{W})}, \label{bridge1}
    \label{eq_bridge}
\end{eqnarray}

for any functions $g({\bm \Theta},\bm{W})$ and $h({\bm \Theta},\bm{W})$. These functions are the so-called proposal distribution and bridge function, respectively. The proposal distribution should approximate the posterior distribution and ensure adequate overlap with it. Furthermore, the function $h(\cdot)$ acts as a bridge that connects the two distributions. Therefore, it must be compatible with both distributions, namely that the function is chosen so it can be meaningfully evaluated with respect to both the proposal distribution and the posterior distribution, effectively "bridging" them. Simple manipulation of \eqref{bridge1} leads %us 
to the following identity:
\begin{eqnarray}
    m(\bm{y}) 
    &=& \frac{\int h({\bm \Theta},\bm{W})g({\bm \Theta},\bm{W})f(\bm{y}|{\bm \Theta},\bm{W})\pi({\bm \Theta},\bm{W})d({\bm \Theta},\bm{W})}{\int h({\bm \Theta},\bm{W})g({\bm \Theta},\bm{W})\pi({\bm \Theta},\bm{W}|\bm{y})d({\bm \Theta},\bm{W})} \nonumber \\
    &=& \frac{E_{g({\bm \Theta},\bm{W})}\left[ h({\bm \Theta},\bm{W})f(\bm{y}|{\bm \Theta},\bm{W})\pi({\bm \Theta},\bm{W}) \right]}{E_{\pi({\bm \Theta},\bm{W}|\bm{y})}\left[ h({\bm \Theta},\bm{W})g({\bm \Theta},\bm{W})\right]}.  \label{bridge}
\end{eqnarray}    

Hence, the marginal likelihood can be estimated using a Monte Carlo estimator based on the identity in \eqref{bridge}. This Monte Carlo marginal likelihood estimator is given by 
\begin{equation}
\begin{aligned}
    \widehat{m}(\bm{y}) = \frac{\frac{1}{T_2} \sum_{t=1}^{T_2} h({\bm \Theta}^{*(t)},{\bm W}^{*(t)})f(\bm{y}|{{\bm \Theta}}^{*(t)},{\bm W}^{*(t)})\pi({{\bm \Theta}}^{*(t)},{\bm W}^{*(t)})}
    {\frac{1}{T_1} \sum_{t=1}^{T_1} h({\widetilde{\bm \Theta}}^{(t)},\widetilde{\bm W}^{(t)})g({\widetilde{\bm \Theta}}^{(t)},\widetilde{\bm W}^{(t)})},
\end{aligned}
\end{equation}
where $({\widetilde{\bm \Theta}}^{(t)},\widetilde{\bm W}^{(t)})$, for $t=1,\dots,T_1$, and $({\bm {\Theta}}^{*(t)},\,{\bm W}^{*(t)})$, for $t=1,\dots,T_2$, are samples from the posterior distribution $\pi(\bm{\Theta},\bm{W}|\bm{y})$ and from the proposal distribution $g({\bm \Theta},\bm{W})$, respectively. The proposal distribution $g(\cdot)$ must be close to the target posterior distribution. Function $h(\cdot)$ plays the role of the bridge that links the two distributions. According to  \cite{meng1996simulating}, the optimal function under the mean square error is given by
 $$
 h({\bm \Theta},\bm{W}) = \frac{T_2}{T_2g({\bm \Theta},\bm{W}) + T_1\pi({\bm \Theta},\bm{W}|\bm{y})}. 
 $$ 

By replacing this optimal bridge function in \eqref{bridge}, the bridge sampling indicator, after some manipulations, simplifies to 

\begin{equation}
\begin{aligned}
    \widehat{m}_o(\bm{y}) = \frac{\frac{1}{T_2} \sum_{t=1}^{T_2} \frac{f(\bm{y}|{{\bm \Theta}}^{*(t)},{\bm W}^{*(t)})\pi({{\bm \Theta}}^{*(t)},{\bm W}^{*(t)})}{T_2g({{\bm \Theta}}^{*(t)},{\bm W}^{*(t)}) + T_1\pi({{\bm \Theta}}^{*(t)},{\bm W}^{*(t)}|\bm{y})}}
    {\frac{1}{T_1} \sum_{t=1}^{T_1} \frac{g({\widetilde{\bm \Theta}}^{(t)},\widetilde{\bm W}^{(t)})}{T_2g({\widetilde{\bm \Theta}}^{(t)},\widetilde{\bm W}^{(t)}) + T_1\pi({\widetilde{\bm \Theta}}^{(t)},\widetilde{\bm W}^{(t)}|\bm{y})}}. \label{bridge_o}
\end{aligned}
\end{equation}
For more details and a thorough explanation of this method, see \citet{gronau2017tutorial}.

Finally, the iterative expression of \eqref{bridge_o} is implemented to obtain the optimal bridge sampling estimator, as described by \citet[][pag. 837]{meng1996simulating}, where an initial guess of the marginal likelihood is updated until convergence is achieved based on a predefined tolerance level. The Bayes theorem is then applied to the joint posterior distribution, and by performing some algebraic manipulations, the following expression for the marginal likelihood $\widehat{m}_o(\bm{y})^{(z+1)}$ at iteration $z+1$ is obtained:

\begin{equation}
\begin{aligned}
    \widehat{m}_o(\bm{y})^{(z+1)} = \frac{\frac{1}{T_2} \sum_{t=1}^{T_2} \frac{S({{\bm \Theta}}^{*(t)},{\bm W}^{*(t)})}{T_2  \widehat{m}_o(\bm{y})^{(z)} + T_1S({{\bm \Theta}}^{*(t)},{\bm W}^{*(t)})}}
    {\frac{1}{T_1} \sum_{t=1}^{T_1} \frac{1}{T_2  \widehat{m}_o(\bm{y})^{(z)} + T_1 S({\widetilde{\bm \Theta}}^{(t)},\widetilde{\bm W}^{(t)})}},  
    \label{bridge_o_iterative}
\end{aligned}
\end{equation}
%(page 837 \cite{meng1996simulating}).
where $S({\bm \Theta},\bm W) = \frac{f(\bm{y}|\bm{ \Theta},\bm{W}\pi(\bm{\Theta},\bm {W})}{g(\bm{\Theta},\bm {W})}$. 

Furthermore, to obtain a reliable estimate of the marginal likelihood, \cite{overstall2010default} proposed dividing the posterior samples from the MCMC procedure (e.g., the Gibbs sampling as in the current work) into two parts: (a) the first part is used to specify the parameters of the proposal distribution, and (b) the second part is used in \eqref{bridge_o_iterative} to compute the numerator of the bridge sampling estimator.

\subsection{Monte Carlo Error of Models per Writer}
\label{mce_per_writer}
\renewcommand\thefigure{\thesection.\arabic{figure}}    
\renewcommand\thetable{\thesection.\arabic{table}}    

\setcounter{figure}{0}
\setcounter{table}{0}

This section compares the Monte Carlo error (MCE) of the marginal likelihood estimator for the considered models. To this end, the marginal likelihood estimation obtained with the prior setup of the implemented model, the $M_2$, $M_3$, and $M_5$,$M_6$ for each writer $i=1,\dots,m$ of the whole dataset ${\bm{\mathcal{D}}}_i$.  Specifically, the estimate of the logarithmic marginal likelihood is obtained, for each writer, from a total number of 10 MCMC runs with 2000 iterations after discarding an additional 1000 iterations as burn-in. Finally, the prior parameters are elicited using data from the remaining writers $(m-i)$. The mean of the log-marginal likelihood estimates and the standard deviation of the MCE are reported in Tables \ref{niw_mce_per_writer} and \ref{manova_mce_per_writer}, for the Normal model with prior setups $M_2$, $M_3$, and the MANOVA model with prior setups $M_5$, $M_6$, respectively. 

Two main results can be observed from Tables \ref{niw_mce_per_writer} and \ref{manova_mce_per_writer}. First, the bridge sampling estimator systematically produces small MCEs, as expected according to the scientific literature \citep{sinharay2005empirical, ardia2012comparative}. Second, when comparing the MCEs of the considered models, the Normal-LogNormal-LKJ prior in the Bayesian Normal model yields smaller MCEs. For the Bayesian MANOVA, however, no clear pattern emerges; the results depend on the writer, and the differences are very balanced.

\begin{table}[!ht]
\centering
\begin{tabular}{|C{1.5cm}|C{2cm}|C{2cm}|C{2cm}|C{2cm}|}
 \hline
 & \multicolumn{4}{c|}{\textbf{Normal}} \\

 \textbf{Writer} & \multicolumn{2}{c}{\textbf{Inverse-Wishart}}  & \multicolumn{2}{c|}{\textbf{LogNormal-LKJ}} \\
 \hhline{~----}
 & \textbf{log-ML} & \textbf{MCE}  & \textbf{log-ML} & \textbf{MCE} \\ \hline
   1 & -1380.3 & 0.109  & -1362.8 & 0.065 \\ \hline
   2 & -1298.4 & 0.086  & -1270.4 & 0.078 \\ \hline
   3 & -1639.4 & 0.084  & -1604.6 & 0.094 \\ \hline
   4 & -1869.8 & 0.082  & -1821.1 & 0.081 \\ \hline
   5 & -1021.8 & 0.075  & -1027.5 & 0.073 \\ \hline
   6 & -1378.2 & 0.094  & -1359.7 & 0.070 \\ \hline
   7 & -1120.7 & 0.075  & -1116.6 & 0.080 \\ \hline
   8 & -1010.6 & 0.087  & -1007.8 & 0.098 \\ \hline
   9 & ~\,-969.4  & 0.129  & ~\,-966.4 & 0.114 \\ \hline
   10 & -1319.9 & 0.105 & -1296.2 & 0.077 \\ \hline
   11 & -1851.4 & 0.120 & -1813.8 & 0.109 \\ \hline
   12 & -1501.5 & 0.122 & -1485.9 & 0.055 \\ \hline
   13 & -1366.0 & 0.096 & -1349.6 & 0.092 \\ \hline
\end{tabular}
\caption{Log-marginal likelihood  ($\log-ML$) estimates (mean) and Monte Carlo errors (standard deviation) for Bayesian  Normal  model with two different prior setups ($M_2$, $M_3$).}
\label{niw_mce_per_writer}
\end{table}

\begin{table}[!ht]
\centering
\begin{tabular}{|C{1.5cm}|C{2cm}|C{2cm}|C{2cm}|C{2cm}|}
 \hline
& \multicolumn{4}{c|}{\textbf{MANOVA}} \\

 \textbf{Writer} & \multicolumn{2}{c}{\textbf{Inverse-Wishart}}  & \multicolumn{2}{c|}{\textbf{LogNormal-LKJ}} \\
 \hhline{~----}
 & \textbf{log-ML} & \textbf{MCE}  & \textbf{log-ML} & \textbf{MCE} \\ \hline
   1 & -1361.3 & 0.142 & -1346.2 & 0.152 \\ \hline
        2 & -1125.1 & 0.155 & -1105.7 & 0.104 \\ \hline
        3 & -1523.9 & 0.126 & -1491.8 & 0.150 \\ \hline
        4 & -1776.7 & 0.267 & -1731.5 & 0.163 \\ \hline
        5 & ~\,-818.2 & 0.111 & ~\,-824.0 & 0.162 \\ \hline
        6 & -1266.4 & 0.234 & -1246.5 & 0.134 \\ \hline
        7 & ~\,-963.3 & 0.201 & ~\,-964.3 & 0.112 \\ \hline
        8 & ~\,-878.8 & 0.116 & ~\,-879.1 & 0.153 \\ \hline
        9 & ~\,-817.0 & 0.175 & ~\,-821.2 & 0.168 \\ \hline
        10 & -1274.7 & 0.116 & -1254.1 & 0.164 \\ \hline
        11 & -1666.5 & 0.182 & -1632.8 & 0.144 \\ \hline
        12 & -1308.4 & 0.184 & -1288.4 & 0.207 \\ \hline
        13 & -1207.1 & 0.178 & -1195.3 & 0.085 \\ \hline
\end{tabular}
\caption{Log-marginal likelihood  ($\log-ML$) estimates (mean) and Monte Carlo errors (standard deviation) for Bayesian  MANOVA model with two different prior setups  ($M_5$, $M_6$).}
\label{manova_mce_per_writer}
\end{table}

\clearpage
 \newpage
\section{Model Comparisons}
\label{model_comparisons_appendix}
\renewcommand\thefigure{\thesection.\arabic{figure}}    
\renewcommand\thetable{\thesection.\arabic{table}}    

\setcounter{figure}{0}
\setcounter{table}{0}

Table \ref{lr_of_conjugate_lkj_models} presents the estimated Bayes factors $\rm{BF}_{1,3}({\bf X}_i)$ and $\rm{BF}_{4,6} ({\bf X}_i)$ (in log scale) per writer. Specifically, this involves comparing the conjugate Normal-Inverse-Wishart and Normal-LogNormal-LKJ prior structure.

\begin{table}[!ht]
\begin{center}
 \small
\begin{tabular}{|c|c|c|c||c|c|c|} 
 \hline
  & \multicolumn{3}{c||}{\textbf{Normal}} & \multicolumn{3}{c|}{\textbf{MANOVA}}\\
  & \multicolumn{3}{c||}{\textbf{Inverse-Wishart vs LogNormal-LKJ}} & \multicolumn{3}{c|}{\textbf{Inverse-Wishart vs LogNormal-LKJ}}\\
 \textbf{Writer} & \multicolumn{3}{c||}{$\log BF_{1,3}^*$ } & \multicolumn{3}{c|}{$\log BF_{4,6}^*$}\\
\hhline{~------}

 & \multicolumn{1}{c|}{\textbf{Value}} & \multicolumn{1}{c|}{\textbf{Sign}} & \multicolumn{1}{c||}{\textbf{Interpretation}} &
 \multicolumn{1}{c|}{\textbf{Value}} & \multicolumn{1}{c|}{\textbf{Sign}} & \multicolumn{1}{c|}{\textbf{Interpretation}}\\
 \hline
    1 & -17.51 & - & Extreme & -13.51 & - & Extreme \\ \hline
        2 & -33.05 & - & Extreme & -23.05 & - & Extreme \\ \hline
        3 & -38.23 & - & Extreme &  -37.76 & - & Extreme \\ \hline
        4 & -50.95 & - & Extreme & -47.93 & - & Extreme\\ \hline
        5 & ~\,~\,7.24 & + & Extreme & ~\,10.33 & + & Extreme \\ \hline
        6 & -19.80 & - & Extreme & -17.71 & - & Extreme \\ \hline
        7 & ~\,-6.31 & - & Extreme & ~\,-1.58 & - & Substantial \\ \hline
        8 & ~\,-5.68  & - & Extreme & ~\,-1.66 & - & Substantial \\ \hline
        9 & ~\,-3.11  & - & Strong & ~\,~\,9.39 & + & Extreme\\ \hline
        10 & -17.61 & - & Extreme & ~\,-6.36 & - &  Extreme\\ \hline
        11 & -34.16 & - & Extreme & ~\,-0.15 & - & Bare Mention \\ \hline
        12 & -19.21  & - & Strong & -20.85 & - & Extreme\\ \hline
        13 & -28.03 & - & Extreme & -27.17 & - & Extreme \\ 
  \hline
  \multicolumn{7}{l}{\footnotesize 
        \it $^{*}$Bridge Sampling estimate}
\end{tabular}
\end{center}
\caption{Logarithmic Bayes factors (per writer) comparing the conjugate Normal-Inverse-Wishart and Normal-LogNormal-LKJ prior setups for the Bayesian Normal and MANOVA models.}
\label{lr_of_conjugate_lkj_models}
\end{table}

\section{Sensitivity Analysis of Prior Elicitation}
\label{section_sensitivity_analysis_appendix}

\subsection{Subsampling of Background Data Examples}
\label{bootstrap_section_examples}

In this section, we present more examples of the implementation of subsampling to the background data associated with randomly selected cases from Section \ref{models_efficiency}. To isolate the effect of prior elicitation, we utilize a single random data split for both the same-writer and different-writer experiments, thereby minimizing potential confounding effects arising from data partitioning. This analysis focuses on the Bayesian MANOVA model, which has been identified as the most effective and recommended approach for this context. For each case, we perform 30 iterations of subsampling with replacement in 50\% of the background writers' data in order to ensure the robustness of the results.

\begin{figure}[!ht]
    \centering
    \includegraphics[width=1\textwidth]{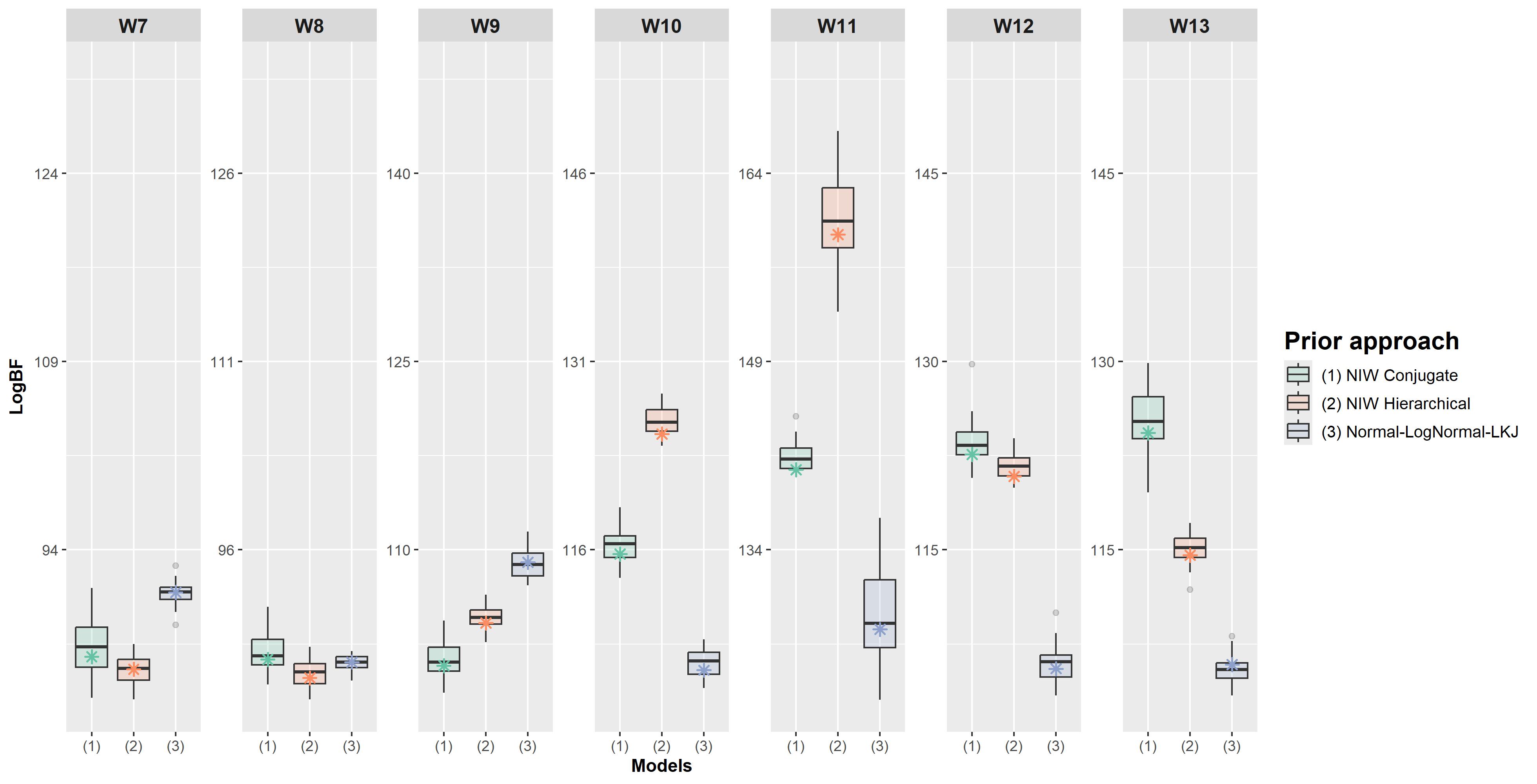}
   \par\vspace{0.2cm} % optional spacing
    \parbox{\textwidth}{%
        \footnotesize\it
        \hspace{0.5cm} $^*$indicates the $\log\mathrm{BF}$ using the complete background dataset for each case.
    }
    \caption{
    Boxplots of Logarithmic Bayes factors ($\log\mathrm{BF}$) for handwriting evaluation for the same writer scenarios over different subsamples of background data for the Bayesian MANOVA approach. }
    \label{same_source_bootstrap2}
\end{figure}

\begin{figure}[!ht]
    \centering
    \includegraphics[width=1\textwidth]{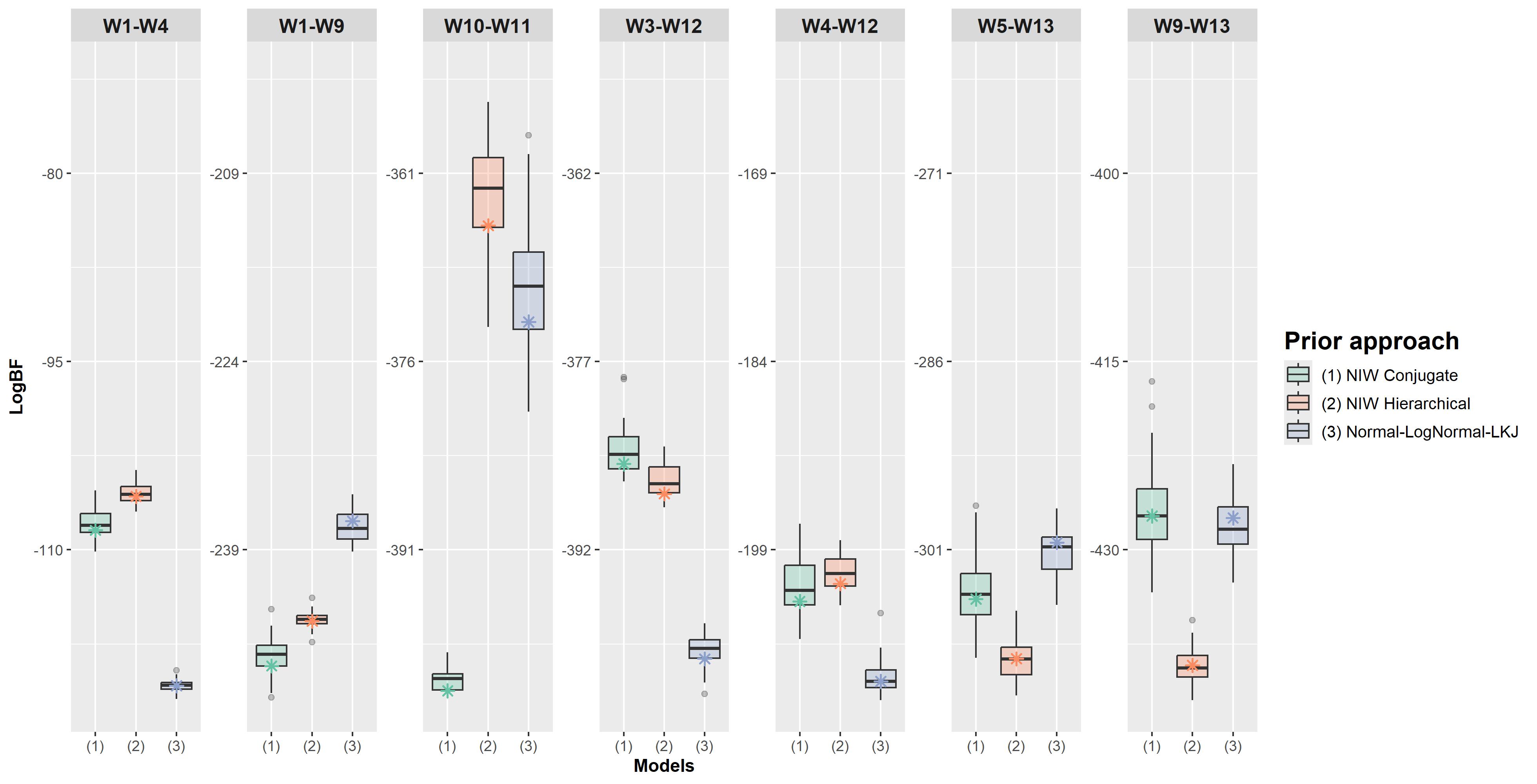}
  \par\vspace{0.2cm} % optional spacing
    \parbox{\textwidth}{%
        \footnotesize\it
        \hspace{0.5cm} $^*$indicates the $\log\mathrm{BF}$ using the complete background dataset for each case.
    }
    \caption{
    Boxplots of Logarithmic Bayes factors ($\log\mathrm{BF}$) for handwriting evaluation for the different writers scenarios over different subsamples of background data for the Bayesian MANOVA approach. }
    \label{different_source_bootstrap2}
\end{figure}

\newpage
\clearpage
\subsection{Sensitivity of the Inverse-Wishart’s Degrees of Freedom}
\label{dof_iw_appendix}

\renewcommand\thefigure{\thesection.\arabic{figure}}    
\renewcommand\thetable{\thesection.\arabic{table}}    

\setcounter{figure}{0}
\setcounter{table}{0}

In this section, a sensitivity analysis is performed to investigate the effect of the choice of the degrees of freedom of the Inverse-Wishart distribution, which models the within-writer variability. 
The specification of this prior parameter is of primary concern, since it has a large impact on the resulting Bayes factor values. 

The sensitivity analysis was performed for values of the degrees of freedom ranging from the minimum value of 11 ($p+2=11$) to 50, with a step size of 10, that is, $\nu \in \{ 11,20,30,40,50\}$. The Bayes factor has therefore been calculated for each value of $\nu$, for all comparisons between characters from the same or different writers, using all models and marginal likelihood estimation methods. 10 sub-samples have been drawn for each pair of writers, following the procedure described in Section~\ref{models_efficiency}.

\begin{figure}[!ht]
\centering
    \includegraphics[width=1\textwidth]{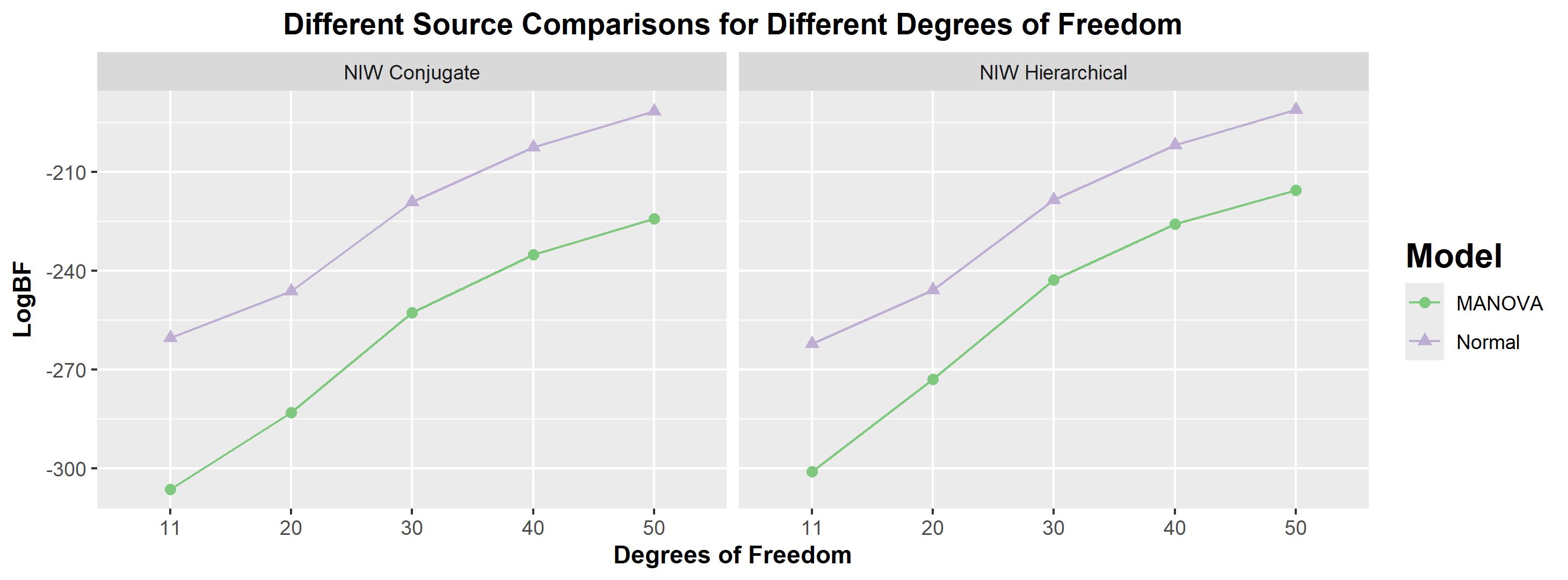}
    \caption{Average logarithmic Bayes factor ($\log\rm{BF}$) for handwriting examination for different writers' comparisons over different degrees of freedom. The Normal model has been implemented using all character types jointly.}
\label{ds_logbf_dof}
\end{figure}

\begin{figure}[!ht]
\centering
    \includegraphics[width=1\textwidth]{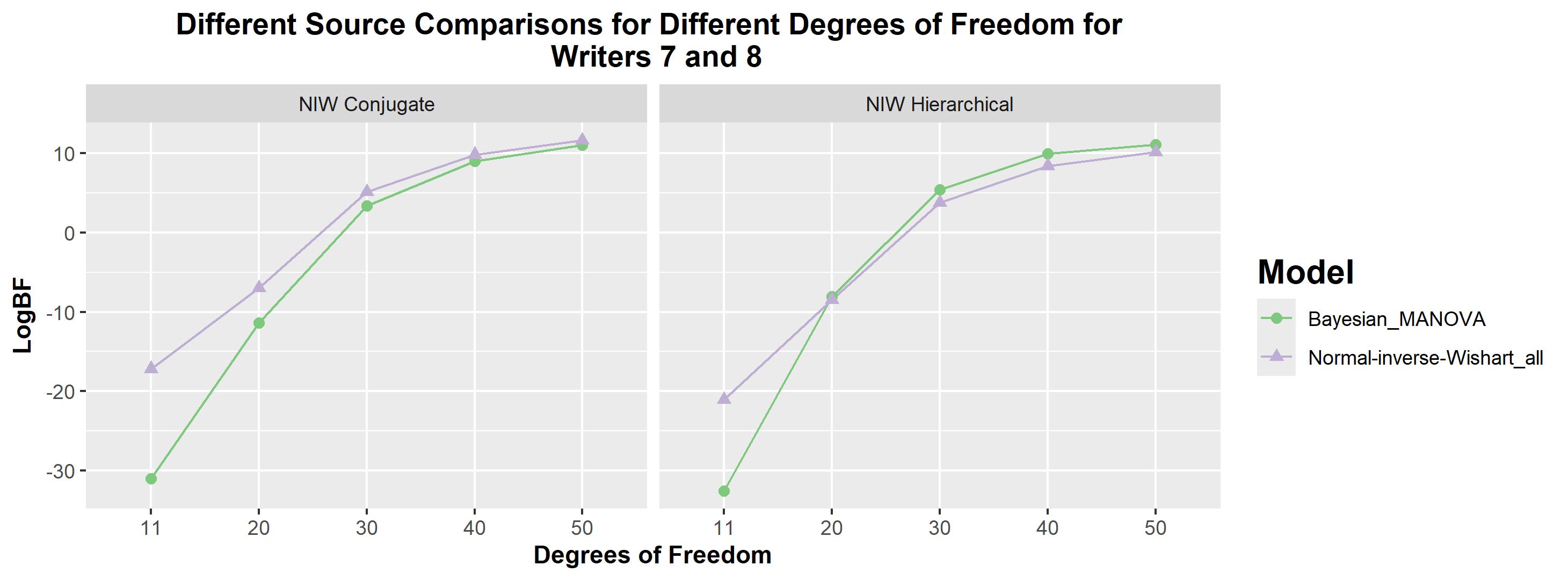}
    \caption{
    Average logarithmic Bayes factor ($\log BF$) for handwriting examination for comparing Writers~7 and 8~over  different degrees of freedom. The normal model has been implemented using all character types jointly.}
\label{ds_logbf_dof_writer_7_8}
\end{figure}

Figure \ref{ds_logbf_dof} presents the average of the Bayes factors (in log scale) for the different choices of degrees of freedom. As expected, the Bayes factor is quite sensitive to the choice of degrees of freedom, which has an incremental effect on its value. For different writers comparisons, an increase of one degree of freedom in the inverse-Wishart distribution produces an increase of about 2.2 units in the log-Bayes factors (with $R^2=0.97$). Hence, the support provided by the evidence in favor of the hypothesis that the compared material originates from the same writer is (falsely) increased. This is not surprising, as the degrees of freedom play an important role in the elicitation process of the Inverse-Wishart distribution. As the degrees of freedom parameter $\nu$ increases, the prior distribution of $\bm{W}$ becomes more concentrated around a smaller region of $\mathbb{R}^p$, as discussed by \cite{gaborini2021bayesian}. Since for larger values of the degrees of freedom, the variability of the variance-covariance matrix is reduced, it can reasonably be expected that, for large values of $\nu$, the resulting log-BF will be bounded by the case where the variance-covariance matrix is constant and fixed at the prior mean of the Inverse-Wishart distribution. In support of this, one only has to look at Figure \ref{ds_logbf_dof_writer_7_8}, where the average logarithmic BFs obtained from the comparison between writers~7 and~8 are shown. As can easily be seen from the Mahalanobis distance in Figure~\ref{mahalanobis}, these writers indeed show great similarities, and it is difficult to discriminate their style. For this pair of writers, the evidence in favour of the wrong hypothesis (i.e., $H_1$) still increases with degrees of freedom, but after the value of 40, it seems to be close to reaching its maximum. Moreover, it can be observed that while for small values of the degrees of freedom (lower than 30), the Bayes factor correctly supports hypothesis $H_2$ (i.e., the compared handwritten material originate from different writers), for larger values of the degrees of freedom, the evidence is in favour of the wrong hypothesis $H_1$  (i.e., the compared handwritten material originate from the same writer).

\newpage
\subsection{Sensitivity of the parameter of the LKJ distribution}
\label{eta_lkj_appendix}

In this section, we present a series of experiments involving different writers, exploring the influence of the LKJ distribution's $\eta$ parameter. The parameter $\eta$ was assigned values of 1, 2, 5, 10, and 20, allowing for a clear view of its effect on the writers' comparison. Only values greater than 1 were considered, as these correspond to prior beliefs favoring the identity matrix. This approach is motivated by results from Fourier analysis, which establish that Fourier coefficients are uncorrelated. Consequently, employing higher values of $\eta$ emphasizes prior structures in which variables are uncorrelated within the covariance matrix.

\begin{figure}[!ht]
\centering
    \includegraphics[width=1\textwidth]{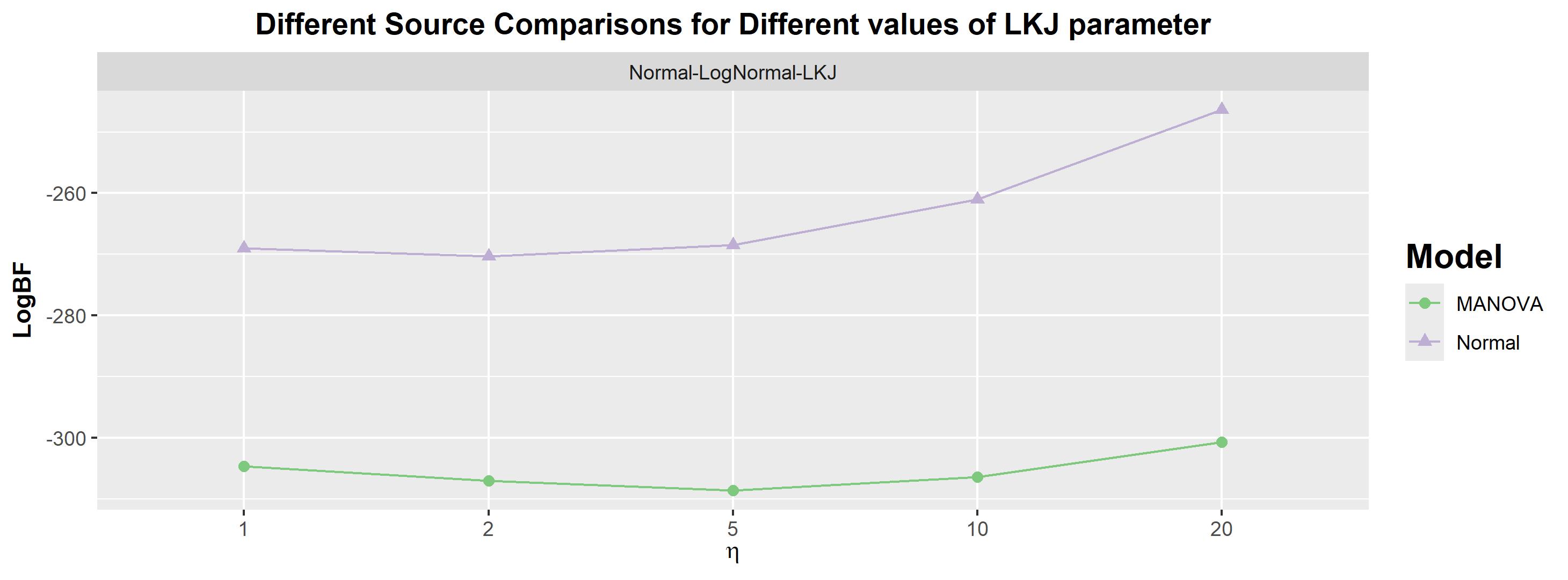}
    \caption{Average logarithmic Bayes factor ($\log\rm{BF}$) for handwriting %authentication ($\log BF_a$) for comparisons of different writers 
    examination for different writer comparisons over different $\eta$ values of the LKJ distribution. The Normal model has been implemented using all character types jointly.}
\label{ds_logbf_eta}
\end{figure}

\begin{figure}[!ht]
\centering
    \includegraphics[width=1\textwidth]{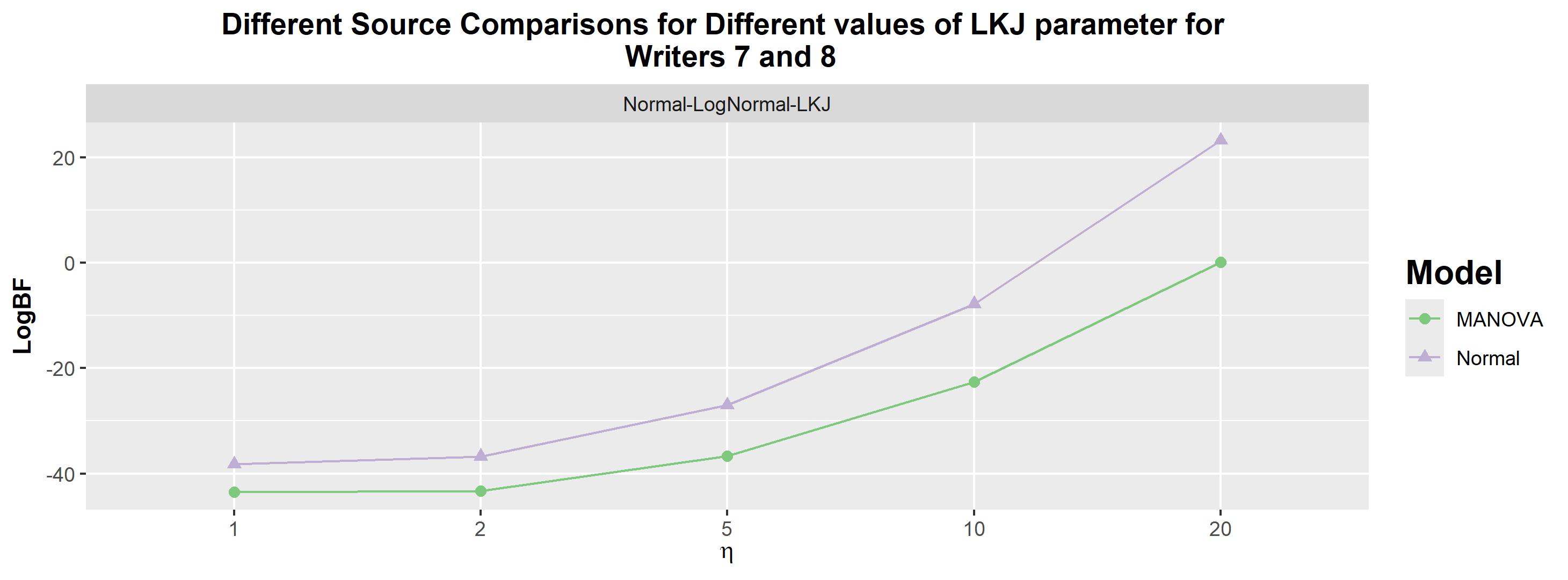}
    \caption{
    Average logarithmic Bayes factor ($\log BF$) for handwriting %authentication ($\log BF_a$) 
    examination for comparing Writers~7 and 8~over different $\eta$ values of LKJ distribution. The Normal model has been implemented using all character types jointly.}
\label{ds_logbf_eta_writer_7_8}
\end{figure}

Figure \ref{ds_logbf_eta} illustrates how the Log Bayes Factor (LogBF) varies with different values of the LKJ parameter $\eta$ for the Bayesian Normal and MANOVA models. The results show that as the LKJ parameter $\eta$ increases, the LogBF values for both models also increase. This indicates that higher values of $\eta$, which correspond to stronger prior beliefs favoring the identity matrix, tend to produce more positive LogBF values. This can affect the decisions, as we can observe in Figure \ref{ds_logbf_eta_writer_7_8}, which compares writers 7 and 8, where the LogBF values change from negative to positive as $\eta$ increases, and from value $\eta = 20$, the evidence supports the incorrect hypothesis. This demonstrates that the choice of prior, specifically the value of $\eta$ in the LKJ distribution, can substantially affect the outcome of writers' comparison.

\newpage
\section{DATA AND CODE}

All data used in this article have been kindly provided to the authors by Dr. Raymond Marquis. Due to confidentiality reasons, we cannot publicly provide access to the actual dataset of this study. For this reason, we provide an example for the IAM handwriting database and which can be found in the Git repository \url{https://tinyurl.com/lampis-tzai-github-paper}.

%For this reason, we provide only the code included in the R library BayesMuCoSoT and it can be found in the Git repository \url{https://lampis-tzai.github.io/BayesMuCoSoT/}. More specifically, in the Git repository, you can find the library site, where a web page of the library with theory and many examples is available.

\bibliographystyle{agsm}
\bibliography{library}

\end{document}